\documentclass[prd,superscriptaddress,preprintnumbers,tightenlines,nofootinbib, eqsecnum]{revtex4-2}

\usepackage{amsmath}
\usepackage{amsfonts}
\usepackage{amssymb}
\usepackage{physics}
\usepackage{bm}
\usepackage{mathrsfs}
\usepackage{graphicx}
\usepackage[colorlinks]{hyperref}
\usepackage[usenames]{color}
\usepackage{stackengine}
\usepackage{tikz}
\usepackage{diagbox}
\usepackage{mathtools}
\DeclarePairedDelimiter{\ceil}{\lceil}{\rceil}
\definecolor{darkgreen}{rgb}{0,0.5,0}
\hypersetup{urlcolor=darkgreen}
\usepackage{empheq}
\usepackage{ulem}
\usepackage{orcidlink}
\usepackage{multirow}
\usepackage{adjustbox}
 \usepackage{float}

\hypersetup{
 unicode=false,          
 pdftoolbar=true,        
 pdfmenubar=true,        
 pdffitwindow=false,     
 pdfstartview={FitH},    
 pdftitle={Constants of motion and fundamental frequencies for elliptic orbits at fourth post-Newtonian order},    
 pdfauthor={David Trestini},     
 pdfnewwindow=true,      
 colorlinks=true,       
 linkcolor=red,          
 citecolor=cyan,        
 filecolor=magenta,      
 urlcolor=darkgreen,           
 linktocpage=true
}

\allowdisplaybreaks

\DeclareSymbolFontAlphabet{\mathrsfs}{rsfs}
\DeclareMathAlphabet{\mathcal}{OMS}{cmsy}{m}{n}

\newcommand{\be}{\begin{equation}}
\newcommand{\ee}{\end{equation}}
\newcommand{\bse}{\begin{subequations}}
\newcommand{\ese}{\end{subequations}}
\newcommand{\ba}{\begin{align}}
\newcommand{\ea}{\end{align}}
\newcommand{\nn}{\nonumber}
\newcommand{\p}{\partial}

\renewcommand{\dd}{\mathrm{d}}
\newcommand{\di}{\mathrm{i}} 
\newcommand{\de}{\mathrm{e}} 
\newcommand{\dM}{\mathrm{M}}

\newcommand{\dI}{\mathrm{I}}

\newcommand{\cA}{\mathcal{A}}
\newcommand{\cB}{\mathcal{B}}
\newcommand{\cC}{\mathcal{C}}
\newcommand{\cD}{\mathcal{D}}

\makeatletter
\g@addto@macro\bfseries{\boldmath}
\makeatother

\defcitealias{BBT22}{paper~I}
\defcitealias{T24_QK}{paper~II}
\begin{document}
\interfootnotelinepenalty=10000
\title{Constants of motion and fundamental frequencies for elliptic orbits \\ at fourth post-Newtonian order}

\author{David \textsc{Trestini}\,\orcidlink{0000-0002-4140-0591}}\email{david.trestini@southampton.ac.uk}

\affiliation{School of Mathematical Sciences and STAG Research Centre, University of Southampton, Southampton SO17 1BJ, United Kingdom}

\date{\today}

\begin{abstract}
In the case of nonspinning compact binary systems on quasi-elliptic orbits, I obtain the conservative map between the constants of motion (energy and angular momentum) and the fundamental (radial and azimuthal) frequencies at the fourth post-Newtonian order, including both instantaneous and tail contributions. This map is expressed in terms of an enhancement function of the eccentricity, which is appropriately resummed to ensure accuracy for any eccentricity; in particular, I recover known results for circular orbits. In order to obtain this map, the local dynamics are expressed using an action-angle formulation. The tail term is treated as a perturbation, which is first localized in time, then Delaunay-averaged. Both operations require a contact transformation of the phase-space variables, which I explicitly control. Using the first law of binary black hole mechanics, I  then obtain the orbit-averaged redshift invariant for eccentric orbits at fourth post-Newtonian order; when properly accounting for the tail contributions, it perfectly agrees with analytical self-force at postgeodesic order~\mbox{\href{https://doi.org/10.1103/PhysRevD.106.044004}{[Phys. Rev. D 106, 044004 (2022)]}}. Finally, I use these results to reexpress the fluxes of energy and angular momentum obtained at third post-Newtonian order in~\mbox{\href{https://doi.org/10.1103/PhysRevD.77.064035}{[Phys. Rev. D 77, 064035 (2008)]}} and~\mbox{\href{https://doi.org/10.1103/PhysRevD.80.124018}{[Phys. Rev. D 80, 124018 (2009)]}} in terms of fundamental frequencies.
\end{abstract}


\maketitle
\tableofcontents

\section{Introduction}
\label{sec:intro}

Compact binaries typically arise from two formation channels: isolated formation or dynamical formation. In the isolated formation scenario, one of the stars in the stellar binary collapses into a black hole, which is then engulfed into a common envelope formed by the remaining star. Due to drag forces, the orbits then tighten significantly, until the remaining star also collapses into a black hole, forming a binary black hole.  The binary then evolves in isolation, gradually inspiralling and shedding its eccentricity due to gravitational-wave emission, such that it is essentially quasi-circular by the time it enters the LIGO-Virgo-KAGRA (LVK) frequency band. This scenario has motivated gravitational wave modelers to focus on quasi-circular binaries, and this has been mostly sufficient for the purposes of the LVK observatories. However, a certain number of binaries exhibit hints of eccentricity~\cite{Romero-Shaw:2025vbc,Morras:2025xfu,Planas:2025plq,Jan:2025fps,Kacanja:2025kpr}, and were probably formed dynamically~\cite{Zevin:2021rtf,Rodriguez:2017pec,DallAmico:2023neb,Samsing:2020tda, Romero-Shaw:2019itr}. In this scenario, a widely separated, eccentric black hole binary gets `hardened' by its interaction with a third body, leading to a more compact, highly eccentric binary. Such systems are expected to be common in future gravitational wave detectors such as Einstein Telescope, Cosmic Explorer or LISA~\cite{DallAmico:2023neb,Saini:2023wdk, Bonetti:2018tpf}. Moreover, LISA will detect many extreme and intermediate mass ratio inspirals (EMRIs and IMRIs), which will generically exhibit high eccentricity~\cite{LISAConsortiumWaveformWorkingGroup:2023arg}. IMRIs, and to a lesser extent EMRIs, can be modeled by hybridizing black hole perturbation theory (and gravitational self-force) information at leading (and subleading) orders in the mass ratio with post-Newtonian information at higher orders in the mass ratio~\cite{Honet:2025gge, Honet:2025lmk}. Analyzing signals from highly eccentric binaries using quasi-circular waveform models can lead to significant biases in the source parameters~\cite{Divyajyoti:2025cwq}, so there is now a strong motivation to improve post-Newtonian gravitational waveform models for eccentric orbits.   

In this work, I will focus on the conservative problem, ignoring dissipative radiation-reaction effects, which can be added separately in a second step. In the center-of-mass frame, a compact binary can be described gauge-invariantly in two ways. The first way is through the Noetherian constants of motion arising from the symmetries of the background Minkowski space-time. For nonspinning particles, there is no spin precession, so one can restrict to the orbital plane and characterize the binary by its energy $E$, associated with time translations, and the norm of its angular momentum $J$, associated with rotations. This description is useful because these are the quantities that enter flux-balance laws, up to small dissipative corrections, called Schott terms, that can be added separately~\cite{Trestini:2025nzr}. Such flux-balance laws are critical to control the evolution of the system under radiation-reaction at high post-Newtonian orders. The second way to describe the system is through the fundamental frequencies of motion, here the radial frequency $n$ (often denoted $\Omega_r$ in a self-force context) and the azimuthal frequency $\omega$ (or $\Omega_\phi$). These are useful because they are associated with observable effects such as the pericenter advance. Thus, in order to control the frequency (and phase) evolution at high post-Newtonian order, which is the main observable in a gravitational-wave signal, one needs to understand the link between the constants of motion and the fundamental frequencies.

In this work, I will establish this mapping at 4PN order, extending previous results at 4PN order for circular orbits~\cite{Damour:2014jta, Damour:2015isa, Damour:2016abl, Bernard:2016wrg, Bernard:2017bvn, Bernard:2017ktp}. For elliptic orbits, this relation was also previously  obtained at 2PN in~\cite{Damour:1988mr,Schafer:1993pkg} and at 3PN in~\cite{Memmesheimer:2004cv}. More recently, Ref.~\cite{Cho:2021oai} obtained the local contribution to this relation at 4PN, but did not fully account for the hereditary tail contribution. Here, I completely control the tail contribution and obtain the full 4PN result, valid for arbitrary eccentricity. Note that the expression of the Schott term for eccentric orbits, which will be needed for the 4PN phasing for elliptic orbits, is left to future work. Moreover, note that partial results have been obtained at higher PN orders~\cite{Bini:2020nsb, Bini:2020hmy, Bini:2020wpo}. 
I also study some applications of these results. First, I deduce from this map the 4PN redshift invariant for eccentric orbits using the first law of binary mechanics~\cite{LeTiec:2011ab, LeTiec:2015kgg,Blanchet:2017rcn,Blanchet:2012at}. This invariant is extremely useful to make contact with gravitational self-force, and I found perfect agreement at 4PN with the postgeodesic order results of~Ref.~\cite{Munna:2022gio}; notably, this agreement is found analytically and for arbitrary eccentricity. I was also able to re-express the 3PN fluxes of energy and angular momentum in terms of the fundamental frequencies; for this, I apply this map to the results of Refs.~\cite{Arun:2007rg,Arun:2007sg,Arun:2009mc}, which obtain these fluxes in terms of energy and angular momentum.

The structure of the paper is as follows. At the end of this introduction, I introduce relevant notation, then summarize and point to the various relations obtained in this work. In Sec.~\ref{sec:dynamics}, I provide reminders about the 4PN equations of motion, and how it derives from a Hamiltonian. In particular, I discuss the fundamental aspects of the  localization of the hereditary piece of the Hamiltonian, which will prove essential to my approach.  In Sec.~\ref{sec:local}, I derive the action-angle formulation of the \textit{local} 4PN dynamics. 
Sec.~\ref{sec:tails} is then devoted to the inclusion of the corrections due to the hereditary tail.
The hereditary contribution is expressed in terms of an enhancement function $\Lambda_0(e)$, for which I propose a very accurate resummation.
Using the full Hamiltonian  in action-angle form, I then straightforwardly obtain in Sec.~\ref{sec:frequencies} the map between the fundamental frequencies, the action variables and the conserved energy and angular momentum.
In Sec.~\ref{sec:redshift}, I use this map and the first law of binary black hole mechanics~\cite{LeTiec:2011ab, LeTiec:2015kgg, Blanchet:2017rcn} to obtain the expression of the 4PN (orbit-averaged) redshift invariant on eccentric orbits. 
In Sec.~\ref{sec:circular}, I study the circular limit of these results by requiring the radial action to vanish. 
Finally,  using the results of Refs.~\cite{Arun:2007rg,Arun:2007sg,Arun:2009mc}, I obtain in Sec.~\ref{sec:flux} the gauge invariant expressions of the orbit-averaged 3PN fluxes in terms of fundamental frequencies. I end with a discussion in Sec.~\ref{sec:discussion}.

\subsection{Notations}
\label{subsec:notations}

Spatial vectors are denoted in bold font, and their spatial components are denoted with Latin indices; e.g. $\bm{u}$ and $u^i$. Conversely, spacetime indices are denoted with Greek letters, $u^\mu = (u^0, u^i)$. The Kronecker delta is $\delta_{ij}$ and the Levi-Civita symbol is $\epsilon_{ijk}$. The $n$-th derivative of a function of time $f(t)$ is denoted $f^{(n)}(t) = \dd^n f / \dd t^n$. Orbit-averages are denoted $\langle f \rangle$, whereas the oscillatory piece is $\widetilde{f}$ such that $f = \langle f \rangle+\widetilde{f}$.

Two compact objects of masses $m_1$ and $m_2$ are considered. The total mass is  $m=m_1+m_2$,  the reduced mass is $\mu = m_1 m_2/m$ and the symmetric mass ratio is $\nu = m_1 m_2 / m ^2 = \mu/m \in\  ]0, 1/4]$.
Moreover, I follow the conventions of Ref.~\cite{Blanchet:2013haa} for the masses: I choose to label the masses such that $m_1 \ge m_2$, and thus define the relative mass difference as \mbox{$\delta = (m_1 - m_2)/m = \sqrt{1-4 \nu} \ge 0$}. The ordinary (small) mass ratio is then $\epsilon=m_2/m_1$.

Each object is associated (in ADM coordinates) with the position vectors $\bm{y}_1(t)$ and $\bm{y}_2(t)$. Similarly, for $A\in\{1,2\}$, one defines the velocity $\bm{v}_A = \dd \bm{y}_A/\dd t$, the acceleration $\bm{a}_A = \dd \bm{v}_A/\dd t$ and the jerk $\bm{b}_A = \dd \bm{a}_A/\dd t$; the \textit{relative} separation is then $\bm{x}_{12} = \bm{y}_1-\bm{y}_2$, the relative velocity is $\bm{v}_{12} = \bm{v}_1-\bm{v}_2$  and the relative acceleration is $\bm{a}_{12} = \bm{a}_1-\bm{a}_2$. The norm of the separation vector is denoted by $r_{12} = |\bm{x}_{12}|$, alongside the unit vector $\bm{n}_{12} = \bm{x}_{12}/r_{12}$. 

When expressing quantities in the center-of-mass frame, the ``$12$'' label is typically dropped. One then also typically describes the motion using spherical coordinates $(r,\phi,\theta)$. In the orbital plane (characterized by $\theta = \pi/2$), the separation vector is decomposed as $\bm{x}=(r\cos \phi, r\sin\phi, 0)$. The conjugate momentum associated with $\bm{x}$ is denoted $\bm{p}$, whereas the momenta associated with $(r,\phi)$ are $(p_r, p_\phi)$. 

The radial action is denoted $I_r$, the azimuthal action $I_\phi$, the angular momentum $J$ and the energy $E$ (without the rest mass $m$). Note that here, one indeed has $J = I_\phi$. I also introduce the Delaunay variable \mbox{$I_{r\phi} = I_r+I_\phi$}. All these conserved quantities have reduced counterparts. The reduced energy is given by \mbox{$\varepsilon=-2E/(m\nu c^2)$} and the reduced angular momentum is \mbox{$j = -2J^2 E/(G^2 m^5 \nu^3)$}; they are such that $\varepsilon = \mathcal{O}(1/c^2)$ and $j = \mathcal{O}(1)$. The reduced action variables are given by \mbox{$i_r = I_r/(G m^2 \nu)$}, \mbox{$i_{\phi} = I_{\phi}/(G m^2 \nu)$}, and \mbox{$i_{r\phi} = I_{r\phi}/(G m^2 \nu)$}.

In the case of planar motion, one is interested in the two fundamental frequencies: the radial frequency $n$ and the azimuthal frequency $\omega$. These are associated with the radial period \mbox{$P=2\pi/n$} and the periastron advance \mbox{$K = \omega/n$}. I also introduce the very useful parameters \mbox{$x=(G m \omega/c^3)^{2/3}$} and \mbox{$\iota = 3x/(K-1)$},  first introduced by Blanchet in Ref.~\cite{Arun:2007sg}. These parameters are chosen such that at leading post-Newtonian order, $x \sim \varepsilon$ and $\iota \sim j$. In the context of comparisons with gravitational self-force, I also introduce $y = (G m_1 \omega/c^3)^{2/3}$ and $\lambda= 3 y/(K-1)$. In the Newtonian problem, I often use the semi-major axis $a$ and the eccentricity $e$.


\subsection{Summary of results}
\label{subsec:summary}

For the reader's convenience,  references to the various 4PN-accurate maps derived in this paper are provided here. For various quantities expressed in terms of $(i_{r\phi},i_r)$, $(\varepsilon,j)$ or $(x,\iota)$ --- refer to Table.~\ref{tab:summary_eccentric}. For the circular links between these quantities, refer to Table.~\ref{tab:circular_link}. Lengthy results in this paper are provided in machine readable form in the Supplemental Material~\cite{Supplemental}.

\begin{table}[hbt!]
    \centering
    \begin{tabular}{|c|c|c|c|}\hline
     \diagbox{Quantity}{In terms of} & $(i_{r\phi},i_\phi)$ & $(\varepsilon,j)$ & $(x,\iota)$ \\ \hline
     $\varepsilon$ &~\eqref{eq:H_inTermsOf_irphi_iphi_sum_loc_log_hered} & &~\eqref{seq:varepsilon_inTermsOf_x_iota_sum_loc_log_hered} \\ 
     $\varepsilon^\text{loc}$/$\varepsilon^\text{log}$/$\varepsilon^\text{hered}$ &~\eqref{eq:H_loc_inTermsOf_irphi_iphi}/\eqref{seq:H_log_inTermsOf_irphi_iphi}/\eqref{seq:H_hered_inTermsOf_irphi_iphi} & &~\eqref{seq:varepsilon_loc_inTermsOf_x_iota}/\eqref{seq:varepsilon_log_inTermsOf_x_iota}/\eqref{seq:varepsilon_hered_inTermsOf_x_iota}\\ \hline
     $j$ &~\eqref{eq:iphi_inTermsOf_varepsilon_j} & &~\eqref{seq:j_inTermsOf_x_iota_sum_loc_log_hered}  \\ 
     $j^\text{loc}$/$j^\text{log}$/$j^\text{hered}$ & & &~\eqref{seq:j_loc_inTermsOf_x_iota}/\eqref{seq:j_log_inTermsOf_x_iota}/\eqref{seq:j_hered_inTermsOf_x_iota}  \\ \hline
     $i_r$ & &~\eqref{seq:ir_inTermsOf_varepsilon_j_sum_loc_log_hered} & \\ 
     ${i_r}^\text{loc}$/${i_r}^\text{log}$/${i_r}^\text{hered}$ & &~\eqref{eq:ir_loc_inTermsOf_varepsilon_j} /~\eqref{seq:irphi_log_inTermsOf_varepsilon_j} /~\eqref{seq:irphi_hered_inTermsOf_varepsilon_j} & \\ \hline
     $i_{r\phi}$ & &~\eqref{seq:irphi_inTermsOf_varepsilon_j_sum_loc_log_hered}  & \\ 
     ${i_{r\phi}}^\text{loc}$/${i_{r\phi}}^\text{log}$/${i_{r\phi}}^\text{hered}$ & &~\eqref{eq:irphi_loc_inTermsOf_varepsilon_j} /~\eqref{seq:irphi_log_inTermsOf_varepsilon_j} /~\eqref{seq:irphi_hered_inTermsOf_varepsilon_j}  & \\ \hline 
     $n$ &~\eqref{seq:n_inTermsOf_irphi_iphi_sum_loc_log_hered} &~\eqref{seq:n_inTermsOf_varepsilon_j_sum_loc_log_hered} &~\eqref{eq:x_iota_inTermsOf_n_omega} \\ 
     $n^\text{loc}$/$n^\text{log}$/$n^\text{hered}$ &~\eqref{seq:n_loc_inTermsOf_irphi_iphi}/~\eqref{seq:n_log_inTermsOf_irphi_iphi} /\eqref{seq:n_hered_inTermsOf_irphi_iphi} &~\eqref{seq:n_loc_inTermsOf_varepsilon_j}/~\eqref{seq:n_log_inTermsOf_varepsilon_j} /\eqref{seq:n_hered_inTermsOf_varepsilon_j} & \\ \hline
     $\omega$ &~\eqref{seq:omega_inTermsOf_irphi_iphi_sum_loc_log_hered} &~\eqref{seq:omega_inTermsOf_varepsilon_j_sum_loc_log_hered} &~\eqref{eq:x_iota_inTermsOf_n_omega} \\ 
     $\omega^\text{loc}$/$\omega^\text{log}$/$\omega^\text{hered}$ &~\eqref{seq:omega_loc_inTermsOf_irphi_iphi}/~\eqref{seq:omega_log_inTermsOf_irphi_iphi} /\eqref{seq:omega_hered_inTermsOf_irphi_iphi} &~\eqref{seq:omega_loc_inTermsOf_varepsilon_j}/~\eqref{seq:omega_log_inTermsOf_varepsilon_j} /\eqref{seq:omega_hered_inTermsOf_varepsilon_j} & \\  \hline 
     $P$ & &~\eqref{seq:P_inTermsOf_varepsilon_j_sum_loc_log_hered} & \\ 
     $P^\text{loc}$/$P^\text{log}$/$P^\text{hered}$ & &~\eqref{seq:P_loc_inTermsOf_varepsilon_j}/~\eqref{seq:P_log_inTermsOf_varepsilon_j} /\eqref{seq:P_hered_inTermsOf_varepsilon_j} & \\  \hline 
     $K$ & &~\eqref{seq:K_inTermsOf_varepsilon_j_sum_loc_log_hered}  & \\ 
     $K^\text{loc}$/$K^\text{log}$/$K^\text{hered}$ & &~\eqref{seq:K_loc_inTermsOf_varepsilon_j}/~\eqref{seq:K_log_inTermsOf_varepsilon_j} /\eqref{seq:K_hered_inTermsOf_varepsilon_j} & \\  \hline 
     $x$ & &~\eqref{seq:x_inTermsOf_varepsilon_j_sum_loc_log_hered} & \\ 
     $x^\text{loc}$/$x^\text{log}$/$x^\text{hered}$ & &~\eqref{seq:x_loc_inTermsOf_varepsilon_j}/~\eqref{seq:x_log_inTermsOf_varepsilon_j} /\eqref{seq:x_hered_inTermsOf_varepsilon_j} &  \\  \hline 
     $\iota$ & &~\eqref{seq:iota_inTermsOf_varepsilon_j_sum_loc_log_hered} & \\ 
     $\iota^\text{loc}$/$\iota^\text{log}$/$\iota^\text{hered}$ & &~\eqref{seq:iota_loc_inTermsOf_varepsilon_j}/~\eqref{seq:iota_log_inTermsOf_varepsilon_j} /\eqref{seq:iota_hered_inTermsOf_varepsilon_j} & \\  \hline 
     $\langle z_1 \rangle$ & &~\eqref{eq:redshift_inTermsOf_varepsilon_j_sum_loc_log_hered} &~\eqref{eq:redshift_inTermsOf_x_iota_sum_loc_log_hered} \\ 
     $\langle z_1^\text{loc} \rangle$/$\langle z_1^\text{log} \rangle$/$\langle z_1^\text{hered} \rangle$ & &~\eqref{seq:redshift_loc_inTermsOf_varepsilon_j}/\eqref{seq:redshift_log_inTermsOf_varepsilon_j}/\eqref{seq:redshift_hered_inTermsOf_varepsilon_j}&~\eqref{seq:redshift_loc_inTermsOf_x_iota}/\eqref{seq:redshift_log_inTermsOf_x_iota}/\eqref{seq:redshift_hered_inTermsOf_x_iota}\\  \hline
      $\langle\mathcal{F} \rangle$ & &~\eqref{seq:energy_flux_inTermsOf_varepsilon_j} &~\eqref{seq:energy_flux_inTermsOf_x_iota} \\ \hline
      $\langle\mathcal{G} \rangle$ & &~\eqref{seq:angular_momentum_flux_inTermsOf_varepsilon_j} &~\eqref{seq:angular_momentum_flux_inTermsOf_x_iota} \\ \hline
    \end{tabular}
    \caption{\label{tab:summary_eccentric}Summary of the expressions obtained in the case of eccentric orbits for various quantities in terms of (i) action variables $(i_{r\phi}, i_\phi)$; (ii) the reduced energy and angular momentum $(\varepsilon,j)$; and (iii) the dimensionless frequencies $(x,\iota)$.}
\end{table}

\begin{table}[h!]
    \centering
    \begin{tabular}{|c||c|c|c|c|}
    \hline
        Quantity & Full result & Local part & Logarithmic part & Hereditary part \\ \hline \hline
        $j_\text{circ}(\varepsilon)$ &~\eqref{eq:jcirc_inTermsOf_varepsilon} &~\eqref{seq:jcirc_loc_inTermsOf_varepsilon}
 &~\eqref{seq:jcirc_log_hered_inTermsOf_varepsilon}
 &~\eqref{seq:jcirc_log_hered_inTermsOf_varepsilon}
 \\ \hline
        $\iota_\text{circ}(x)$ &~\eqref{seq:iota_circ_inTermsOf_x} &~\eqref{seq:iota_circ_loc_inTermsOf_x} &~\eqref{seq:iota_circ_log_hered_inTermsOf_x} &~\eqref{seq:iota_circ_log_hered_inTermsOf_x} \\ \hline
        $K_\text{circ}(x)$ &~\eqref{seq:K_circ_inTermsOf_x} &~\eqref{seq:K_circ_loc_inTermsOf_x} &~\eqref{seq:K_circ_log_hered_inTermsOf_x} &~\eqref{seq:K_circ_log_hered_inTermsOf_x} \\ \hline
        $\langle z_1^\text{circ}\rangle(\varepsilon) $&~\eqref{seq:redshift_circ_inTermsOf_varepsilon}  & & &  \\ \hline
        $\langle z_1^\text{circ}\rangle(x)$ &~\eqref{seq:redshift_circ_inTermsOf_x} & & & \\ \hline
    \end{tabular}
    \caption{Summary of the circular links obtained in Sec.~\ref{sec:circular}.}
    \label{tab:circular_link}
\end{table}

\section{Hamiltonian formulation for the 4PN dynamics of compact binaries}
\label{sec:dynamics}

\subsection{The 4PN equations of motion and their derivation from an action principle}
\label{subsec:EOM}

The most straightforward description of the motion of a compact binary is through the equations of motion. In a given coordinate system, at any time (described by coordinate time $t$), each particle (labeled $A$) is assigned a position vector $\bm{y}_A(t)$ and an associated velocity vector $\bm{v}_A(t)$. Generically, post-Newtonian motion is then constrained by an equation of the form
\begin{align}
    \label{eq:acceleration_functional}
    \bm{a}_A \equiv \frac{\dd \bm{v}_A}{\dd t} &= \mathfrak{F}[\bm{y}_1, \bm{y}_2, \bm{v}_1, \bm{v}_2] \,,
\end{align}
where $\bm{a}_A$ denotes the acceleration of particle $A\in\{1,2\}$ and $\mathfrak{F}$ is some \textit{functional} of the phase-space trajectories \mbox{$t \mapsto \bm{y}_1(t)$}, \mbox{$t \mapsto \bm{y}_2(t)$}, \mbox{$t \mapsto \bm{v}_1(t)$} and \mbox{$t \mapsto \bm{v}_2(t)$}. It is sufficient to determine $\bm{a}_1$, because $\bm{a}_2$ is then determined by switching the labels $1 \leftrightarrow 2$. Up to 3.5PN order, $\mathfrak{F}$ is in fact an ordinary function of the positions and velocities, such that the equations of motion~\eqref{eq:acceleration_functional} reduce to an ordinary differential equation on the phase-space vector $(\bm{y}_1,\bm{y}_2,\bm{v}_1,\bm{v}_2)$. However, it was first shown in~\cite{Blanchet:1987wq} that, at 4PN, $\mathfrak{F}$ contains a piece which can be expressed as an integral over the past history of the binary (thus preserving causality), and~\eqref{eq:acceleration_functional} then becomes an \textit{integro-differential} equation. This feature is referred to as \textit{hereditary} or \textit{non-local-(in-time)}, and the particular integral appearing is named a \textit{tail}. 
In ADM coordinates, the equations of motion read
%
\begin{align}\label{eq:a1_split_inst_tail}
    \bm{a}_1 = \bm{a}_{1}^\text{loc} + \bm{a}_{1}^{\text{tail}} \,,
\end{align}
where $\bm{a}_{1}^\text{loc}$ is some complicated instantaneous function of the phase-space variables, and where the tail contribution reads~\cite{Bernard:2017ktp}
\begin{align}\label{eq:a1_tail}
 a_{1,\text{tail}}^i &= - \frac{8 G^2 \dM}{5 c^8} \Bigg\{ y_1^j \int_{0}^{\infty}\dd\tau\, \ln\left(\frac{c \tau}{2 b_0}\right)\,\dI_{ij}^{(7)}(t-\tau)   - y_1^i  \bigg[\dI_{ij}^{(3)}\, \ln \left(\frac{r_{12}}{b_0}\right)\bigg]^{(3)} + \frac{n_{12}^i}{4m_1 r_{12}} \dI_{jk}^{(3)}\dI_{jk}^{(3)} \Bigg\}\,,
\end{align}
where $\dM = m + \mathcal{O}(1/c^2)$ is the ADM mass and $b_0$ is an arbitrary constant length scale which cancels out in the full acceleration.
Here, $\dI_{ij}^{(n)}$ is the $n$-th time-derivative of the source quadrupole moment, which is given at leading (Newtonian) order by
\begin{align}\label{eq:Iij_N_expression}
    \dI_{ij} = \sum_{A\in\{1,2\}} m_A y_A^{\langle i} y_A^{j \rangle} + \mathcal{O}\left(\frac{1}{c^2}\right) \,.
\end{align}
When taking time derivatives of the quadrupole moment, accelerations appear, which can be straightforwardly order-reduced using the Newtonian equations of motion (one of course neglects 5PN corrections). 
Both the instantaneous and tail pieces of the acceleration contain conservative contributions (leading to effects such as the periastron advance) and dissipative contributions (leading to the inspiral of the binary). In the instantaneous piece, the conservative contributions come with an integer PN order (Newtonian, 1PN, 2PN, 3PN, and 4PN) whereas the dissipative pieces come with a half-integer PN order (2.5PN and 3.5PN). In the tail piece, such a split does not occur, but these two contributions can be disentangled by dividing this integral in time-even and time-odd contributions, which read~\cite{Bernard:2017ktp}  
\begin{widetext}
\begin{subequations}\label{eq:a1_tail_cons_diss}\begin{align}
\label{seq:a1_tail_cons}
  a_{1,\text{tail, cons}}^i &= - \frac{4 G^2 \dM}{5 c^8} \Bigg\{ y_1^j \int_{0}^{\infty}\dd\tau\, \ln\left(\frac{c \tau}{2 b_0}\right)\,\left[\dI_{ij}^{(7)}(t-\tau) - \dI_{ij}^{(7)}(t+\tau) \right]      - 2 y_1^i \bigg[\dI_{ij}^{(3)}\, \ln \left(\frac{r_{12}}{b_0}\right)\bigg]^{(3)}+ \frac{n_{12}^i}{2m_1 r_{12}} \dI_{jk}^{(3)}\dI_{jk}^{(3)} \Bigg\}\,,\\*
\label{seq:a1_tail_diss}
     a_{1,\text{tail, diss}}^i &= - \frac{4 G^2 \dM}{5 c^8} y_1^j \int_0^{+\infty} \dd\tau\, \ln\left(\frac{c\tau}{2 b_0}\right) \left[\dI_{ij}^{(7)}(t-\tau) + \dI_{ij}^{(7)}(t+\tau) \right] \,. 
\end{align}\end{subequations}
\end{widetext}
Since this work is only concerned with conservative effects, I will only consider the conservative acceleration
\begin{align}\label{eq:a1_cons_split_inst_tail}
    \bm{a}_{1}^{\text{cons}}& = \bm{a}_{1}^\text{inst, cons} +  \bm{a}_{1}^\text{tail, cons} \,,
\end{align}
where $ \bm{a}_{1}^\text{inst, cons}$ is simply obtained by keeping only the integer PN orders (i.e., even powers of $1/c$) in $ \bm{a}_{1}^\text{inst}$.

Although the 4PN tail contribution was initially obtained \textit{via} the metric arising from an asymptotic matching between the near and far zone~\cite{Blanchet:1985sp,Blanchet:1987wq}, it was later found that the conservative acceleration, including the tail term, derives from an action principle. This action can either be constructed in the ADM formalism~\cite{Damour:2014jta} (where the hereditary part is guessed from the equations of motion) or derived entirely from scratch using the Fokker action~\cite{Bernard:2015njp, Bernard:2016wrg, Bernard:2017ktp,Bernard:2017bvn,Marchand:2017pir} or EFT methods~\cite{Foffa:2011np,Galley:2015kus}. The full action for the relative motion in the center-of-mass frame can be split as $S = S_\text{loc} + S_\text{tail}$. The
local piece initially depends not only on the positions and velocities, but also on the accelerations $\bm{a}_A$ and even the jerks $\bm{b}_A = \dd \bm{a}_A/\dd t$. However, terms that depend on $\bm{b}_A$ are eliminated as total derivatives in the Lagrangian~\cite{Bernard:2015njp}, and terms which are at least quadratic in the acceleration are eliminated using the ``double-zero'' method~\cite{Damour:1985mt, Barker:1980kef}, such that one is left with an action which is at most linear in the accelerations --- this last residual dependence in the accelerations is eliminated by suitable coordinate shifts~\cite{Bernard:2015njp}. Thus, the local action finally reads 
\begin{subequations}\label{eq:S_inst_and_S_tail}
\begin{align}
\label{seq:S_inst}
    S_\text{loc} &= \int_{-\infty}^{+\infty} \dd t \,  L_\text{loc}\Bigl(\bm{x}(t), \bm{v}(t)\Bigr) \,. 
\end{align}
The tail piece then reads
\begin{align}
 \label{seq:S_tail}
      S_\text{tail} &= \frac{G^2 \dM}{5c^8}\  \mathrm{Pf}_{\frac{2r_{12}(t)}{c}} \int_{-\infty}^{+\infty} \dd t \int_{-\infty}^{+\infty}  \frac{\dd t' }{|t-t'|} \dI_{ij}^{(3)}(t)\dI_{ij}^{(3)}(t') \,,
\end{align}
\end{subequations}
where  $\mathrm{Pf}_{\tau_0}$ denotes the Hadamard \textit{partie finie}
regularization\footnote{\label{footnote:Pf}
The \textit{partie finie}~\cite{Hadamard,Blanchet:2000nu, Damour:2014jta}  is defined in terms of a scale $\tau_0$, and it is introduced to make the action explicitly symmetric in $t \leftrightarrow t'$. Here, it will suffice to know that, for any $f(t)$ which tends sufficiently fast to $0$ in the $t \rightarrow \pm \infty$ limit, one has the identity~\cite{Bernard:2015njp}
\begin{align*}
    \mathrm{Pf}_{\tau_0}  \int_{- \infty}^{+\infty} \dd t' \frac{f(t')}{|t-t'|} = \int_0^{+\infty} \dd \tau \ln\left(\frac{\tau}{\tau_0}\right)\bigg[f^{(1)}(t-\tau) - f^{(1)}(t+\tau)\bigg]\,.
\end{align*}
}
of the integral associated with the timescale $\tau_0$  and where $\dI_{ij}^{(3)}$ here denotes the third time derivative of $\dI_{ij}$. 
For now, these derivatives should be performed without replacing the accelerations by the equations of motion, namely  \mbox{$\dI_{ij}^{(3)} = 2\sum_A m_A (3 v_A^{\langle i} a_A^{j\rangle} + y_A^{\langle i} b_A^{j \rangle})$}. One can then vary~\eqref{seq:S_tail} with respect to the first particle, namely \mbox{$\bm{y}_1(t) \rightarrow \bm{y}_1(t) + \delta \bm{y}_1(t)$}, \mbox{$\bm{v}_1(t) \rightarrow \bm{v}_1(t) + \delta \bm{v}_1(t)$}, \mbox{$\bm{a}_1(t) \rightarrow \bm{a}_1(t) + \delta \bm{a}_1(t)$}, etc.; using the identity of Footnote~\ref{footnote:Pf}, it follows that
\begin{subequations}
\label{eq:delta_S_tail}
\begin{align}\label{seq:delta_S_tail_a}
\delta S_\text{tail} &= \frac{2 G^2 \dM}{5c^8}  \int_{-\infty}^{+\infty} \dd t\ \Biggl\{ -\frac{\delta r_{12}(t)}{r_{12}(t)}  \dI_{ij}^{(3)}(t)  \dI_{ij}^{(3)}(t) - 2 \ \delta \dI_{ij}^{(3)}\!(t)\,  \dI_{ij}^{(3)}(t)  \ln\left( \!\frac{r_{12}(t)}{b_0} \!\right)  +  \delta \dI_{ij}^{(3)}\!(t) \  \mathrm{Pf}_{\frac{2 b_0}{c}}    \int_{-\infty}^{+\infty}  \frac{\dd t' }{|t-t'|} \,\dI_{ij}^{(3)}(t')    \Biggr\} \nonumber\\*
&= \frac{2 G^2 \dM}{5c^8}  \int_{-\infty}^{+\infty} \dd t\ \Biggl\{  -\frac{\delta r_{12}(t)}{r_{12}(t)}  \dI_{ij}^{(3)}(t)  \dI_{ij}^{(3)}(t)  +  2 \, \delta\dI_{ij} \frac{\dd^3}{\dd t^3}\bigg[\dI_{ij}^{(3)}(t) \ln \left(\frac{r_{12}(t)}{b_0}\right)\bigg]  \nonumber\\*
& \qquad\qquad\qquad\qquad\qquad - \delta \dI_{ij} \int_0^{+\infty}\dd \tau\, \ln\left(\frac{c\tau}{2b_0}\right)\bigg[ \dI_{ij}^{(7)}(t-\tau) - \dI_{ij}^{(7)}(t+\tau) \bigg]\Biggr\}\,,
\end{align}
where   the second line has been obtained by integrating by parts. Now, using $\delta r_{12} = \delta y_1^k n_{12}^k$ and $\delta\dI_{ij} = 2 m_1 \delta y_1^{\langle i} y_1^{j \rangle}$, the variation of the action can be rewritten as 
\begin{align}\label{seq:delta_S_tail_b}
\delta S_\text{tail} &= \frac{2 G^2 \dM\,m_1}{5c^8}  \int_{-\infty}^{+\infty} \dd t\ \delta y_1^{i} \Biggl\{- \frac{n_{12}^i}{m_1 \,r_{12}}  \dI_{jk}^{(3)}(t)\, \dI_{jk}^{(3)}(t) + 4  y_1^{j}  \frac{\dd^3}{\dd t^3}\bigg[\dI_{ij}^{(3)}(t) \ln \left(\frac{r_{12}(t)}{b_0}\right)\bigg]  \nonumber\\
& \qquad\qquad\qquad\qquad\qquad - 2  y_1^{j} \int_0^{+\infty}\dd \tau\, \ln\left(\frac{c\tau}{2b_0}\right)\bigg[ \dI_{ij}^{(7)}(t-\tau) - \dI_{ij}^{(7)}(t+\tau) \bigg] \Biggr\}\,.
\end{align}
\end{subequations}
One recovers the equations of motion by asking that \mbox{$\delta S = \delta S_{\text{loc}} + \delta S_{\text{tail}} = 0$}. Recall that 
\begin{align}\label{eq:S_Newtonian}
    S_\text{loc} = \int_{-\infty}^{\infty}\dd t \,\bigg(\frac{m_1 v_1^2}{2}+ \frac{m_2 v_2^2}{2}+\frac{Gm_1 m_2}{r_{12}}\bigg) + \mathcal{O}\left(\frac{1}{c^2}\right)
\end{align}
and that its variation (with respect to particle $1$) reads
\begin{align}\label{eq:delta_S_Newtonian}
    \delta S_\text{loc} = -  m_1 \int_{-\infty}^{\infty} \dd t \, \delta y_1^i \left(a_1^i + \frac{G  m_2 \,n_{12}^i}{r_{12}^2}\right)  + \mathcal{O}\left(\frac{1}{c^2}\right)\,;
\end{align}
thus, the expression~\eqref{seq:a1_tail_cons} for $\bm{a}_{1,\text{tail}}^\text{cons}$ is exactly recovered from~\eqref{eq:delta_S_Newtonian} and~\eqref{seq:delta_S_tail_b}.

\subsection{Formalism for constructing the 4PN Hamiltonian and treatment of the hereditary piece}
\label{subsec:4PN_Hamiltonian_formalism}
The goal is now to construct a Hamiltonian from the  action \eqref{eq:S_inst_and_S_tail}. 
Recall that initially,  the derivatives of the quadrupole moment entering the tail action~\eqref{seq:S_tail} should be performed without replacing the accelerations by the equations of motion. However, one can subsequently perform a coordinate shift, given explicitly in (5.15) of~\cite{Bernard:2015njp}; the action in the new coordinates can then be obtained by `naively' using the order-reduced expression, given in~(5.13)~of~\cite{Bernard:2015njp}. For convenience, a coordinate transformation to the center-of-mass frame is then performed (this procedure can introduce extra non-localities at 4.5PN order~\cite{Blanchet:2024loi}), in which case the equations of motion can be written in terms of the relative position $\bm{x} =  \bm{y}_1 - \bm{y}_2$, velocity $\bm{v} = \bm{v}_1 - \bm{v}_2$ and acceleration $\bm{a} =\bm{a}_1 - \bm{a}_2$; the relative position vector is decomposed as $r = |\bm{x}|$ and $\bm{n} = \bm{x}/r$.
The Hamiltonian is then extracted from this action by writing~\cite{Bernard:2015njp,Bernard:2016wrg}
\begin{align}\label{eq:S_with_Hamiltonian}
    S = \int_{-\infty}^{+\infty} \dd t \Biggl[ \bm{p}(t) \cdot \bm{v}(t) - H^\text{loc}(\bm{x}(t),\bm{p}(t)) - H^\text{tail}[\bm{x},\bm{p}]\Biggr] \,,
\end{align}
where one immediately identifies $H^\text{loc}(\bm{x},\bm{p}) = \bm{p} \cdot\bm{v}-L_\text{loc}(\bm{x},\bm{v}(\bm{x},\bm{p}))$ and 
\begin{align}\label{eq:H_tail}
    H^\text{tail}[\bm{x},\bm{p}]=-\frac{G^2 \dM}{5c^8} \dI_{ij}^{(3)}(t) \mathcal{T}_{ij}^{(3)}\,.
\end{align}
For any $N \in\mathbb{N}$, I have introduced~\cite{Blanchet:2017rcn} 
\begin{align}\label{eq:Tijn_def}
\mathcal{T}_{ij}^{(N)} &= \mathrm{Pf}_{\frac{2r(t)}{c}}  \int_{-\infty}^{+\infty}  \frac{\dd t' }{|t-t'|} \dI_{ij}^{(N)}(t') \nonumber \\*
 &=- 2\, \dI_{ij}^{(N)} \ln \left(\frac{r(t)}{b_0}\right) + \int_0^{\infty}  \dd \tau\ln\left(\frac{c\tau}{2 b_0}\right)\Big[\dI_{ij}^{(N+1)}(t-\tau) - \dI_{ij}^{(N+1)}(t+\tau)\Big] \,,  
\end{align}
where the order-reduced derivatives of the quadrupole moment were computed at leading order in (3.4) of~\cite{Bernard:2016wrg} and read (after order reduction) 
\begin{subequations}
\label{eq:dt3Iij_dt4Iij_momenta}
\begin{align}\label{seq:dt3Iij_momenta}
\dI_{ij}^{(3)} &= \frac{2G m}{r^2}\left(3 p_r n^{\langle i} n^{j \rangle} - 4 n^{\langle i}p^{j \rangle}\right) \,, \\
\label{seq:dt4Iij_momenta}
\dI_{ij}^{(4)} &= \frac{2 G}{r^3 \nu}\Biggl\{\left[3 \bm{p}^2 - 15 p_r^2 + \frac{G m^3 \nu^2}{r}\right]n^{\langle i} n^{j
 \rangle} + 18 p_r n^{\langle i } p^{j \rangle} - 4 p^{\langle i} p^{j \rangle}\Biggr\} \,.
\end{align}\end{subequations}
In order to obtain (the generalization of) Hamilton's equations, one should again vary the action, but this time with respect to both~$\bm{x}$ and~$\bm{p}$. This leads to the following equations (in polar coordinates):\footnote{\label{footnote:factor_two_convention}In many works of the PN literature~\cite{Bernard:2016wrg, Blanchet:2017rcn, Blanchet:2025agj}, the factor $2$ appearing in~\eqref{eq:Hamilton_equations} is conventionally moved into the definition of the variation formula~\eqref{eq:H_tail_variation_formulas}. Here, I choose conventions which agree with~\cite{Isoyama:2014mja,Fujita:2016igj,Isoyama:2018sib,Blanco:2022mgd, Blanco:2023jxf, Blanco:2024fte,Lewis:2025ydo}; see Footnote~\ref{footnote:factor_two_convention_2}.}
\begin{align}\label{eq:Hamilton_equations}
        \dot{r} &= \frac{\partial H^\text{loc}}{\partial p_r} +2 \  \frac{\delta  H^\mathrm{tail}}{\delta p_r(t)} \,,&
        \dot{\phi} &=  \frac{\partial H^\text{loc}}{\partial p_\phi} + 2 \  \frac{\delta  H^\mathrm{tail}}{\delta p_\phi(t)}  \,,\nonumber\\
        \dot{p}_r &= -\frac{\partial H^\text{loc}}{\partial r} - 2 \  \frac{\delta H^\text{tail}}{\delta r(t)}  \,, &
        \dot{p}_\phi &= -\frac{\partial H^\text{loc}}{\partial \phi}  - 2 \  \frac{\delta H^\text{tail}}{\delta \phi(t)}  \,,
\end{align}
where the notation $\delta/\delta r(t)$ indicates that the  differentiation is with respect to $r(t)$ but not $r(t')$ (and similarly for $\phi$, $p_r$, and $p_\phi$); this arises from the doubling of the phase-space in such pseudo-Hamiltonian systems, or equivalently, from the $t \leftrightarrow t'$ symmetry of the tail action; see, e.g.,~(11)~of~\cite{Blanco:2024fte}. The variational formulas then read${}^\text{\ref{footnote:factor_two_convention}}$
\begin{align}\label{eq:H_tail_variation_formulas}
\frac{\delta H^\text{tail}}{\delta p_r(t)} &= - \frac{G^2 \dM}{5 c^8} \frac{\partial \dI_{ij}^{(3)}}{\partial p_r} \mathcal{T}_{ij}^{(3)} \,, & \frac{\delta H^\text{tail}}{\delta p_\phi(t)} &= - \frac{G^2 \dM}{5 c^8} \frac{\partial \dI_{ij}^{(3)}}{\partial p_\phi} \mathcal{T}_{ij}^{(3)} \,, \nonumber\\*
\frac{\delta H^\text{tail}}{\delta r(t)} &= - \frac{G^2 \dM}{5 c^8} \Bigl[\frac{\partial \dI_{ij}^{(3)}}{\partial r} \mathcal{T}_{ij}^{(3)} - \frac{1}{r}\dI_{ij}^{(3)}\dI_{ij}^{(3)} \Bigr] \,,& \frac{\delta H^\text{tail}}{\delta \phi(t)} &= - \frac{G^2 \dM}{5 c^8 } \frac{\partial \dI_{ij}^{(3)}}{\partial \phi} \mathcal{T}_{ij}^{(3)} \,.
\end{align}
It is worth stressing that the explicit $t \leftrightarrow t'$ symmetry of the action (when formulating it with a \textit{partie finie}) leads to dynamical equations that do not involve the variation of the nonlocal tail term $\mathcal{T}_{ij}^{(3)}$, which greatly simplifies the study of these systems; the resulting equations of motion are still integro-differential, though.
This symmetry is actually always present in \textit{pseudo-Hamiltonian} systems in the sense of Refs.~\cite{Isoyama:2014mja,Fujita:2016igj,Isoyama:2018sib,Blanco:2022mgd, Blanco:2023jxf, Blanco:2024fte,Lewis:2025ydo}, and plays an important role in their study.

The motivation for constructing a Hamiltonian is to find constants of motion, which are conserved under the conservative dynamics. Indeed, if one restricts to the instantaneous (order-reduced) Hamiltonian, one has the usual property
%
\begin{align}
\label{eq:H_inst_pphi_inst_onshell_conservation}
\frac{\dd H^\text{loc}_\text{on-shell}}{\dd t}\Bigg|_{\bm{a}_\text{cons}^\text{loc}} = 0  &&\text{and}&& 
\frac{\dd p_\phi^\text{on-shell}}{\dd t}\Bigg|_{\bm{a}_\text{cons}^\text{loc}} = 0 \,,  
\end{align}
where the Hamiltonian and angular momentum are evaluated after solving for the \textit{instantaneous}, conservative equations of motion, and the time derivative is taken using this same acceleration. However, when instead taking the total time-derivative of the on-shell value of the \textit{total, hereditary} Hamiltonian (where this time  the full conservative equations of motion are used, including the hereditary term), this conservation property is lost, and one instead finds [see (3.8) and (3.18) of~\cite{Bernard:2016wrg}]  
\begin{subequations}
\label{eq:H_J_onshell_nonconservation}
\begin{align}
\label{seq:H_onshell_nonconservation}
\frac{\dd H_\text{on-shell}}{\dd t}\Bigg|_{\bm{a}_\text{cons}} &= \frac{G^2 \dM}{5 c^8} \Biggl\{\dI_{ij}^{(4)}(t) \mathrm{Pf}_{\frac{2r(t)}{c}} \int_{-\infty}^{+\infty} \frac{\dd \tau}{|\tau|}\dI_{ij}^{(3)}(t+\tau) - \dI_{ij}^{(3)}(t) \mathrm{Pf}_{\frac{2r(t)}{c}} \int_{-\infty}^{+\infty} \frac{\dd \tau}{|\tau|}\dI_{ij}^{(4)}(t+\tau)\Biggr\} \,,\\
\label{seq:J_onshell_nonconservation}
\frac{\dd p_{\phi}^\text{on-shell}}{\dd t}\Bigg|_{\bm{a}_\text{cons}} &= \frac{4 G^2 \dM}{5 c^8} \epsilon_{zjk} \dI_{jl}^{(3)} \mathrm{Pf}_{\frac{2r(t)}{c}}\int_{-\infty}^{+\infty} \frac{\dd \tau}{|\tau|} \dI_{kl}^{(3)}(t+\tau) \,,
\end{align}
\end{subequations}
where the first index in the Levi-Civita symbol is the Cartesian $z$-component, which is orthogonal to the orbital plane. 
The goal is thus to \textit{localize} the Hamiltonian: the hereditary Hamiltonian should be transformed into an ordinary local Hamiltonian, such that the conserved energy is given in the usual manner by its on-shell value. This procedure is subtle, but it is now well understood, and can be addressed in different ways.

One way of looking at the problem, advocated by Refs.~\cite{Damour:2015isa, Damour:2016abl}, is to notice that higher-order Hamiltonians,  which include a finite number of derivatives of the momenta, are not conserved on shell either. Such higher-order Hamiltonians have been studied in the context of the (local) 2PN and 3PN equations of motion~\cite{Schafer:1984mr,Damour:1985mt,Damour:1990jh,Damour:1999cr,Damour:2016abl}. They give rise to generalized Hamiltonian equations, which are expressed in terms of functional (rather than ordinary) derivatives. \textit{A priori}, it is not allowed to replace the derivatives of the momenta inside the Hamiltonian using the equations of motion: such a `naively order-reduced' Hamiltonian would give rise to incorrect equations of motion. The way to address the problem is then to go back to the action (or Lagrangian) and notice that there exists a point transformation such that the action becomes ordinary (i.e., without higher-order derivatives) when expressed in terms of the new variables~\cite{Damour:1985mt}. This translates to a contact transformation on the phase-space variables, such that the Hamiltonian in the new variables is ordinary, and the energy corresponds to its on-shell value.  Crucially, it was shown that the functional expression of the ordinary Hamiltonian in terms of the new phase-space variables is simply given by `naively' order-reducing the higher-order Hamiltonian, with the additional information that the resulting reduced Hamiltonian should be interpreted as a function of the new variables. 
One  can then address nonlocal actions by Taylor-expanding the non-locality, effectively transforming the nonlocal action into a local action, which formally contains an infinite tower of derivatives of the coordinates. It can then be argued that the order-reduction procedure described previously carries through for an infinite tower of derivatives of the variables, which gives rise to a contact transformation which is \textit{a priori} very complicated. However, it does not need to be controlled, since through the same argument, one concludes that the localized Hamiltonian in terms of the new variables is given by `naively' localizing the Hamiltonian, with the additional information that the resulting localized Hamiltonian should be interpreted as a function of the new variables.
To illustrate this procedure more precisely, I introduce the toy action
\begin{subequations}\label{eq:S_toy}
\begin{align}\label{seq:S_toy_nonlocal}
    S_\text{toy}&=\frac{1}{2}\int_{-\infty}^\infty\dd t \Bigg[\dot{x}^2+ \int_{0}^\infty\dd \tau \ G\big(x(t), x(t-\tau),x(t+\tau)\big) \Bigg] \,,
\end{align}
where $G$ is some arbitrary function. One could derive from this action some nonlocal equations of motion, like in Sec.~\ref{subsec:EOM}, but here I will do something else. Deliberately forgetting about convergence problems, e.g. by assuming that they can be dealt with by a regulating kernel $\mu(\tau)$ inside the integral~\cite{Damour:2016abl}, one performs a Taylor expansion around $\tau=0$ and finds
\begin{align}\label{seq:S_toy_localized}
    S_\text{toy}&= \frac{1}{2}\int_{-\infty}^\infty\dd t \Bigg[\dot{x}^2+ \int_{0}^\infty\dd \tau \ \widehat{G}\big[x(t), \dot{x}(t), \ddot{x}(t),\ldots; \tau\big] \Bigg] \,.
\end{align}
\end{subequations}
Note that the explicit expression of $\widehat{G}$ is also very complicated, and I do not attempt to control it explicitly here.
I then introduce
\begin{align}
    \widehat{F}\big[x(t), \dot{x}(t), \ddot{x}(t), \ldots\big] =  \int_{0}^\infty\dd \tau \ \widehat{G}\big[x(t), \dot{x}(t), \ddot{x}(t),\ldots; \tau\big]
\end{align}
and find that its variation reads  
\begin{align}\label{eq:delta_F_hat}
     \delta\widehat{F} &= \sum_{N=0}^{\infty}  \frac{\p \widehat{F}}{\p {x}^{(N)}} \delta x^{(N)}  = \delta x\sum_{N=0}^{\infty} (-)^N\frac{\dd^N}{\dd t^N}\bigg[\frac{\p \widehat{F}}{\p x^{(N)}}\bigg]  + \frac{\dd}{\dd t}\Bigg\{\sum_{N=1}^{\infty}\sum_{M=0}^{N-1} (-)^M \frac{\dd^M}{\dd t^M}\bigg[\frac{\p \widehat{F}}{\p x^{(N)}}\bigg] \delta x^{(N-1-M)}\Bigg\}\,,
\end{align}
where  the $N$-th time derivative of $x(t)$ is denoted as $x^{(N)}$.
I then vary the action \eqref{seq:S_toy_localized}, use
the identity \eqref{eq:delta_F_hat}, integrate by parts, discard boundary terms associated with total derivatives, and finally find that the (formally) localized equations of motion read
\begin{align}
    \ddot{x} = \frac{1}{2}\sum_{N=0}^{\infty} (-)^N\frac{\dd^N}{\dd t^N}\bigg[\frac{\p \widehat{F}}{\p x^{(N)}}\bigg] \,,
\end{align}
where the right-hand side depends on $x$, $\dot{x}$, $\ddot{x}$, $\dddot{x}$, etc. These equations of motion could have been equivalently obtained by Taylor expanding the nonlocal equations of motion obtained like in Sec.~\ref{subsec:EOM}. Due to this structure, these localized equations of motion are unusable in practice, and it is not guaranteed that the infinite sum converges. However, in formal analogy to the case of a \textit{finite} number of higher-order derivatives, one postulates the existence of a point transformation that removes all higher-order derivatives in the action \eqref{seq:S_toy_localized}. Following this analogy, one concludes that the associated `naively order-reduced' hereditary Hamiltonian is a correct description of the motion in some new phase-space variables. However, one does not easily control in this way the transformation between the old and new phase-space variables;  but as we will see, this is not necessary if one is only interested in linking the fundamental frequencies to the energy and angular momentum.

Another way of looking at the problem, advocated by Refs.~\cite{Isoyama:2014mja,Fujita:2016igj,Isoyama:2018sib,Blanco:2022mgd, Blanco:2023jxf, Blanco:2024fte,Lewis:2025ydo}, is to realize that the hereditary action gives rise to a pseudo-Hamiltonian system. This means that the phase-space variables are doubled in the right-hand side of the associated Hamilton's equations; one then takes derivatives  with respect to only one of the variables, and only at the end are the doubled phase-space variables set to coincide. One then finds that the `naively order-reduced' Hamiltonian leads to the correct equations of motion \textit{in the old variables}, but for a \textit{perturbed, noncanonical symplectic form}. This perturbation to the symplectic form was worked out explicitly. One can then perform an explicitly prescribed \textit{noncanonical} transformation of the phase-space variables, such that in these \textit{new variables}, the equations of motion are driven by the `naively order-reduced' Hamiltonian and a canonical symplectic form. At the end of the day, this second approach leads to the exact same prescription for the localized Hamiltonian as that of Ref.~\cite{Damour:2015isa, Damour:2016abl}, but avoids the formal Taylor expansion and now also allows for an explicit control of the associated transformation of the phase-space variables, which could be useful for the construction of a 4PN quasi-Keplerian parametrization of the motion in ADM or modified harmonic coordinates (generalizing the local parametrization of~\cite{Cho:2021oai}). I have explicitly applied this approach in Appendix~\ref{app:shift_localization}.  

Finally, note that one can  also construct conserved quantities by directly evaluating the right-hand sides of~\eqref{eq:H_J_onshell_nonconservation}. From these, one then obtains corrections $\delta H$ and $\delta p_\phi$ which should be added to the on-shell values to obtain the truly conserved energy $E = H_\text{on-shell} + \delta H$ and angular momentum $J = p_\phi^\text{on-shell} + \delta p_\phi$. This method was employed in Sec.~IV of~Ref.~\cite{Bernard:2016wrg}, and was explicitly shown to yield identical results for the conserved quantities in the case of circular orbits; see also  \cite{Blanchet:2017rcn}. Due to practical considerations, this is not the method that will be employed here, but I would of course expect it to yield identical results.

 \subsection{Obtaining the local equations of motion from the local Hamiltonian}
\label{subsec:Hamiltonian_loc}

I will now focus on the treatment of the \textit{local} equations of motion and conserved quantities at 4PN. 
The local piece of the rescaled Hamiltonian (in ADM coordinates) is given to the 4PN order in Eqs.~(5.13) and (5.17) of Ref.~\cite{Damour:2014jta} in terms of rescaled variables. Namely, one should do the following replacements when going from the notations of that reference (which we refer to by the acronym `DJS': Damour, Jaranowski and Schäfer) to my notations:
\begin{align}\label{eq:rescaling_DJS}
    H_\text{DJS}^\text{loc}& \rightarrow \frac{H^\text{loc}}{\mu}\,, & r_{\text{DJS}} &\rightarrow \frac{r}{G m} \,,&
    \bm{p}_{\text{DJS}}^2 & \rightarrow \frac{1}{\mu^2}\left( p_r^2 + \frac{p_\phi^2}{r^2} \right) \,, &
    (\bm{n}\cdot \bm{p})_{\text{DJS}} &\rightarrow \frac{p_r}{\mu} \,.
\end{align} 
The local Hamiltonian in terms of $(r,\phi, p_r, p_\phi)$ is given by 
\begin{equation}\label{eq:H_loc_pr_pphi}
H^\text{loc} = \frac{p_r^2}{2   m \nu} + \frac{p_\phi^2}{2 m \nu\, r^2} -  \frac{G m^2 \nu}{r} + \cdots
\end{equation}
where the higher-order terms  are given in Ref.~\cite{Damour:2014jta} in terms of rescaled variables; see also the Supplemental Material~\cite{Supplemental} for the complete expression. 
One then solves for $({p}_r, {p}_\phi)$ in the system of equations 
\begin{equation}\label{eq:rDot_phiDot_system}
\left\{
\begin{aligned}
\dot{r} &= \left[\frac{\partial H^\text{loc}}{\partial p_r}\right](p_r, p_\phi, r) \\ 
\dot{\phi} &= \left[\frac{\partial H^\text{loc}}{\partial p_\phi}\right](p_r, p_\phi, r)  
\end{aligned} \quad.
\right.
\end{equation}
This is done iteratively, order by order in the PN expansion. One finds that, on shell, these relations read
\begin{subequations}
\label{eq:pr_pphi_loc_inTermsOf_r_phi_rDot_phiDot}
\begin{align}
\label{seq:pr_loc_inTermsOf_r_phi_rDot_phiDot} 
p_r^{\text{loc}} &= m \nu \dot{r}   \Biggl\{1+\frac{1}{c^2}\Biggl[\frac{\dot{r}^2+r^2 \dot{\phi}^2}{2} \left(1-3 \nu \right)  +\frac{G m}{r} (3+2\nu )  \Biggr] + \dots \Biggr\} \,, \\
\label{seq:pphi_loc_inTermsOf_r_phi_rDot_phiDot} 
p_\phi^{\text{loc}}  &= m \nu r^2 \dot{\phi}\Biggl\{1+\frac{1}{c^2}\Biggl[\frac{\dot{r}^2+r^2 \dot{\phi}^2}{2} \left(1-3 \nu \right)  +\frac{G m}{r} (3+\nu )   \Biggr] + \cdots \Biggr\} \,,
\end{align}
\end{subequations}
where the complete local 4PN expressions have been relegated to the Supplemental Material~\cite{Supplemental}.
The (local) energy and angular momentum are then obtained, respectively, as the on-shell value of the (local) Hamiltonian and the  momentum conjugate to the angle $\phi$, namely 
\begin{subequations}
\label{eq:E_J_loc_inTermsOf_H_pphi_loc}
\begin{align}
\label{seq:E_loc_inTermsOf_H_loc}
 E^{\text{loc}}  &= H^\text{loc}\left[r, p_r^\text{loc}(\dot{r}, \dot{\phi}, r), p_\phi^\text{loc}(\dot{r}, \dot{\phi}, r)\right]\,, \\
\label{seq:J_loc_inTermsOf_pphi_loc}
J^{\text{loc}}  &= p_\phi^{\text{loc}} (\dot{r}, \dot{\phi}, r)\,.
\end{align}\end{subequations}
Using the expressions for the conjugate momenta~\eqref{eq:pr_pphi_loc_inTermsOf_r_phi_rDot_phiDot} and appropriately PN-expanding, I find that 
\begin{subequations}
    \label{eq:E_J_loc_inTermsOf_r_phi_rDot_phiDot}
\begin{align}
\label{seq:E_loc_inTermsOf_r_phi_rDot_phiDot}
E^{\text{loc}}  &= m \nu  \, \Biggl\{\frac{\dot{r}^2}{2}+\frac{r^2 \dot{\phi}^2}{2}-\frac{G m}{r}+ \cdots \Biggr\} \,,\\*
\label{seq:J_loc_inTermsOf_r_phi_rDot_phiDot}
J^{\text{loc}} &= m \nu  r^2  \dot{\phi} \,\Biggl\{1+\frac{1}{c^2}\Biggl[\frac{\dot{r}^2+r^2 \dot{\phi}^2}{2} \left(1-3 \nu \right)  +\frac{G m}{r} (3+\nu )  \Biggr]+ \cdots\Biggr\} \,,
\end{align}
\end{subequations}
where the complete local 4PN expressions have been relegated to the Supplemental Material~\cite{Supplemental}. 
Note that these expressions differ from Eqs.~(3.1)~and~(3.2)~of~Ref.~\cite{Bernard:2017ktp}, simply because the latter reference works in harmonic coordinates rather than ADM coordinates.
Using the fact that $(E^{\text{loc}}, J^{\text{loc}})$ are constants of motion (for the \textit{local} equations of motion), one can finally solve for $\{\dot{r}^2, \dot{\phi}\}$ in the previous expression~\eqref{eq:E_J_loc_inTermsOf_r_phi_rDot_phiDot}. They have a polynomial structure in $s = 1/r$,
\begin{align}\label{eq:rDot_phiDot_loc_inTermsOf_r}
 \big[\dot{r}^2\big]^{\text{loc}}  &= \mathcal{R}(1/r) &&\text{and}&
\big[\dot{\phi}\big]^{\text{loc}}  &= \mathcal{S}(1/r) \,,
\end{align}
where the polynomials read
\begin{subequations}
\label{eq:R_S_inTermsOf_s}
\begin{align}
\label{seq:R_inTermsOf_s}
\mathcal{R}(s) &= A + 2 B s  + C s^2 + D_1 s^3 + D_2 s^4 + D_3 s^5 + D_4 s^6 + D_5 s^7 + D_6 s^8 + D_7 s^9 + \mathcal{O}(10)\,,\\
\label{seq:S_inTermsOf_s}
\mathcal{S}(s) &= F s^2 + I_1 s^3 + I_2 s^4 + I_3 s^5 + I_4 s^6 + I_5 s^7 + I_6 s^8 + I_7 s^9 + \mathcal{O}(10)\,,
\end{align}
\end{subequations}
in which ignorable terms that are 5PN or higher are denoted by $\mathcal{O}(10)$.
The coefficients $(A,B,C,D_n,F,I_n)$ are functions of energy and angular momentum, and I provide them in Eq.~\eqref{eq:ABCDFI_inTermsOf_varepsilon_j} (see also the Supplemental Material~\cite{Supplemental}) in terms of the reduced energy and angular momentum
\begin{align}\label{eq:varepsilon_j_def}
    \varepsilon= - \frac{2E}{m \nu c^2}&& \text{and} &&j = - \frac{2 J^2 E}{G^2 m^5 \nu^3} \,,
\end{align}
which are defined such that $\varepsilon = \mathcal{O}(1/c^2)$ and $j = \mathcal{O}(1)$. They have the following PN scalings: $A$, $B$, $C$, and $F$ are of Newtonian order $\sim \mathcal{O}(1)$; $D_1$ and $I_1$ are of 1PN order $\sim \mathcal{O}(1/c^2)$; $D_2$, $I_2$, $D_3$, and $I_3$ are of 2PN order $\sim \mathcal{O}(1/c^4)$; $D_4$, $I_4$, $D_5$, and $I_5$ are of 3PN order $\sim \mathcal{O}(1/c^6)$; and $D_6$, $I_6$, $D_7$, and $I_7$ are of 4PN order $\sim \mathcal{O}(1/c^8)$. I find perfect agreement for these coefficients with Ref.~\cite{Cho:2021oai}; see Eqs.~(9)~and~(16) in that reference, as well as the associated Supplemental Material. I also agree with the 3PN result of Eq.~(A1)~of~Ref.~\cite{Memmesheimer:2004cv} for $A$, $B$, $C$, $D_n$.
Finally, injecting~\eqref{eq:rDot_phiDot_loc_inTermsOf_r} into~\eqref{seq:pr_loc_inTermsOf_r_phi_rDot_phiDot}, I find that $p_r^2$ can be expressed (on shell) as a polynomial in $r$, namely
\begin{align}\label{eq:cal_I_def}
\big[p_r^2\big]^{\text{loc}} &= \mathcal{I}(1/r) \,,
\end{align}
where the polynomial reads
\begin{align}\label{eq:cal_I_of_s_poly}
\mathcal{I}(s) =  \cA + 2\cB s + \cC s^2 + \cD_1 s^3 + \cD_2 s^4 + \cD_3 s^5 + \cD_4 s^6 + \cD_5 s^7 + \cD_6 s^8 + \cD_7 s^9 + \mathcal{O}(10)
\end{align}
and where the coefficients are given in terms of $(\varepsilon,j)$ in Eq.~\eqref{eq:cal_ABCD_inTermsOf_varepsilon_j}; see also the Supplemental Material~\cite{Supplemental}. They are in agreement\footnote{To find agreement, one must first specify the values  $\omega_\text{static}=0$ and $\omega_\text{kinetic}=41/24$, which were determined after the publication of~Ref.~\cite{Damour:1999cr}. Moreover, note the following typo in Eq.~(B1f)~of~Ref.~\cite{Damour:1999cr}: the first term of that equation (i.e., the 2PN piece) should be~$\frac{1}{c^4} \frac{3}{4
}\nu^2 j^4$; compare to the correct 2PN-accurate expression in Eq.~(3.4f)~of~Ref.~\cite{Damour:1988mr}.} at 3PN with Eq.~(B1)~of~Ref.~\cite{Damour:1999cr}. They have the following PN scalings: $\mathcal{A}$, $\mathcal{B}$, and $\mathcal{C}$ are of Newtonian order $\sim \mathcal{O}(1)$; $\mathcal{D}_1$ is of 1PN order $\sim \mathcal{O}(1/c^2)$; $\mathcal{D}_2$ and $\mathcal{D}_3$ are of 2PN order $\sim \mathcal{O}(1/c^4)$; $\mathcal{D}_4$ and $\mathcal{D}_5$ are of 3PN order $\sim \mathcal{O}(1/c^6)$; and $\mathcal{D}_6$ and $\mathcal{D}_7$ are of 4PN order $\sim \mathcal{O}(1/c^8)$. Note that these coefficients, in calligraphic script, differ from those in italic script of Eq.~\eqref{eq:ABCDFI_inTermsOf_varepsilon_j}.

 \section{Action-angle formulation of the local dynamics}
\label{sec:local}

In this section, I will obtain the local Hamiltonian in terms of action variables. In the process, I obtain the expressions for the (local) action variables in terms of the conserved (local) energy and angular momentum, which are themselves expressed in terms of the original coordinates. To avoid clutter, I will often omit the `loc' label on the right-hand side of such relations.

\subsection{Action variables in the local sector}
\label{subsec:actions_loc}

I will assume that the reader is familiar with the Hamilton-Jacobi formalism applied to the Kepler problem; see, e.g., Sec.~10.8 of~Ref.~\cite{Goldstein}. In the absence of spin effects, the local, 4PN motion  is fully separable in spherical coordinates, which makes it straightforward to define the usual set of action variables (with dimensions $[\text{mass}]\times[\text{length}]^2\times[\text{time}]^{-1}$): 
\begin{align} \label{eq:Ir_Itheta_Iphi_def_loc}
 I_r^{\text{loc}} = \frac{1}{2\pi} \oint \dd r \, p_r^{\text{loc}} &\,, &I_\theta^{\text{loc}} = \frac{1}{2\pi} \oint \dd \theta \, p_\theta^{\text{loc}} &\,,& I_\phi^{\text{loc}} = \frac{1}{2\pi} \oint \dd \phi \, p_\phi^{\text{loc}} \,.
\end{align}
One then switches to the Delaunay variables~\cite{Delaunay_Lune_1,Delaunay_Lune_2} to better address the degeneracies of the Kepler problem; these read, in Goldstein's notation~\cite{Goldstein} 
\begin{align}\label{eq:I1_I2_I3_def_loc}
I_1^{\text{loc}} = I_\phi^{\text{loc}} & \,, & I_2^{\text{loc}} = I_\phi^{\text{loc}} + I_\theta^{\text{loc}} & \,, & I_3^{\text{loc}} =I_r^{\text{loc}} + I_\phi^{\text{loc}} + I_\theta^{\text{loc}} \,,
\end{align}
which are such that the Newtonian Hamiltonian depends only on the action variable $I_3^\text{loc}$. 
Since spins are neglected, the motion is planar: $I_\theta^{\text{loc}} = 0$ and $\theta = \pi/2$. Restricting to the orbital plane, the action-angle formulation of the motion only involves two pairs of dynamical action-variables~\cite{Bini:2020hmy},
\begin{align}\label{eq:Iphi_I_rphi_def_loc}
I_\phi^{\text{loc}} &= I_2^{\text{loc}} &\text{and} & & I_{r\phi}^{\text{loc}} &= I_r^{\text{loc}} + I_\phi^{\text{loc}} = I_3^{\text{loc}} \,.
\end{align}
Note that $I_\phi^\text{loc} = p_\phi^\text{loc} = J^\text{loc}$ is simply the angular momentum. The only non-trivial quantity to compute in terms of energy and angular momentum is the (local) radial action. After changing variables from $r$ to $s = 1/r$ for convenience, it follows that the (local) radial action can be rewritten using Eqs.~\eqref{eq:Ir_Itheta_Iphi_def_loc} and~\eqref{eq:cal_I_def} as 
\begin{align}\label{eq:Ir_loc_contour_integral}
I_r^{\text{loc}} = \frac{1}{2\pi}\oint  \frac{\dd s}{s^2} \sqrt{\mathcal{I}(s)} \,.
\end{align}
Thanks to Eq.~\eqref{eq:cal_I_of_s_poly}, this can be computed \textit{à la} Sommerfeld using complex analysis~\cite{Sommerfeld, Goldstein, Damour:1988mr}. First, perform (under the integral sign) the PN expansion $\sqrt{\mathcal{I}(s)} = \sqrt{ \cA + 2\cB s + \cC s^2} \sum_k \alpha_k s^k $, where the coefficients $\alpha_k$ depend on the coefficients $\cA$, $\cB$, etc., and can be straightforwardly determined from the PN expansion. One is then left with the computation of  the following master integrals:
\begin{align}\label{eq:Ipq_integral_def}
\mathcal{I}_{p,q} &= \frac{1}{2\pi} \oint \dd s\, s^{p-2} \left(\cA+ 2 \cB s + \cC s^2\right)^{1/2-q},
\end{align}
where $(p,q) \in \mathbb{N}^2$; recall that $\cA<0$, $\cB>0$, and \mbox{$\cC<0$} for bound orbits. Using the residue theorem, one finally finds that~\cite{Trestini:2024zpi}
\begin{align}\label{eq:Ipq_integral_formula}
\mathcal{I}_{p,q} &= \big[p=0\big](-1)^q(1-2q)\cB(-\cA)^{-1/2-q} + \big[p=1\big] (-1)^{q+1} (-\cA)^{1/2-q}  \nonumber  \\*
& + \big[p\ge 2q\big](-1)^{q+1}(-\cC)^{1/2-q} \sum_{k=\ceil{\frac{p}{2}-q}}^{p-2q} \frac{\Gamma\left(\frac{1}{2}-q+1\right) \cA^{p-k-2q}\, (2\cB)^{2k+2q-p}\,\cC^{-k}}{\Gamma\left(\frac{1}{2}-q-k+1\right) (2k+2q-p)!(p-k-2q)!}\,,
 \end{align}
 where $\Gamma(x)$ is the gamma function, $\ceil{x}$ is the ceiling of $x$, and $\big[\mathcal{P}\big]$ is the Iverson bracket \cite{IversonBracketWolfram,Iverson1962} defined such that $[\mathcal{P}] = 1$ if $\mathcal{P}$ is true, and $[\mathcal{P}] = 0$ otherwise. This formula, which first appeared in Eq.~(A4) of Ref.~\cite{Trestini:2024zpi}, reproduces all the integrals obtained for particular values of $(p,q)$ in Eq.~(B2) of Ref.~\cite{Damour:1999cr}. It would be interesting to extend this formula to the case where there are logarithmic terms in $r$, see Appendix C of \cite{Bernard:2016wrg}.
One then finds that the radial action can be expressed at 4PN order in terms of the coefficients $\cA$, $\cB$, etc. as follows:
\begin{align}\label{eq:Irloc_inTermsOf_ABCD}
 I_r^{\text{loc}} &= \frac{\cB}{\sqrt{-\cA}}-\sqrt{-\cC}+\frac{\cB \cD_1}{2 (-\cC)^{3/2}}+\frac{15 \cB^2 \cD_1^2-3 \cA \cC \cD_1^2-12 \cB^2 \cC \cD_2+4 \cA \cC^2 \cD_2+20 \cB^3 \cD_3-12 \cA \cB \cC \cD_3}{16 (-\cC)^{7/2}} \nonumber\\
 & +\frac{1}{32 (-\cC)^{11/2}}\biggl[105 \cB^3 \cD_1^3-35 \cA \cB \cC \cD_1^3-140 \cB^3 \cC \cD_1 \cD_2+60 \cA \cB \cC^2 \cD_1 \cD_2+315 \cB^4 \cD_1 \cD_3-210 \cA \cB^2 \cC \cD_1 \cD_3 \nonumber\\*
 &\qquad\qquad\qquad +15 \cA^2 \cC^2 \cD_1 \cD_3-70 \cB^4 \cC \cD_4+60 \cA
   \cB^2 \cC^2 \cD_4-6 \cA^2 \cC^3 \cD_4+126 \cB^5 \cD_5-140 \cA \cB^3 \cC \cD_5+30 \cA^2 \cB \cC^2 \cD_5\biggr] \nonumber\\
 &+\frac{1}{1024 (-\cC)^{15/2}}
\biggl[15015 \cB^4 \cD_1^4-6930 \cA \cB^2 \cC \cD_1^4+315 \cA^2 \cC^2 \cD_1^4-27720 \cB^4 \cC \cD_1^2 \cD_2+15120 \cA \cB^2 \cC^2 \cD_1^2 \cD_2 \nonumber\\*
 &\qquad\qquad\qquad -840 \cA^2 \cC^3 \cD_1^2 \cD_2+5040 \cB^4 \cC^2 \cD_2^2-3360 \cA \cB^2 \cC^3 \cD_2^2+240 \cA^2 \cC^4 \cD_2^2+72072 \cB^5 \cD_1^2
   \cD_3 \nonumber\\*
 &\qquad\qquad\qquad -55440 \cA \cB^3 \cC \cD_1^2 \cD_3+7560 \cA^2 \cB \cC^2 \cD_1^2 \cD_3-22176 \cB^5 \cC \cD_2 \cD_3+20160 \cA \cB^3 \cC^2 \cD_2 \cD_3 \nonumber\\*
 &\qquad\qquad\qquad -3360 \cA^2 \cB \cC^3 \cD_2 \cD_3 +24024 \cB^6 \cD_3^2-27720 \cA \cB^4 \cC \cD_3^2+7560 \cA^2 \cB^2 \cC^2 \cD_3^2-280 \cA^3 \cC^3 \cD_3^2 \nonumber\\*
 &\qquad\qquad\qquad -22176 \cB^5 \cC \cD_1 \cD_4   +20160 \cA \cB^3 \cC^2 \cD_1 \cD_4-3360 \cA^2 \cB \cC^3 \cD_1 \cD_4+48048 \cB^6 \cD_1 \cD_5-55440 \cA \cB^4 \cC
   \cD_1 \cD_5  \nonumber\\*
 &\qquad\qquad\qquad +15120 \cA^2 \cB^2 \cC^2 \cD_1 \cD_5  -560 \cA^3 \cC^3 \cD_1 \cD_5-7392 \cB^6 \cC \cD_6+10080 \cA \cB^4 \cC^2 \cD_6-3360 \cA^2 \cB^2 \cC^3 \cD_6 \nonumber\\*
 &\qquad\qquad\qquad  +160 \cA^3 \cC^4 \cD_6 +13728 \cB^7 \cD_7  -22176 \cA \cB^5 \cC \cD_7+10080 \cA^2 \cB^3 \cC^2 \cD_7-1120 \cA^3 \cB \cC^3 \cD_7\biggr] \,.
 \end{align}

It will now prove useful to introduce the rescaled action variables (with dimensions $[\text{length}]^{-1}[\text{time}]$ )
\begin{align}\label{eq:rescaled_actions_def_loc}
i_\phi^{\text{loc}} =  \frac{I_\phi^{\text{loc}}}{G m^2 \nu} & \,,&   i_r^{\text{loc}} = \frac{I_r^{\text{loc}}}{G m^2 \nu}  & \,,&  i_{r\phi}^{\text{loc}} = \frac{I_{r\phi}^{\text{loc}}}{G m^2 \nu} \,,
\end{align}
where $I_{r\phi}^{\text{loc}} = I_r^{\text{loc}}+I_\phi^{\text{loc}}$. Note that $\{c \, i_\phi^{\text{loc}}, c \, i_r^{\text{loc}},  c \,i_{r\phi}^{\text{loc}}\}$ are dimensionless quantities;   the reduced action variables are given the dimension of an inverse velocity in order to keep $c$ as the PN order counting parameter. 
The expressions for the coefficients $(\mathcal{A}, \mathcal{B}, \mathcal{C}, \mathcal{D}_n)$, provided in Eq.~\eqref{eq:cal_ABCD_inTermsOf_varepsilon_j}, are then plugged back into Eq.~\eqref{eq:Irloc_inTermsOf_ABCD}.  This results in the following expression for the radial action at 4PN in terms of the (local) energy and angular momentum:
\begin{align}\label{eq:ir_loc_inTermsOf_varepsilon_j}
 i_r^{\text{loc}} &= \frac{1}{c \sqrt{\varepsilon }}  \Biggl\{1-\sqrt{j}+\varepsilon  \Biggl[-\frac{15}{8}+\frac{\nu}{8} +\frac{3}{\sqrt{j}} \Biggr]+\varepsilon^2 \Biggl[\frac{35}{128} +\frac{15}{64} \nu +\frac{3}{128} \nu^2 +\frac{1}{\sqrt{j}}\biggl(-\frac{15}{4}+\frac{3}{2} \nu \biggr)  +\frac{1}{j^{3/2}}\biggl(\frac{35}{4}-\frac{5}{2}\nu \biggr)\Biggr] \nonumber\\*
 & \qquad+\varepsilon^3 \Biggl[\frac{21}{1024}-\frac{105}{1024} \nu +\frac{15}{1024} \nu^2+\frac{5}{1024}  \nu^3 +\frac{1}{\sqrt{j}}\biggl(\frac{15}{16}-\frac{15}{16} \nu +\frac{3}{4} \nu^2 \biggr) \nonumber\\*
 & \qquad\qquad +\frac{1}{j^{3/2}}\biggl(-\frac{105}{4}+ \nu\Bigl(\frac{109}{3}-\frac{41}{128} \pi^2\Bigr)  -\frac{15}{4} \nu^2 \biggr)  +\frac{1}{j^{5/2}}\biggl(\frac{231}{4}+ \nu\left(-\frac{125}{2}+\frac{123}{128} \pi^2\right)  +\frac{21}{8} \nu^2\biggr)\Biggr] \nonumber\\
& \qquad +\varepsilon^4 \Biggl[\frac{99}{32768}-\frac{105}{8192} \nu +\frac{105}{16384} \nu^2+\frac{15}{8192} \nu^3+\frac{35}{32768} \nu^4 +\frac{1}{\sqrt{j}}\biggl(-\frac{15}{32} \nu^2 + \frac{3}{8}\nu^3 \biggr) \nonumber\\
&\qquad\qquad +\frac{1}{j^{3/2}}\biggl(\frac{1575}{64} + \nu \Bigl(-\frac{20323}{288}+\frac{35569}{24576} \pi^2\Bigr)  +  \nu^2\Bigl(\frac{4045}{96}-\frac{205}{512}\pi^2\Bigr)-\frac{15}{4} \nu^3 \biggr)  \nonumber\\
&\qquad\qquad +\frac{1}{j^{7/2}}\biggl(\frac{32175}{64}+  \nu \Bigl(-\frac{248057}{288}+\frac{425105}{24576} \pi^2\Bigr)
   + \nu^2 \Bigl(\frac{18925}{96}-\frac{1025}{256} \pi^2\Bigr) -\frac{45}{16}\nu^3 \biggr)  \nonumber\\
&\qquad\qquad +\frac{1}{j^{5/2}}\biggl(-\frac{9009}{32}+ \nu\Bigl(\frac{293413}{480}-\frac{51439}{4096} \pi^2\Bigr)  + \nu^2\Bigl(-\frac{7013}{32}+\frac{123}{32} \pi^2 \Bigr)+\frac{105}{16}\nu^3\biggr)\Biggr] + \mathcal{O}(\varepsilon^5)\Biggr\} \,,
\end{align}
where I wrote $(\varepsilon,j)$ instead of $(\varepsilon^\text{loc},j^\text{loc})$ to avoid clutter. This local expression is in perfect agreement at 4PN with~Refs.~\cite{Bini:2020wpo, Bini:2020nsb} after proper conversion from EOB variables to PN variables;\footnote{\label{footnote:EOB_conversion} In Ref.~\cite{Bini:2020wpo}, consider Eqs.~(13.20)--(13.22) alongside Table X.  One can immediately discard $I_9$ at 4PN, but should include $I_5$ and $I_7$ despite their dependence on some undetermined 5PN coefficients, $a_6^{\nu^2}$ and $\bar{d}_5^{\nu^2}$, which do not enter the final 4PN-expanded result. First re-express their Eq.~(9.2) in terms of $\gamma_\text{BDG}$ and $j_\text{BDG}$ using $h_\text{BDG}=\sqrt{1+2\nu(\gamma_\text{BDG}-1)}$ [see their Eq.~(6.2)]. Then, setting $G=c=m=1$ for the purpose of the conversion, use the relations $\gamma_\text{BDG}=1- \frac{\varepsilon}{2}+\frac{\varepsilon^2 \nu}{8}$ and $j_\text{BDG} = \sqrt{j/\varepsilon}$ [see their Eqs.~(13.2)--(13.5)]. One then exactly recovers Eq.~\eqref{eq:ir_loc_inTermsOf_varepsilon_j} upon re-expansion in small $\varepsilon$.}  note that the radial action has a remarkably simple structure when expressed in EOB variables.
Up to the conventional factor $m \nu^2$, this expression is also in agreement\footnote{In Eq.~(4.2c) of~\cite{LeTiec:2015kgg}, $\nu^4$ should in fact be $\nu^3$.} with \cite{LeTiec:2015kgg} at 3PN order. The azimuthal action variable, by definition, is simply given by 
\begin{align}\label{eq:iphi_inTermsOf_varepsilon_j}
    i_\phi &= \frac{1}{c} \sqrt{\frac{j}{\varepsilon}} \,.
\end{align}
The other Delaunay variable is then simply related to the radial action by
\begin{align}\label{eq:irphi_loc_inTermsOf_varepsilon_j}
    i_{r\phi}^{\text{loc}} &= i_{r}^{\text{loc}} + \frac{1}{c} \sqrt{\frac{j}{\varepsilon}} \,,
\end{align}
such that $i_{r\phi}^{\text{loc}}$ differs from $i_{r}^{\text{loc}}$ only by its Newtonian coefficient. The leading-order behavior of each action variable is thus
\begin{align}\label{eq:irphi_ir_inTermsOf_varepsilon_j_leading_order}
i_{r\phi}= \frac{1}{c\sqrt{\varepsilon}}\Bigl(1+ \mathcal{O}(\varepsilon)\Bigr)  && \text{and} &&
  i_r= \frac{1-\sqrt{j}}{c\sqrt{\varepsilon}}\Bigl(1+ \mathcal{O}(\varepsilon)\Bigr) \,.
\end{align}

\subsection{Local Hamiltonian in terms of action variables}
\label{subsec:Hamiltonian_actions_loc}

In order to obtain the Hamiltonian in terms of the action variables, it suffices to solve (iteratively) for the energy $\varepsilon$ in Eqs.~\eqref{eq:ir_loc_inTermsOf_varepsilon_j} and~\eqref{eq:irphi_loc_inTermsOf_varepsilon_j}; note that the angular momentum is trivially related to $i_\phi$ by~\eqref{eq:iphi_inTermsOf_varepsilon_j}. At 4PN, the local Hamiltonian (or energy) reads
\begin{align}\label{eq:H_loc_inTermsOf_irphi_iphi}
H^\text{loc}(i_{r\phi},i_\phi) &= -\frac{m \nu}{2 i_{r\phi}^{\,2}}  \Biggl\{1 +\frac{1}{c^2}\Biggl[\frac{1}{i_{r\phi}^{\,2}}\biggl(-\frac{15}{4}+\frac{\nu}{4}\biggr)+\frac{6}{i_\phi i_{r\phi}}\Biggr]  \nonumber\\
&  +\frac{1}{c^4}\Biggl[\frac{1}{i_{r\phi}^{\,4}}\biggl(\frac{145}{8}-\frac{15}{8} \nu +\frac{\nu ^2}{8}\biggr)+\frac{1}{i_\phi i_{r\phi}^{\,3}}\biggl(-\frac{105}{2}+6 \nu \biggr)+\frac{27}{i_\phi^{\,2} i_{r\phi}^{\,2}}+\frac{1}{i_\phi^{\,3} i_{r\phi}}\biggl(\frac{35}{2}-5 \nu \biggr)\Biggr] \nonumber\\
& +\frac{1}{c^6}\Biggl[\frac{1}{i_{r\phi}^{\,6}}\biggl(-\frac{6363}{64}+\frac{805}{64} \nu -\frac{45}{32} \nu ^2+\frac{5}{64} \nu^3 \biggr)+\frac{1}{i_\phi i_{r\phi}^{\,5}}\biggl(\frac{825}{2}-75 \nu +6 \nu ^2\biggr)+\frac{1}{i_\phi^{\,2} i_{r\phi}^{\,4}}\biggl(-450+\frac{135}{2}\nu \biggr) \nonumber\\*
&\qquad +\frac{1}{i_\phi^{\,3} i_{r\phi}^{\,3}}\biggl(-\frac{303}{4}+\nu\left(\frac{1427}{12}-\frac{41}{64} \pi ^2\right)  -10 \nu ^2 \biggr)+\frac{1}{i_\phi^{\,4} i_{r\phi}^{\,2}}\biggl(\frac{315}{2}-45 \nu \biggr) \nonumber\\*
&\qquad +\frac{1}{i_\phi^{\,5} i_{r\phi}}\biggl(\frac{231}{2}+\nu\left(-125+\frac{123}{64}\pi^2\right)  +\frac{21}{4}\nu ^2\biggr)\Biggr]  \nonumber\\
&+ \frac{1}{c^8}\Biggl[\frac{1}{i_{r\phi}^{\,8}}\biggl(\frac{75303}{128}-\frac{10713}{128} \nu +\frac{1545}{128} \nu ^2-\frac{75}{64} \nu ^3+\frac{7}{128} \nu ^4\biggr)+\frac{1}{i_\phi i_{r\phi}^{\,7}}\biggl(-\frac{50703}{16}+\frac{5745}{8} \nu -\frac{195}{2} \nu ^2+6 \nu ^3\biggr) \nonumber\\
&\qquad +\frac{1}{i_\phi^{\,2} i_{r\phi}^{\,6}}\biggl(\frac{85365}{16}-\frac{10395}{8} \nu +\frac{945}{8} \nu^2\biggr)  \nonumber\\
&\qquad +\frac{1}{i_\phi^{\,3} i_{r\phi}^{\,5}}\biggl(-\frac{46275}{32}+\nu \left(-\frac{59639}{72}+\frac{124129}{12288} \pi ^2\right) +\nu ^2\left(\frac{1547}{6}-\frac{41}{32} \pi ^2\right) -15 \nu ^3\biggr)  \nonumber\\
&\qquad +\frac{1}{i_\phi^{\,4} i_{r\phi}^{\,4}}\biggl(-\frac{21435}{8}+ \nu\left(\frac{8035}{4}-\frac{615}{64} \pi ^2\right) -\frac{375}{2} \nu ^2\biggr)  \nonumber\\
&\qquad +\frac{1}{i_\phi^{\,5} i_{r\phi}^{\,3}}\biggl(-\frac{7749}{16}+\nu\left(\frac{467473}{240}-\frac{80959}{2048} \pi ^2\right)  +\nu^2\left(-\frac{8643}{16}+\frac{1107}{128} \pi ^2\right) +\frac{63}{4} \nu ^3\biggr)  \nonumber\\
&\qquad +\frac{1}{i_\phi^{\,6} i_{r\phi}^{\,2}}\biggl(\frac{20307}{16}+\nu\left(-\frac{5025}{4}+\frac{1107}{64} \pi ^2\right)  +66 \nu ^2\biggr)  \nonumber\\
&\qquad +\frac{1}{i_\phi^{\,7} i_{r\phi}}\biggl(\frac{32175}{32}+\nu\left(-\frac{248057}{144}+\frac{425105}{12288} \pi ^2\right)  +\nu^2\left(\frac{18925}{48}-\frac{1025}{128} \pi ^2\right) -\frac{45}{8}\nu ^3\biggr)\Biggr] + \mathcal{O}\left(\frac{1}{c^{10}}\right)\Biggr\}\,,\nonumber \\
&
\end{align}
where again, I wrote $(i_{r\phi},i_\phi)$ instead of $(i_{r\phi}^{\text{loc}}, i_\phi^{\text{loc}})$ to avoid clutter.
By construction, the Hamiltonian does not depend on the angle variables. This expression agrees with Eq.~(3.13)~of~Ref.~\cite{Damour:1988mr} at~2PN, with Eq.~(4.18)~of~Ref.~\cite{Damour:1999cr} at~3PN (once the correct ambiguity parameters are chosen, namely $\omega_\mathrm{static} = 0$ and $\omega_\mathrm{kinetic}= 41/24$), and finally, with Eq.~(13.27) and Table XI of Ref.~\cite{Bini:2020wpo} once the EOB Hamiltonian is appropriately converted to a PN Hamiltonian.$^{\text{\ref{footnote:EOB_conversion}}}$

\subsection{Angle variables in the local problem}

\label{subsec:angles_loc}

Having constructed the (local) \textit{action} variables $(I_{r\phi}^{\text{loc}}, I_\phi^{\text{loc}})$, one now wants to determine the (local) \textit{angle} variables $(\ell^{\text{loc}}, g^{\text{loc}})$, with the requirement that they must be conjugate to the action variables and grow linearly in time, namely 
\begin{subequations}\label{eq:local_angles_and_frequencies_defs}
\begin{align}
\label{seq:ell_and_n_loc_defs}
\frac{\dd \ell^\text{loc}}{\dd t} &= n^\text{loc}  = \frac{\partial H^\text{loc} }{\partial I_{r\phi}^\text{loc}}\bigg|_{I_\phi^\text{loc} } = \frac{1}{G m^2 \nu} \frac{\partial H^\text{loc} }{\partial i_{r\phi}^\text{loc}}\bigg|_{i_{\phi}^\text{loc} }  \,,\\
\label{seq:g_and_omega_minus_n_loc_defs}
\frac{\dd g^\text{loc}}{\dd t}  &= \omega^\text{loc}  - n^\text{loc}  =  \frac{\partial H^\text{loc} }{\partial I_\phi^\text{loc}}\bigg|_{I_{r\phi}^\text{loc}}  =  \frac{1}{G  m^2 \nu} \frac{\partial H^\text{loc} }{\partial i_\phi^\text{loc}}\bigg|_{i_{r\phi}^\text{loc}}  \,.
\end{align}
\end{subequations}
Here, I have introduced the (local) radial frequency $n^\text{loc}$ and  azimuthal frequency $\omega^\text{loc}$. Thus, the (local) action-angle variables are the standard Delaunay variables\footnote{Here, I follow the conventions of \cite{Damour:2015isa}, but one could also have alternatively worked with another combination of these variables introduced by Poincaré, which are better adapted to taking the limit of circular orbits: $(\lambda=\ell + g, \varpi=-g, I_{r\phi}, I_r)$; see Sec. IV~of \cite{Damour:2016abl}.}
of celestial mechanics: $\ell^\text{loc}$ is the  \textit{mean anomaly}   and $g^\text{loc}$ is the \textit{argument of the periastron}.  Note that at Newtonian order, the problem is degenerate, since $\omega=n+\mathcal{O}(1/c^2)$ and $g$ is a constant at this order (there is no periastron precession).

In order to construct the angle variables, one introduces the characteristic function~\cite{Goldstein} 
\begin{align}\label{eq:characteric_function_loc}
W^\text{loc}(r,\phi; I_{r\phi}^{\text{loc}}, I_\phi^{\text{loc}}) =  I_\phi^{\text{loc}} \, \phi + \int^r \dd r' \sqrt{\mathcal{I}(1/r';  I_{r\phi}^{\text{loc}}, I_\phi^{\text{loc}})} \,,
\end{align}
which is the type-2 generating function for the transformation between $(r,\phi,p_r^{\text{loc}},p_\phi^{\text{loc}})$ and the action-angle variables $(\ell^{\text{loc}}, g^{\text{loc}}, I_{r\phi}^{\text{loc}}, I_\phi^{\text{loc}})$. Note that $\mathcal{I}(1/r'; I_{r\phi}^{\text{loc}},I_\phi^{\text{loc}})$ was defined by Eq.~\eqref{eq:cal_I_of_s_poly}, where $J^{\text{loc}}= I_\phi^{\text{loc}} = p_\phi^{\text{loc}}$ and where $E^{\text{loc}}$ should be expressed in terms of the action variables using Eq.~\eqref{eq:H_loc_inTermsOf_irphi_iphi}.
By construction of the action variables, one recovers $p_r^{\text{loc}} = \partial W^{\text{loc}}/\partial r$ and $p_\phi^{\text{loc}} = \partial W^{\text{loc}} / \partial \phi$. The angle variables are then defined as \mbox{$\ell^{\text{loc}}= \partial W^{\text{loc}}/\partial I_{r\phi}^{\text{loc}}$} and \mbox{$g^{\text{loc}} = \partial W^{\text{loc}}/\partial I_{\phi}^{\text{loc}}$}, thus ensuring that they are conjugate to the action variables. The angle variables have thus been expressed in terms of $(r,\phi, I_{r\phi}^{\text{loc}}, I_\phi^{\text{loc}})$, or equivalently, in terms of $(r,\phi, E^{\text{loc}}, J^{\text{loc}})$; however, this expression is still plagued with an integral, preventing it from being entirely explicit. In the local sector, this was solved explicitly up to 4PN~\cite{Cho:2021oai, Memmesheimer:2004cv, Schafer:1993pkg, DamourDeruelle86}; the solution is called the \textit{quasi-Keplerian parametrization}, and is the 4PN generalization of the Keplerian parametrization of Sec.~\ref{subsec:localizing}.  Alternatively, one can invoke the periodic nature of the motion and expand it into a Fourier series, namely 
\begin{align} \label{eq:r_phi_inTermsOf_Delaunay_loc}
r &= \sum_{(p,q)\in\mathbb{Z}^2} {}_{p,q}\cC_r(I_{r\phi}^\text{loc},I_\phi^\text{loc}) \de^{\di (p \, \ell^\text{loc} +q \,g^\text{loc})}\ ,   &&&
\phi &= \sum_{(p,q)\in\mathbb{Z}^2} {}_{p,q}\cC_{\phi}( I_{r\phi}^\text{loc},I_\phi^\text{loc}) \de^{\di (p\,\ell^\text{loc} +q \, g^\text{loc})}\ ,
\end{align}
where $ {}_{p,q}\cC_r(I_{r\phi}^\text{loc},I_\phi^\text{loc}) $ and ${}_{p,q}\cC_{\phi}( I_{r\phi}^\text{loc},I_\phi^\text{loc})$ are some coefficients to be determined.

\section{Tail corrections to the local Hamiltonian in action-angle variables}
\label{sec:tails}

The objective is now to study the full problem, including the tails. This will be made possible by the existence of action variables associated with the full Hamiltonian. Restricting to the motion in the orbit plane ($\theta = \pi/2$), these are expressed in terms of the one-form $\Theta =p_r \, \dd r + p_\phi \,\dd \phi$ as~\cite{Arnold:1989who, Schmidt:2002qk, Witzany:2024ttz} 
\begin{align}
  I_k = \frac{1}{2 \pi} \oint_{\mathcal{C}_k} \Theta \,,
\end{align}
where $k \in \{r,\phi\}$ and  $\mathcal{C}_k$ is a closed curve on the $2$-torus in the 4-dimensional phase space $(r,\phi, p_r,p_\phi)$ defined by the equations of motion. For definiteness, suppose that $p_r$ and $p_\phi$ can be written as functions of $(r,\phi)$. The action variables are independent of the specific choice of curve; they only depend on the homotopy class of the curve, which is specified by the label $k \in \{r,\phi\}$. This fact makes action variables gauge-invariant and thus extremely useful objects. However, in the absence of separability, it is very difficult (or sometimes impossible) to compute these explicitly, unless they are obtained as a perturbation of a separable system. Thus,   the tail contributions are treated as  perturbations to the actions~\eqref{eq:rescaled_actions_def_loc} associated with the local, separable problem. Just like in the local case, the motion is planar, so $p_\theta = 0$  and one can rotate the frame such that  $\theta = \pi/2$; the $\theta$ coordinate is thus ignorable. One then naturally introduces the Delaunay variable $I_{r\phi} = I_r + I_\phi$ as well as the rescaled action variables
\begin{align}\label{eq:rescaled_actions_def}
i_\phi =  \frac{I_\phi}{G m^2 \nu} & \,,&   i_r = \frac{I_r}{G m^2 \nu}  & \,,&  i_{r\phi} = \frac{I_{r\phi}}{G m^2 \nu} \,,
\end{align}
which will differ from their local counterparts~\eqref{eq:rescaled_actions_def_loc} only by small 4PN contributions due to the tails. Note that $I_\phi$ and $J$ are always trivially related ($I_\phi = J$ or $I_\phi^\text{loc} = J^\text{loc}$), but that $I_\phi \neq  I_\phi^\text{loc}$. 

I will closely follow the method laid out in Ref.~\cite{Blanco:2024fte}, which establishes precisely how to localize nonlocal perturbations to a local Hamiltonian. One key difference with the usual treatment of Chapter 12 of~\cite{Goldstein} is that the contact transformation from the unperturbed variables to the perturbed variables is not canonical anymore, due to the nonlocal nature of the perturbation.  Thus, the local Delaunay variables that have been constructed are still valid variables to describe the problem, but they are not action-angle variables with respect to the perturbed Hamiltonian. They are not even canonical anymore: the perturbed system in these variables also acquires a perturbation to the (canonical) symplectic form.

I will now describe the various contact transformations of the phase-space variables to recover an action-angle formulation for the perturbed problem.

\subsection{Splitting the tail term into a logarithmic and hereditary contribution}
\label{subsec:split_log_hered}

The local action-angle variables that have been constructed (associated with the local Hamiltonian $H^\text{loc}$) will now be denoted by $(\ell^\text{loc}, g^\text{loc}, I_{r\phi}^\text{loc}, I_\phi^\text{loc})$. However, it does not matter whether one uses the local or complete variables when working at leading order, so I will drop the `loc' tag in that case.  At Newtonian order, the Keplerian parameters are also introduced as functions of the action variables
%
\begin{align}\label{eq:a_e_inTermsOf_Irphi_Iphi}
a(I_{r\phi},I_\phi) &= \frac{I_{r\phi}^2}{G m^3 \nu^2} \,, &&&
e(I_{r\phi},I_\phi) &= \sqrt{1-\frac{I_\phi^2}{I_{r\phi}^2}} \,.
\end{align}
%
The tail Hamiltonian is then split into a `logarithmic' and a `hereditary' part: $H^\text{tail} = H^\text{log} + H^\text{hered}$. The individual pieces are defined as 
\begin{subequations}
\label{eq:H_tail_split_log_hered}
\begin{align}
\label{seq:H_log_def}
 H^\text{log} &=  \frac{2 G^2 m}{5 c^8} \dI_{ij}^{(3)}\dI_{ij}^{(3)} \ln\left(\frac{r}{\eta}\right) \,,\\*
 \label{seq:H_hered_def}
 H^\text{hered} &= - \frac{G^2 m}{5 c^8}\  \mathrm{Pf}_{\frac{2 \eta }{c}}  \int_{-\infty}^{+\infty}  \frac{\dd t' }{|t-t'|} \dI_{ij}^{(3)}(t)  \dI_{ij}^{(3)}(t') \nonumber \\*
&= - \frac{G^2 m}{5 c^8}  \dI_{ij}^{(3)}(t) \int_{0}^\infty \dd\tau \ln\left(\frac{c \tau}{2 \eta}\right)\left[\dI_{ij}^{(4)}(t-\tau)-\dI_{ij}^{(4)}(t+\tau)\right] \,,  
\end{align}
\end{subequations}
where I have introduced the scale
\begin{align}
\label{eq:def_eta_scale}
\eta  = \frac{1}{4}\exp(-\gamma_\mathrm{E}) \sqrt{\frac{c^2 a^3}{G m }} \,
\end{align}
and where the ADM mass $\dM$ has been replaced (at this order) by the total mass $m$.
The scale \eqref{eq:def_eta_scale} cancels out in the sum $H^\text{log}+H^\text{hered}$  and was chosen to simplify the expression of the localized hereditary Hamiltonian, see~Eq.~\eqref{eq:H_hered_localized_expr}.  
The logarithmic Hamiltonian $H^\text{log}$ is a non-hereditary perturbation to $H^\text{loc}$, and can be treated straightforwardly using time-independent canonical perturbation theory, as described in Section 12.4 of~\cite{Goldstein}; see also App. C of \cite{Bernard:2016wrg} for its treatment \textit{à la} Sommerfeld using complex analysis. At this point, one only needs to know that $(\ell^\text{loc}, g^\text{loc}, I_{r\phi}^\text{loc}, I_\phi^\text{loc})$ are a set of canonical variables for the Hamiltonian $H^\text{loc} + H^\text{log}$, but are not action-angle variables anymore (the local action variables undergo small 4PN oscillations under the flow generated by $H^\text{loc} + H^\text{log}$). I will not yet construct the canonical transformation that goes from $(\ell^\text{loc}, g^\text{loc}, I_{r\phi}^\text{loc}, I_\phi^\text{loc})$ to some action-angle variables for $H^\text{loc} + H^\text{log}$; instead,  I will  perform this transformation at the end in Sec.~\ref{subsec:averaging}, to obtain an action-angle formulation for the full dynamics $H^\text{loc} + H^\text{log} + H^\text{hered}$.
 
\subsection{Localizing the hereditary Hamiltonian}
\label{subsec:localizing}

The idea is now to perturb $H^\text{loc} + H^\text{log}$ with the nonlocal-in-time, perturbation Hamiltonian $H^\text{hered}$. As discussed at the end of Sec.~\ref{subsec:4PN_Hamiltonian_formalism}, it is licit to `naively order-reduce' the Hamiltonian and compute the on-shell value of the tail piece of the 4PN Hamiltonian, but the resulting motion will be described in terms of new phase-space variables. For my purposes, the map between old and new variables will not be needed. Nonetheless, for completeness, I have worked out this map explicitly in App.~\ref{app:shift_localization}. In order to proceed with the `naive order-reduction',  I will make use of the explicit map between $(r,\phi, p_r,p_\phi)$ and $(\ell,g,I_{r\phi},I_\phi)$ at Newtonian order: the Keplerian parametrization. This is in fact the truncation at Newtonian order of the 4PN (local) quasi-Keplerian parametrization of Ref.~\cite{Cho:2021oai}; the latter is the contact transformation which was implicitly used in Sec.~\ref{sec:local}. I also introduce the eccentric anomaly $u(\ell,I_{r\phi},I_\phi)$, which is defined implicitly by Kepler's equation~\cite{Boetzel:2017zza}:
\begin{align}\label{eq:Kepler_equation_Newtonian}
\ell &= u - e \sin u \,.
\end{align}
The Keplerian parametrization  reads 
\begin{subequations}
\label{eq:r_phi_pr_phi_InTermsOf_a_e_u_Newtonian}
\begin{align}
\label{seq:r_InTermsOf_a_e_u_Newtonian}
r &= a(1- e \cos u) \,,\\*
\label{seq:phi_InTermsOf_a_e_u_Newtonian}
\phi &= g+u + 2 \arctan\Bigl[\frac{\beta(e) \sin u}{1-\beta(e) \cos u}\Bigr] \,, \\*
\label{seq:pr_InTermsOf_a_e_u_Newtonian}
p_r &= m \nu \ \sqrt{\frac{G m }{a}} \    \frac{e \sin u}{1-e\cos u} \,,\\*
\label{seq:pphi_InTermsOf_a_e_u_Newtonian}
p_\phi &= I_\phi\,,
\end{align} 
\end{subequations}
where it should be recalled that $(a,e)$ are given by in terms of the action variables by~\eqref{eq:a_e_inTermsOf_Irphi_Iphi} and where \mbox{$\beta(e)=\tfrac{e}{1+\sqrt{1-e^2}}$.} The equations of motion are then given at Newtonian order by $\dd \ell/ \dd t = n$, where the Newtonian expression for the radial frequency reads 
\begin{align}
\label{eq:n_inTermsOf_Irphi_Iphi_Newtonian}
n &=  \sqrt{\frac{G m}{a^3}}=\frac{G^2 m^5 \nu^3}{I_{r\phi}^{3}} \,,
\end{align}
and by $\dd g / \dd t= 0$. Thus, $g$ is a constant, and the reference frame can always be rotated such that $g=0$. Taking a time derivative of the Kepler equation, one also finds that $\dd u /\dd t = n/(1-e \cos u)$. 

The tail piece of the Hamiltonian is expressed in terms of the quadrupole moment, which reads at Newtonian order
 \begin{align}\label{eq:Iij_matrix}
\dI_{ij} = m \nu r^2 \left(n^i n^j-\frac{1}{3}\delta^{ij}\right) =  m \nu r^2  \begin{pmatrix} \cos^2 \phi - \frac{1}{3} && \cos\phi \sin\phi && 0 \\ \cos\phi \sin\phi && \sin^2 \phi - \frac{1}{3} && 0 \\
0 && 0 && -\frac{1}{3}\end{pmatrix}_{ij} \ .
\end{align}
Using the Keplerian parametrization~\eqref{eq:r_phi_pr_phi_InTermsOf_a_e_u_Newtonian}, all components of the quadrupole moment can now be implicitly expressed in terms of the Delaunay variables $(\ell, g, I_{r\phi}, I_\phi)$, through the eccentric anomaly $u(\ell, I_{r\phi},I_\phi)$.  Using the equations of motion, the $N$-th time derivatives $\dI_{ij}^{(N)}$ are  found to be expressible (after order reduction) as functions of  the Delaunay variables. I finally obtain
\begin{align}\label{eq:dt3Iijdt3Iij_expression}
    \dI_{ij}^{(3)}\dI_{ij}^{(3)} &= \frac{G^3 m^5 \nu^2}{3a^5}\biggl[\frac{88(1-e^2)}{(1-e\cos u)^6} + \frac{16}{(1-e\cos u)^5} - \frac{8}{(1-e\cos u)^4}\biggr] \,,
\end{align}  
such that the logarithmic piece of the Hamiltonian, after replacing $\eta$ by its expression \eqref{eq:def_eta_scale}, reads in Delaunay variables
\begin{align}\label{eq:H_log_expression}
    H^\mathrm{log} &= \frac{2 G^2 m}{5 c^8}  \, \dI_{ij}^{(3)}\dI_{ij}^{(3)}\biggl[\ln (1-e \cos u) + \frac{1}{2}\ln\left(\frac{G m }{c^2a}\right) + 2\ln 2 + \gamma_\mathrm{E}\biggr] \,.
\end{align}
Note that I have not localized anything for the logarithmic piece --- this is simply the result of the transformation from physical phase-space variable to action angles. Now, in order  to localize the hereditary integral explicitly,  some more tools are required. Namely, the periodic nature of a Keplerian orbit needs to be invoked (\textit{via} the Delaunay variables) and  its Fourier decomposition is introduced. Hence,   the components of the quadrupole moment are decomposed into a Fourier series as
\begin{equation}\label{eq:Iij_Fourier_expansion}
    \dI_{ij}  =   \sum_{p\in\mathbb{Z}} {\,}_p\widetilde{\dI}_{ij} \de^{\di p \ell} = \mathcal{I}_2 \sum_{p\in\mathbb{Z}} {\,}_p\widehat{\dI}_{ij} \de^{\di p \ell},
\end{equation}
where I have used $g=0$ and introduced an overall normalization factor 
\begin{align}\label{eq:I2_expression}
\mathcal{I}_2 = m \nu a^2 = \frac{I_{r\phi}^4}{G^2 m^5 \nu^3}\,,
\end{align} such that $ {\,}_p\widetilde{\dI}_{ij} = \mathcal{I}_2  {\,}_p\widehat{\dI}_{ij} $. The various Fourier coefficients  are obtained \textit{via} the formula
\begin{align}\label{eq:pIij_tilde_def}
{}_p \widetilde{\dI}_{ij} &= \frac{1}{2\pi} \int_0^{2\pi} \dd \ell  \, \de^{-\di p \ell} \, \dI_{ij}(\ell, I_{r\phi}, I_\phi)
\end{align}
and their (normalized) expressions (for $p \in \mathbb{Z}^*$) are given by Ref.~\cite{Loutrel:2016cdw}
\begin{subequations}
\label{eq:pIij_hat_expr}
\begin{align}
\label{seq:pIxx_hat_expr}
    {}_p\widehat{\dI}_{xx} &= - \frac{2}{3} \frac{3-e^2}{e^2} \frac{J_p(p e)}{p^2} + \frac{2(1-e^2)}{e} \frac{J_p'(p e)}{p} \,,\\
\label{seq:pIxy_hat_expr}
    {}_p\widehat{\dI}_{xy} &= 2 \di \sqrt{1-e^2} \biggl[- \frac{1-e^2}{e^2} \frac{J_p(p e)}{p} + \frac{1}{e} \frac{J_p'(p e)}{p^2}\biggr] \,,\\
\label{seq:pIyy_hat_expr}
    {}_p\widehat{\dI}_{yy} &= \frac{2}{3} \frac{3-2e^2}{e^2} \frac{J_p(pe)}{p^2} -  \frac{2(1-e^2)}{e}\frac{J_p'(pe)}{p} \,,\\
\label{seq:pIzz_hat_expr}
    {}_p\widehat{\dI}_{zz} &= \frac{2}{3} \frac{J_p(pe)}{p^2} \,,
\end{align}\end{subequations}
where $J_p(x)$ are the Bessel functions of the first kind and  $J'_p(x)$ are their derivatives (see \cite{Arun:2007rg, Bernard:2016wrg, Blanchet:2025agj} for alternative forms). The  $(x,z)$ and $(y,z)$ components are vanishing, as well as all the $p=0$ coefficients. The other components are obtained using the symmetry in the indices, e.g.~\mbox{${}_p\widehat{\dI}_{yx} = {}_p\widehat{\dI}_{xy}$}. Finally, once can check the trace-free character of the quadrupole moment: \mbox{$ {}_p\widehat{\dI}_{xx} +  {}_p\widehat{\dI}_{yy}+  {}_p\widehat{\dI}_{zz} = 0$}.
To compute time derivatives, one can use $\dd / \dd t = \dot{\ell} (\partial/\partial \ell) + \dot{g}  (\partial/\partial g)$, which reduces to $\dd / \dd t = n (\partial/\partial \ell)$ at this order.    The $N$-th derivative of the (order-reduced) quadrupole moment then reads  
\begin{equation}\label{eq:dtNIij_from_Fourier}
    \dI_{ij}^{(N)}  =   \mathcal{I}_2 \sum_{p\in\mathbb{Z}} (\di p n)^N {\,}_p \widehat{\dI}_{ij} \de^{\di p \ell} \,,
\end{equation}
where  the equations of motion $\dd \ell/\dd t= n$ were used, in accordance with the discussion at the beginning of Sec.~\ref{subsec:4PN_Hamiltonian_formalism}. 
The hereditary tail Hamiltonian~\eqref{eq:H_tail} is now naturally expressible in terms of the Delaunay variables at time $t$ and also at time $t' = t- \tau$. Note that I have \textit{not} localized anything at this point; I have  only performed a contact transformation and a local order reduction.  The great advantage of working with Delaunay (action-angle) variables is that their time-evolution (under the Newtonian Hamiltonian) is extremely simple: $I_{r\phi}(t-\tau) = I_{r\phi}(t)$,  $I_{\phi}(t-\tau) = I_{\phi}(t)$,   $\ell(t-\tau) = \ell(t) - n \tau$, $g(t-\tau) = g(t) - (\omega -n) \tau = 0$. However, one must recall that Hamilton's equations~\eqref{eq:Hamilton_equations} are integro-differential, such that one is \textit{a priori} not allowed to express the Delaunay variables at time $t'$ in terms of the Delaunay variables at time $t$ using the equations of motion. Nonetheless, as discussed in Sec.~\ref{subsec:4PN_Hamiltonian_formalism}, it is in fact licit to perform the latter replacements inside the integral in the Hamiltonian if one simultaneously performs a small 4PN contact transformation on the phase space-variables. This was first shown on a toy example in Ref.~\cite{Damour:2016abl}, but the general expression of the shift was worked out for any pseudo-Hamiltonian in (47) of~\cite{Blanco:2024fte} (see also (19) of~\cite{Blanco:2022mgd}), and I compute it explicitly in this case in Appendix~\ref{app:shift_localization}. I find that the new set of variables $(\ell', g', I_{r\phi}', I_\phi')$ are related to their local counterparts by
\begin{subequations}
\label{eq:ell_g_Irphi_Iphi_prime_inTermsOf_loc}
\begin{align}
\label{seq:ell_prime_inTermsOf_loc}
\ell' &= \ell^\text{loc}  -\frac{2\di G^2 m}{5 c^8} \sum_{p+q\neq 0} \frac{p^3 q^3}{p+q}  n^5 \Biggl\{   \frac{\partial ({}_p \widetilde{\dI}_{ij})}{\partial I_{r\phi}} \ln \left|\frac{p}{q}\right| + \frac{1}{n}\frac{\partial n}{\partial I_{r\phi}} \Bigg[1 + \left(2+ \frac{q}{p+q}\right)\ln \left|\frac{p}{q}\right| \Bigg] {}_p \widetilde{\dI}_{ij}  \Biggr\}\  {}_q \widetilde{\dI}_{ij} \de^{\di (p+q)\ell}  \,, \\
\label{seq:g_prime_inTermsOf_loc}
g' &=g^\text{loc}  -\frac{2\di G^2 m}{5 c^8} \sum_{p+q\neq 0} \frac{p^3 q^3}{p+q}  n^5   \ln \left|\frac{p}{q}\right| \frac{\partial ({}_p \widetilde{\dI}_{ij})}{\partial I_{\phi}}     {\ }_q \widetilde{\dI}_{ij} \ \de^{\di (p+q)\ell} \,, \\
\label{seq:Irphi_prime_inTermsOf_loc}
I_{r\phi}' &= I_{r\phi}^\text{loc}    -\frac{2 G^2 m}{5 c^8} \sum_{p+q\neq 0} \frac{p^4 q^3}{p+q} \ln \left|\frac{p}{q}\right| n^5 {\ }_p \widetilde{\dI}_{ij} {\ }_q \widetilde{\dI}_{ij} \ \de^{\di (p+q)\ell} \,, \\
\label{seq:Iphi_prime_inTermsOf_loc}
I_{\phi}' &= I_{\phi}^\text{loc}  \,,
\end{align}
\end{subequations}
where one should recall that $n$ and ${}_p \widetilde{\dI}_{ij} = \mathcal{I}_2 \ {}_p \widehat{\dI}_{ij}$ are expressed in terms of the action variables $I_{r\phi}$ and $I_\phi$; see Eqs.~\eqref{eq:n_inTermsOf_Irphi_Iphi_Newtonian},~\eqref{eq:I2_expression} and~\eqref{eq:pIij_hat_expr}.

As announced, the resulting localized Hamiltonian is obtained by naively replacing the equations of motion inside the hereditary Hamiltonian\footnote{\label{footnote:factor_two_convention_2}In (62) of~\cite{Blanco:2024fte}, (21) of~\cite{Blanco:2022mgd} or (247) of~\cite{Lewis:2025ydo}, there is a factor $1/2$ which is canceled by the factor $2$ of~\eqref{eq:Hamilton_equations}; see also Footnote~\ref{footnote:factor_two_convention}.}, with the additional information that the expression for $H^\text{loc}$ provided in Eq.~\eqref{eq:H_loc_inTermsOf_irphi_iphi} should now be interpreted as a function of $(I_{r\phi}', I_\phi')$ rather than a function of $( I_{r\phi}^\text{loc}, I_\phi^\text{loc})$. 
The Hamiltonian then depends only of the phase-space variables at time $t$ (not at $t' < t$) and on integrals over a dummy (time-like) variable~$\tau$. The latter are computed using the usual formula~\cite{GradshteynRyzhik, Blanchet:2013haa, Arun:2004ff, Blanchet:2023sbv}
\begin{align}\label{eq:integration_formula_tail}
    \int_0^\infty \dd\tau \ln\left(\frac{c \tau}{2 b_0}\right)\de^{\di n p \tau} = - \frac{1}{pn}\left(\frac{\pi}{2} \mathrm{sign}(p) + \di\gamma_{\mathrm{E}} + \di \ln\left(\frac{2|p|n b_0}{c}\right)\right) \,.
\end{align}
Thus, I finally find that the genuinely hereditary piece is given, after localization, by   
\begin{align}\label{eq:H_hered_localized_expr}
    H^\mathrm{hered} &= -\frac{2 G^2 m}{5 c^8} n^6 \left(\mathcal{I}_2\right)^2 \sum_{(p,q)\in\mathbb{Z}^2}p^3 q^3 \ln\left(\frac{|p|}{2}\right)  {\ }_p\widehat{\dI}_{ij} {\ }_q\widehat{\dI}_{ij} \de^{\di(p+q)\ell} \,.
\end{align}

\subsection{Delaunay averaging of the tail Hamiltonian}
\label{subsec:averaging}

The new Hamiltonian is now localized, but depends not only on $(I_{r\phi}', I_\phi')$ but also on $(\ell', g')$.  Thus, $(I_{r\phi}', I_\phi')$ are not constant anymore, and $(\ell', g', I_{r\phi}', I_\phi')$ are not a set of action-angle variables for the perturbed Hamiltonian. To restore the action-angle structure of the Hamiltonian, one performs a so-called Delaunay averaging~\cite{Delaunay_Lune_1,Delaunay_Lune_2}, which was first applied to this problem by Refs.~\cite{Damour:2015isa, Damour:2016abl} following the methods laid out in Ref.~\cite{BrouwerClemence}. This consists in performing a \textit{canonical} transformation on the phase-space variables whose effect is to remove the dependence on the angle variables in the Hamiltonian. Since this dependence manifests itself only in the oscillatory pieces of the Fourier decomposition, one is then left only with the non-oscillatory piece after the procedure. Thus, in practice, the new `Delaunay averaged' Hamiltonian will be obtained \textit{via} naive orbit-averaging. Here, I will follow the treatment laid out in Sec.~12.4 of Ref.~\cite{Goldstein}, and only quote the results in that textbook at leading order in the perturbations (note that my conventions differ by factors of $2\pi$); the reader is invited to refer to it for the detailed derivation.

From~(12.65)~of~\cite{Goldstein}, one learns that the generating function $\mathcal{Y}(\ell', g', I_{r\phi}, I_\phi)$ for the canonical transformation reads
\begin{align}\label{eq:generating_function_Y_def}
\mathcal{Y}(\ell', g', I_{r\phi}, I_\phi) = \ell' I_{r\phi} + g' I_\phi + \delta\mathcal{Y}(\ell', g', I_{r\phi}, I_\phi) + \mathcal{O}\left(\frac{1}{c^{10}}\right) \,,
\end{align}
where $\delta\mathcal{Y}$ must satisfy, according to Eq.~(12.76)~of~Ref.~\cite{Goldstein}, 
\begin{align}\label{eq:contraint_on_delta_Y}
    n' \frac{\partial (\delta\mathcal{Y})}{\partial \ell'} + (\omega'-n') \frac{\partial (\delta\mathcal{Y})}{\partial g'} = \langle H^\text{tail} \rangle  - H^\text{tail}  + \mathcal{O}\left(\frac{1}{c^{10}}\right) \,.
\end{align} 
Here, I have introduced the average of $f$ as $\langle f \rangle = (2\pi)^{-2}\int_0^{2\pi} \dd \ell \int_0^{2\pi} \dd g \ f(\ell,g,I_{r\phi},I_\phi)$.
Neglecting 5PN terms (i.e.,~1PN~terms relatively to the 4PN Hamiltonian) and using  \mbox{$\omega-n =  \mathcal{O}(1/c^2)$}, this simplifies to 
\begin{align}\label{eq:simplified_contraint_on_delta_Y}
    n \frac{\partial (\delta \mathcal{Y})}{\partial \ell}   =   - \widetilde{H}^\text{tail} +  \mathcal{O}\left(\frac{1}{c^{10}}\right) \,,
\end{align} 
where   the oscillatory piece is denoted with a tilde, namely
$\widetilde{f} = f - \langle f \rangle $. 
Moreover, one requires that $\langle \delta \mathcal{Y} \rangle = 0$. One is also free to choose $\delta \mathcal{Y}$ such that it does not depend on $g'$ at this order.
The Hamiltonian in the new (genuinely action-angle) variables will then only depend on $(I_{r\phi}, I_\phi)$, and will be obtained `practically' by removing $(\ell', g')$  from the Hamiltonian in terms of $(\ell', g', I_{r\phi}', I_\phi')$ by orbit-averaging, and then performing the `naive' replacements $I_{r\phi}' \rightarrow I_{r\phi}$ and $I_{\phi}' \rightarrow I_{\phi}$.

I will first obtain the expression of the orbit-averaged Hamiltonian; I will solve for $\delta\mathcal{Y}$ in~\eqref{eq:simplified_contraint_on_delta_Y} in a second step.
The logarithmic Hamiltonian $H^\mathrm{log}$ can be exactly orbit-averaged without the need for an expansion in Fourier series. Recall that the Newtonian problem is degenerate, such that the orbit average of a function of eccentric anomaly $f(u)$ reads    
\begin{align}\label{eq:orbit_average_over_u_formula}
\left\langle f \right\rangle &= \frac{1}{2\pi} \int_{0}^{2\pi} \dd \ell \,  f(u(\ell)) = \frac{1}{2\pi} \int_{0}^{2\pi} \dd u \, (1-e \cos u) f(u)\,,
\end{align}
where I have used $\dd \ell/ \dd u = 1- e \cos (u)$ at this order; see~\eqref{eq:Kepler_equation_Newtonian}.
The orbit averaging procedure is then performed using the formulas provided in  (8.4) and (8.6) of~\cite{Arun:2007sg}, which I reproduce here:  
\begin{subequations}
\label{eq:integration_formula_one_minus_e_cos_u_power_N_with_and_without_log}
\begin{align}
\label{seq:integration_formula_one_minus_e_cos_u_power_N}
\frac{1}{2\pi}\int_0^{2\pi} \frac{\dd u}{(1-e \cos u)^{N}} &= \frac{(-1)^{N-1}}{(N-1)!}\left[\frac{\dd^{N-1}}{\dd y^{N-1}} \left(\frac{1}{\sqrt{y^2 - e^2}}\right)\right]_{y=1} \,,\\
\label{seq:integration_formula_one_minus_e_cos_u_power_N_log}
\frac{1}{2\pi}\int_0^{2\pi} \dd u \ \frac{\ln(1-e\cos u)}{(1-e \cos u)^{N}} &= \frac{(-1)^{N-1}}{(N-1)!} \left[\frac{\dd^{N-1} Y(y;e)}{\dd y^{N-1}}\right]_{y=1} \,,
\end{align}
\end{subequations}
where I have introduced~\cite{Arun:2007sg}
\begin{align}\label{eq:Y_of_y_e_def}
Y(y;e) &= \frac{1}{\sqrt{y^2-e^2}}\Biggl\{\ln\left[\frac{\sqrt{1-e^2}+ 1}{2}\right] + 2 \ln\left[1+\frac{\sqrt{1-e^2}-1}{y+\sqrt{y^2-e ^2}}\right]\Biggr\} \,.
\end{align}
Thus, one finds that
\begin{subequations}
 \label{eq:dt3Iijdt3Iij_averaged_with_and_without_log}
    \begin{align}
    \label{seq:dt3Iijdt3Iij_averaged}
    \left\langle \dI_{ij}^{(3)}\dI_{ij}^{(3)} \right\rangle & = \frac{G^3 m^5 \nu^2(96+292 e^2 + 37 e^4)}{3 a^5(1-e^2)^{7/2}} \,,\\*
    \label{seq:dt3Iijdt3Iij_log_averaged}
    \left\langle \ln\Bigl(1-e\cos(u)\Bigr)\,\dI_{ij}^{(3)}\dI_{ij}^{(3)} \right\rangle & = \frac{G^3 m^5 \nu^2}{36 a^5(1-e^2)^{7/2}}\Biggl[-2408 - 3792 e^2 - 255 e^4 + (2408+2692 e^2) \sqrt{1 - e^2} \nonumber\\*
     &\qquad\qquad\qquad\qquad\quad + (1152 + 3504 e^2 + 444 e^4)\ln\left(\frac{2(1-e^2)}{1+\sqrt{1-e^2}}\right)\Biggr] \,.
    \end{align}
\end{subequations}
Thanks to these expressions, I finally obtain  
\begin{align}
\boxed{
\begin{aligned}
    \left\langle H^\mathrm{log}\right\rangle &= \frac{ G^5 m^6 \nu^2}{a^5 c^8 (1-e^2)^{7/2}}\Biggl\{- \frac{1204}{45} - \frac{632}{15}e^2 - \frac{17}{6}e^4 +  \frac{1204+ 1346 e^2}{45}  \sqrt{1-e^2}  \\*
    &\qquad\qquad\qquad\qquad\quad + \frac{96 + 292 e^2 + 37 e^4}{15} \Biggl[2 \ln\left(\frac{1-e^2}{1+\sqrt{1-e^2}}\right)+\ln\left(\frac{G m}{c^2 a}\right) +6 \ln 2+ 2\gamma_\mathrm{E}\Biggr]\Biggr\} \,.
\end{aligned} 
}
\refstepcounter{equation}\tag{\theequation}\label{eq:H_log_averaged}
\end{align}

The only task left to complete is  the orbit-averaging of the \textit{localized} tail Hamiltonian~\eqref{eq:H_hered_localized_expr}. Since it is already expressed as a Fourier series, the orbit-averaging is straightforward: only terms satisfying $p+q=0$ survive the orbit averaging, and one immediately finds that
\begin{align}\label{eq:H_hered_averaged_inTermsOf_pIij}
    \left\langle H^{\mathrm{hered}}\right\rangle = \frac{2 G^2 m}{5 c^8} n^6 \left(\mathcal{I}_2\right)^2 \sum_{p\in\mathbb{Z}}p^6 \ln \left(\frac{|p|}{2}\right)  {\ }_p\widehat{\dI}_{ij} {\ }_{-p}\widehat{\dI}_{ij} \,.
\end{align}
As in Eq.~(4.1)~of~Ref.~\cite{Munna:2019fjz}, I define the \textit{enhancement function}
\begin{align}
\label{eq:Lambda_0_def}
\boxed{\Lambda_0(e) = \frac{1}{16}\sum_{p=1}^\infty p^{6} \ln \left(\frac{p}{2}\right)  {\ }_p\widehat{\dI}_{ij} {\ }_{-p}\widehat{\dI}_{ij} \,,} \end{align}
where one should recall that the coefficients ${\ }_p\widehat{\dI}_{ij}$ are expressed only in terms of $e$. Replacing the normalization factor and the radial frequency, one then simply finds that
\begin{align}\label{eq:H_hered_averaged_inTermsOf_Lambda_0}
    \boxed{\left\langle H^{\mathrm{hered}}\right\rangle = \frac{64 G^5 m^6 \nu^2}{5 a^5 c^8}   \Lambda_0(e) \,.}
\end{align}
I have compared the small-eccentricity expansion of  $\langle H^{\mathrm{tail}}\rangle = \langle H^{\mathrm{log}}\rangle+ \langle H^{\mathrm{hered}}\rangle$ with Ref.~\cite{Bini:2020wpo} and found perfect agreement at 4PN up to neglected $\mathcal{O}(e^{12})$ terms; see Eqs.~(2.14)--(2.15) and Table I in that reference. I have also compared this small-eccentricity expansion with Eqs.~(52)--(53)~of~Ref.~\cite{Cho:2021oai}. Keeping in mind that the fractions appearing in that work are not genuine fractions but approximations for floating-point numbers~\cite{Sashwat}, I have found \textit{numerical} agreement with that work up to neglected $\mathcal{O}(e^{12})$ terms which are uncontrolled in their approach.

Finally, the oscillatory part of the hereditary Hamiltonian is trivially given by the criteria $p+q\neq 0$, namely
\begin{align}\label{eq:H_hered_oscillatory_expr}
    \widetilde{H}^\mathrm{hered} &= -\frac{2 G^2 m}{5 c^8} n^6 \left(\mathcal{I}_2\right)^2 \sum_{p+q\neq 0}p^3 q^3 \ln\left(\frac{|p|}{2}\right)  {\ }_p\widehat{\dI}_{ij} {\ }_q\widehat{\dI}_{ij} \de^{\di(p+q)\ell} \,.
\end{align}
In order to express explicitly the oscillatory part of the logarithmic piece of the Hamiltonian, one needs to perform the expansion of $\ln(1-e\cos u)$ as a Fourier series, which reads
\begin{align}\label{eq:log_Fourier_expansion}
    \ln(1-e\cos u) = \sum_{p \in \mathbb{Z}} {}_p \widehat{L}  \, \de^{\di p \ell} \,,
\end{align}
where Eq.~\eqref{eq:pIij_tilde_def} leads to the explicit expression
\begin{align}\label{eq:pLhat_def}
    {}_p \widehat{L} = \frac{1}{2\pi} \int_0^{2\pi} \dd u \,  (1-e \cos u)  \, \ln (1-e \cos u)\, \de^{-\di p (u-e\sin u)} \,.
\end{align}
These integrals are \textit{a priori} not expressible in closed form but can be either expanded for small eccentricity or related to the class of extended Bessel functions introduced in Eq.~(23) of Ref.~\cite{Liu:2025tcj}. The oscillatory piece then  reads
\begin{align}\label{eq:H_log_oscillatory_expr}
    \widetilde{H}^\text{log} &= -\frac{2 G^2 m}{5 c^8}  n^6  (\mathcal{I}_2)^2 \Biggl\{ \sum_{p+q+r\neq 0} \!\!\!\!\! p^3 q^3   {\,}_p\widehat{\dI}_{ij} {\,}_q\widehat{\dI}_{ij} {\,}_r{\widehat{L}} \ \de^{\di (p+q+r)\ell} + \left(\frac{1}{2} \ln\left(\frac{G m }{c^2 a}\right) + 2 \ln 2 + \gamma_\text{E} \right) \!\!\sum_{p+q\neq 0} \!\!\!  p^3 q^3  {\,}_p\widehat{\dI}_{ij} {\,}_q\widehat{\dI}_{ij}   \ \de^{\di (p+q)\ell} \Biggr\} \,.
\end{align}
This provides an explicit expression for the oscillatory tail Hamiltonian $\widetilde{H}^\text{tail} = \widetilde{H}^\text{log}  + \widetilde{H}^\text{hered}$, where the individual pieces were computed in Eqs.~\eqref{eq:H_log_oscillatory_expr} and~\eqref{eq:H_hered_oscillatory_expr}. This allows for the  explicit  integration of Eq.~\eqref{eq:simplified_contraint_on_delta_Y} with respect to $\ell$; recalling that $\langle \delta \mathcal{Y} \rangle = 0$ and that $\delta  \mathcal{Y}$ should not depend of $g'$, I find
\begin{align}\label{eq:delta_Y_expr}
    \delta\mathcal{Y} &= -\frac{2 \di G^2 m}{5 c^8}  n^5  (\mathcal{I}_2)^2 \Biggl\{ \sum_{p+q+r\neq 0} \frac{p^3 q^3}{  p+q+r}   {\ }_p\widehat{\dI}_{ij} {\ }_q\widehat{\dI}_{ij} {\ }_r{\widehat{L}} \ \de^{\di (p+q+r)\ell'} \nonumber\\*
    & \qquad\qquad  + \sum_{p+q\neq 0} \left[\frac{1}{2} \ln\left(\frac{G m }{c^2 a}\right) + 2 \ln 2 + \gamma_\text{E} + \ln\left(\frac{|p|}{2}\right)\right] \frac{p^3 q^3}{p+q}  {\ }_p\widehat{\dI}_{ij} {\ }_q\widehat{\dI}_{ij}   \ \de^{\di (p+q)\ell'}   \Biggr\}  \,.
\end{align}
Finally, the canonical transformation associated with the Delaunay averaging is given, according to Eq.~(12.68) and (12.73) of Ref.~\cite{Goldstein}, by
\begin{align}\label{eq:ell_g_Irphi_Iphi_noprime_inTermsOf_prime}
    \ell &= \ell' +  \frac{\partial (\delta \mathcal{Y})}{\partial I_{r\phi}'}\ ,&&&
    g &=g' +  \frac{\partial (\delta \mathcal{Y})}{\partial I_{\phi}'}\ ,\nonumber\\
    I_{r\phi} &= I_{r\phi}' - \frac{\partial (\delta \mathcal{Y})}{\partial \ell'}\ ,&&&
     I_{\phi} & = I_{\phi}'\ .
\end{align}
Since at Newtonian order, the Hamiltonian $H = - G^2 m^5 \nu^3/(2 I_{r\phi}^2)\, + \, \mathcal{O}(1/c^2)$ only depends on $I_{r\phi}$, it is straightforward to verify that the shift $\delta I_{r\phi}$ given in Eq.~\eqref{eq:ell_g_Irphi_Iphi_noprime_inTermsOf_prime} leads to a variation of the Hamiltonian $\delta H = n \,\delta I_{r\phi}$ which exactly cancels out the oscillatory piece of the Hamiltonian given in Eqs.~\eqref{eq:H_log_oscillatory_expr} and~\eqref{eq:H_hered_oscillatory_expr}; only the secular piece remains. The shifts with respect to the other variables have the only purpose of preserving the canonical nature of the contact transformation, which follows automatically from the existence of a generating function.

\subsection{Resummation of the tail enhancement function}
\label{subsec:resumming}

\begin{figure}[b] 

\begin{tikzpicture}
    \node[anchor=south west, inner sep=0] (img) 
        {\includegraphics[width=\textwidth,trim=0 10 0 5,clip]{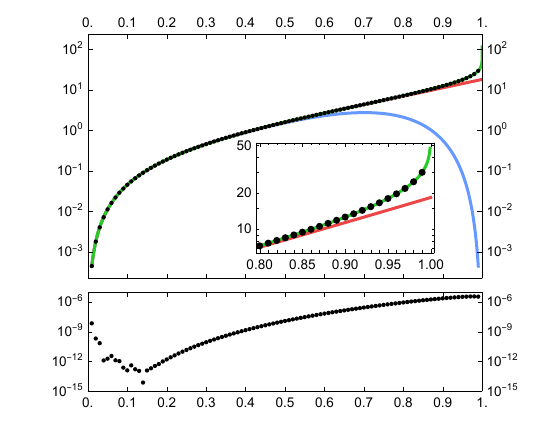}};
    
    \node[anchor=west] at ([xshift=9.25cm, yshift=0cm]img.south west) {\Large $e$};
    
    \node[anchor=west, rotate=90] at ([xshift=1.25cm, yshift=7.25cm]img.south west) {\Large $(1-e^2)^{7/2}\,\Lambda_0(e)$};

    \node[anchor=west, rotate=90] at ([xshift=1.25cm, yshift=0.9cm]img.south west) {\Large ${|\Lambda_0 - \Lambda_0^{[4]}|}/{\left|\Lambda_0\right|}$};
\end{tikzpicture}

\caption{\label{fig:enhancement}
\textit{(Top panel)} Comparison of various estimates for $\Lambda_0(e)$, normalized by the divergent factor $(1-e^2)^{-7/2}$: (i) the black dots are the data points for the reference numerical estimate; (ii) the blue curve corresponds to the small-eccentricity expansion~\eqref{eq:Lambda_0_small_e_expansion}, including terms up to $\mathcal{O}(e^{10})$; (iii) the red curve corresponds to the naive resummation~\eqref{eq:Lambda_0_resummed_naive_truncated_N} with $N = 4$; and (iv) the green curve corresponds to the proposed resummation~\eqref{eq:Lambda_0_resummed_truncated_N} with $N=4$. The small eccentricity expansion becomes fully inadequate for $e \ge 0.5$; whereas both resummations provide a reasonable estimate for larger $e$. \mbox{\textit{(Inset plot in top panel)}}~Same as the top panel, but zooming into the large eccentricity region, namely $0.8 \le e \le 1$. The `naive' resummation is not accurate in this regime, whereas the proposed resummation is extremely accurate. \textit{(Bottom panel)} Relative error of the proposed resummation $\Lambda_0^{[N]}(e)$ for $N=4$ [see Eq.~\eqref{eq:Lambda_0_resummed_truncated_N}], with respect to the numerical estimate for $\Lambda_0(e)$. For any value of the eccentricity, the relative error  always remains smaller than $4\cdot 10^{-6}$. For $e \le 0.15$, the error is dominated by numerical noise, due to the division by $|\Lambda_0(e)|$ which vanishes in the $e\rightarrow 0$ limit. For $e>0.15$, the error is dominated by the inaccuracy of the resummation; this required controlling the numerical estimate very precisely by summing over many modes. For $e=0.2$, $p_\mathrm{max}=20$ modes are enough, whereas for $e=0.99$, $p_\mathrm{max}=15\,000$ modes were required!}
\end{figure}

In order to obtain an explicit expression for $\left\langle H^{\mathrm{hered}}\right\rangle$ (without an unwieldy infinite sum), one approach is to perform a small eccentricity expansion of $\Lambda_0(e)$, defined in \eqref{eq:Lambda_0_def}; thanks to the properties of the Bessel functions, any finite truncation in the eccentricity expansion has the immediate effect of truncating the Fourier sum. Thus, I find that the~$e \rightarrow 0$~expansion of this enhancement function reads 
\begin{align}
\label{eq:Lambda_0_small_e_expansion}
    \Lambda_0(e)&\underset{e \rightarrow 0}{\sim} e^2 \left[-\frac{277}{24} \ln 2+\frac{729}{64}\ln 3\right] + e^4 \left[\frac{11353}{96}\ln 2 -\frac{13851}{256} \ln 3\right] \nonumber \\
   &\quad +e^6 \left[-\frac{21997}{32} \ln 2+\frac{419661}{4096} \ln 3 +\frac{9765625}{36864} \ln 5\right]  +e^8 \left[\frac{5056751}{2304} \ln 2+\frac{26915409}{32768} \ln 3-\frac{419921875}{294912}\ln 5\right] \nonumber \\
   &\quad +e^{10} \left[-\frac{4852988101}{691200} \ln 2-\frac{138733913079}{26214400}\ln
   3+\frac{93681640625}{28311552} \ln 5 +\frac{678223072849}{235929600}  \ln 7 \right]+ \mathcal{O}\left(e^{12}\right);
\end{align}
see Eq.~\eqref{eq:coeff_calA_p} for the general term and the Supplemental Material~\cite{Supplemental} for an implementation in \texttt{Wolfram Mathematica}. However, it is well known that the small eccentricity expansion performs very badly even for moderate eccentricities; see Fig.~\ref{fig:enhancement}. To solve this problem, two groups~\cite{Loutrel:2016cdw, Forseth:2015oua, Munna:2019fjz, Munna:2020juq} have independently investigated the asymptotic properties of similar enhancement functions in the limit where~$e \rightarrow 1$, allowing them to resum the enhancement functions in a way that preserves accuracy for any eccentricity. This is possible because \textit{all} the terms entering the infinite sum are controlled analytically. Following the recipe laid out in Sec.~5.1 of~\cite{Loutrel:2016cdw}, I computed the asymptotic expansion  of~$\Lambda_0(e)$. This relies on the observation that the sum~\eqref{eq:Lambda_0_def} is dominated by the $p\rightarrow \infty$ terms as $e \rightarrow 1$. Thus, the procedure consists in three steps: (i)~compute the uniform asymptotic expansion of each term in the sum as $p \rightarrow \infty$ (this involves expanding Bessel functions, whose uniform asymptotic expansions involve Airy functions, or equivalently, modified Bessel functions of the second kind); (ii)~replace the sum by an integral and compute it explicitly (the integrands all come as quadratic products of modified Bessel functions of the second kind); and (iii) Taylor expand the result as $\epsilon = 1-e^2 \rightarrow 0$. This procedure is in fact well defined only for the first few terms in the expansion; after a certain \textit{breakdown order}, which usually corresponds to terms of order $\mathcal{O}(\epsilon^0)$, an infinite number of integrals contribute to each order and the procedure becomes ill-defined~\cite{Loutrel:2016cdw}. This is not a problem \textit{per se} because the resulting expansion is a (divergent) asymptotic series anyway: for such series, keeping only the few first terms in the expansion usually yields a very good approximation of the original function, but adding more terms can degrade the agreement and eventually lead to a divergence. Thus, there exists an $e$-dependent \textit{optimal truncation} which minimizes the error incurred by the expansion; the resulting optimal expansion is usually called ``superasymptotic'' \cite{Boyd1999}. Here, I have only kept the first three terms in the expansion, and have not investigated what the optimal truncation order was (this order would depend on $e$ anyway), nor whether the neglected~$\mathcal{O}(\epsilon^{-1/2})$ contribution is in fact well defined by this procedure. I find that the asymptotic behavior of the enhancement function  as $e \rightarrow 1$ reads
\begin{align}\label{eq:Lambda_0_superasymptotic}
\Lambda_0(e) \underset{e\rightarrow 1}{\sim}\quad&\frac{1}{(1-e^2)^{7/2}}\Biggl[\frac{65}{3}-\frac{425}{96}\gamma_\text{E} -\frac{425}{24}\ln 2 -\frac{425}{192}  \ln 3 - \frac{425}{64} \ln (1-e^2) \Biggr] \nonumber \\
 + &\frac{1}{(1-e^2)^{5/2}}\Biggl[-\frac{3301}{160}+\frac{61}{16}\gamma_\text{E}+\frac{61}{4} \ln 2 +\frac{61}{32} \ln 3 + \frac{183}{32} \ln \left(1-e^2\right)\Biggr]  \nonumber\\
 + &\frac{1}{(1-e^2)^{3/2}}\Biggl[\frac{31707}{11200}-\frac{37}{96} \gamma _\text{E} -\frac{37}{24} \ln 2 -\frac{37}{192}\ln 3 -\frac{37}{64} \ln \left(1-e^2\right)\Biggr]  + \mathcal{O}\left(\frac{1}{\sqrt{1-e^2}}\right) \,.
\end{align}
Crucially, one notices that factoring out $\ln(1-e^2)$ reveals a prefactor that resums exactly (up to a factor $-3/2$) into the Peters and Mathews enhancement function 
\begin{align}\label{eq:f_of_e_Peters_Mathews}
    f(e) = \frac{1}{16}\sum_{p=1}^\infty \, p^{6}  {\ }_p\widehat{\dI}_{ij} {\ }_{-p}\widehat{\dI}_{ij} = \frac{1}{(1-e^2)^{7/2}} \left(1 + \frac{73}{24}e^2+ \frac{37}{96} e^4\right) \,;
\end{align} 
notice that this infinite sum would be proportional to Eq.~\eqref{eq:Lambda_0_def} if it were not for the $\ln(p/2)$ contribution.
Based on the behavior of similar enhancement functions, see, e.g., (171) and (172) of~\cite{Loutrel:2016cdw}, I will assume that the logarithmic factor is in fact exact, namely that there are no more logarithmic terms at higher orders in the expansion. Thus, I recover the behavior predicted by (1.1), (1.3), (1.5) and (4.3) of Ref.~\cite{Munna:2019fjz}: 
\begin{align}\label{eq:Lambda_0_e_1_behavior_Munna}
    \Lambda_0(e)  \underset{e\rightarrow 1}{\sim}  - \frac{3}{2} f(e) \ln(1-e^2) \,.
\end{align}
This leads to the following proposed resummation:
\begin{align}\label{eq:Lambda_0_resummed}
    \boxed{\Lambda_0(e) = -\frac{3}{2} \biggl[f(e) \ln(1-e^2) + \frac{e^2}{(1-e^2)^{7/2}}\lambda_0(e)\biggr] \,, {\color{white} \Bigg(}  \!\! \!\!  }
\end{align}
where  $\lambda_0(e)$ is \textit{defined} by the relation~\eqref{eq:Lambda_0_resummed}. From Eqs.~\eqref{eq:Lambda_0_small_e_expansion} and \eqref{eq:Lambda_0_superasymptotic}, one reads off: 
\begin{subequations} \label{eq:lambda0_value_01}\begin{align}
     \label{seq:lambda0_value_0}
     \lambda_0(0) &= 1+\frac{277}{36}\ln 2 - \frac{243}{32}\ln 3 \approx  -2.0092 \,,\\
    \label{seq:lambda0_value_1}
    \lambda_0(1)& = -\frac{130}{9}+\frac{425}{144}\gamma_{\text{E}}+\frac{425}{36}\ln 2+\frac{425}{288}\ln 3 \approx -2.9367 \,.
\end{align}\end{subequations}
The value for $e=1$ agrees exactly with an analogous computation for scattering orbits \cite{Bini:2017wfr}; obtaining a complete description for $e\in[0,+\infty)$ will be the object of future work. 
In practice,   $\lambda_0(e)$ will be replaced by its truncated small eccentricity expansion
\begin{align}
\label{eq:lambda_0_expansion}
\lambda_0^{[N]}(e) &= \sum_{p=0}^{N} \alpha_p e^{2p}\,,\end{align}
from which one defines the approximate (or truncated) resummation
\begin{align}\label{eq:Lambda_0_resummed_truncated_N}
    \Lambda_0^{[N]}(e) = -\frac{3}{2} \Biggl[f(e) \ln(1-e^2) + \frac{e^2}{(1-e^2)^{7/2}}\lambda_0^{[N]}(e)\Biggr] \,.
\end{align}
In order to ensure that the same small-eccentricity expansions of $\Lambda_0(e)$ and $\Lambda_0^{[N]}(e)$ coincide, one requires that the coefficients are given (for $N=4$) by
\begin{subequations}
\label{eq:alpha_n_coeffs}
\begin{align}
\label{seq:alpha_0_coeff}
    \alpha_0 &= 1+\frac{277}{36}\ln 2-\frac{243}{32}\ln 3  \,,\\
\label{seq:alpha_1_coeff}
    \alpha_1 &= \frac{85}{24}-\frac{5077}{48}\ln 2+\frac{8019}{128}\ln 3 \,,\\
\label{seq:alpha_2_coeff}
    \alpha_2 &= \frac{215}{96}+\frac{6143}{8}\ln 2-\frac{466479}{2048}\ln 3-\frac{9765625 }{55296}\ln 5 \,, \\
\label{seq:alpha_3_coeff}
    \alpha_3 &= \frac{839}{576}-\frac{5925115}{1728}\ln 2 -\frac{2197287}{16384}\ln 3+\frac{693359375}{442368} \ln 5 \,,\\
\label{seq:alpha_4_coeff}
    \alpha_4 &= \frac{49}{45}+\frac{3105571069}{259200}\ln 2+\frac{66387426093}{13107200} \ln 3-\frac{267587890625}{42467328}\ln 5 -\frac{678223072849}{353894400}\ln 7 \,;
\end{align}\end{subequations}
see Eq.~\eqref{eq:alpha_n_general} for the general term and the Supplemental Material~\cite{Supplemental} for an implementation in \texttt{Wolfram Mathematica}.

Moreover, sparked by a comment in Ref.~\cite{Munna:2019fjz}, which claims that ``it is not necessary to isolate the logarithmic divergence in $\Lambda_2(e)$'' [where $\Lambda_2(e)$ is another enhancement function in the same family as $\Lambda_0(e)$], I also considered the following naive, truncated resummation:
\begin{align}\label{eq:Lambda_0_resummed_naive_truncated_N}
    {\Lambda}_{0,\text{naive}}^{[N]}(e) = \frac{e^2}{(1-e^2)^{7/2}}\sum_{p=0}^{N} \alpha^\text{naive}_p e^{2p}  \,,
\end{align}
where the coefficient $\alpha^\text{naive}_p$ are immediately determined by requiring that the small-eccentricity expansions of $\Lambda_0(e)$ and ${\Lambda}_{0,\text{naive}}^{[N]}(e)$ coincide. 

I now compare these resummations with a numerical estimate for $\Lambda_0(e)$. This numerical estimate was obtained by truncating the sum $\Lambda_0(e)$ to some order $p_\text{max}(e)$, which is determined \textit{ad hoc} by requiring that the relative error between the numerical estimate and the resummation $\Lambda_0^{[4]}(e)$ has converged to a stable value. Note that for better precision, I did not expand the Bessel functions in small eccentricity when computing this numerical estimate~\cite{Liu:2025tcj}. I~obtained this estimate for $e \in \{k/100, k\in[\![1,99]\!]\}$, and the truncation order of Eq.~\eqref{eq:Lambda_0_def} increases with eccentricity: for $e=0.2$, $p_\mathrm{max}=20$ modes are enough, whereas for $e=0.99$, $p_\mathrm{max}=15\,000$ modes were required.  In Fig.~\ref{fig:enhancement}, I compare the following quantities, which are normalized by the divergent factor $(1-e^2)^{-7/2}$: (i)~the numerical estimate;  (ii)~the small eccentricity expansion~\eqref{eq:Lambda_0_small_e_expansion} of $\Lambda_0(e)$, neglecting terms that are $\mathcal{O}(e^{12})$; (iii)~the naive resummation~\eqref{eq:Lambda_0_resummed_naive_truncated_N} with $N = 4$; and (iv)~my proposed resummation~\eqref{eq:Lambda_0_resummed_truncated_N}  with $N = 4$. I find that the small eccentricity expansion completely breaks down for $e \gtrsim 0.5$, but that both resummations perform  well for large eccentricities. When focusing on eccentricities greater than $0.8$, it is moreover clear that the proposed resummation~\eqref{eq:Lambda_0_resummed_truncated_N} is much closer to the numerical value than the naive resummation~\eqref{eq:Lambda_0_resummed_naive_truncated_N}: the relative error on $\Lambda_0(e)$ for the proposed resummation (with $N=4$) remains smaller than $4\cdot 10^{-6}$ for any value of the eccentricity! One concludes that it is very useful to control the logarithmic behavior of the enhancement function, and that the resummed enhancement function~\eqref{eq:Lambda_0_resummed_truncated_N} is both computationally efficient and numerically accurate, even for eccentricities very close to $1$. Note that an analogous comparison was  carried out in Fig.~1~of~Ref.~\cite{Henry:2023tka}; there, the errors were dominated by the numerical noise. 

\begin{figure}[t]

\begin{tikzpicture}
    \node[anchor=south west, inner sep=0] (img) 
        {\includegraphics[width=\textwidth,trim=0 10 0 5,clip]{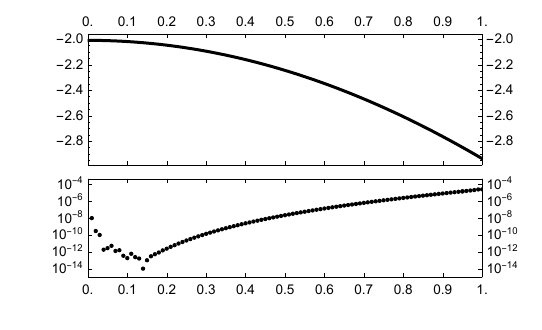}};
    
    \node[anchor=west] at ([xshift=9.25cm, yshift=0cm]img.south west) {\Large $e$};
    
    \node[anchor=west, rotate=90] at ([xshift=1.25cm, yshift=6.2cm]img.south west) {\Large $\lambda_0(e)$};

    \node[anchor=west, rotate=90] at ([xshift=1.25cm, yshift=0.9cm]img.south west) {\Large ${|\lambda_0 - \lambda_0^{[4]}|}/{\left|\lambda_0\right|}$};
\end{tikzpicture}

\caption{\label{fig:enhancement_lambda_0}
\textit{(Top panel)} Numerical plot of $\lambda_0(e)$.  \textit{(Bottom panel)} Relative error of the small eccentricity expansion $\lambda_0^{[N]}(e)$ for $N=4$ [see Eq.~\eqref{eq:lambda_0_expansion}], with respect to the numerical estimate for $\lambda_0(e)$. For any value of the eccentricity, the relative error $\Delta \lambda_0^{[4]}(e) = |\lambda_0^{[4]}(e)-\lambda_0(e)|/|\lambda_0(e)|$ always remains smaller than $\Delta\lambda_0^{[4]}(1) \approx 3\cdot 10^{-5}$. For $e \le 0.15$, the error is dominated by numerical noise.}
\end{figure}

Finally, using the same numerical setup, I performed a similar comparison for $\lambda_0(e)$ itself. This function is finite across the range $e\in[0,1]$ , and its exact value for $e=1$ is known, see Eq.~\eqref{seq:lambda0_value_1}. I thus find that the relative error function, defined as \mbox{$\Delta \lambda^{[N]}_0(e)=|\lambda_0(e) - \lambda_0^{[N]}(e)|/|\lambda_0(e)|$}, is bounded by its value at $e=1$, and that for $N=4$,  $\Delta \lambda^{[4]}_0(1) \approx 3\cdot10^{-5}$. The functions $\lambda_0(e)$ and the associated error $\Delta \lambda^{[4]}_0(e)$ are plotted for reference in Fig.~\ref{fig:enhancement_lambda_0}.

\subsection{Tail Hamiltonian in terms of action variables}
\label{subsec:tail_Hamiltonian}

Using the Newtonian relation between the Keplerian parameters and the action variables~\eqref{eq:a_e_inTermsOf_Irphi_Iphi}, I find that the tail Hamiltonian is expressed in terms of the action variables as follows:
\begin{subequations}
\label{eq:H_log_hered_inTermsOf_irphi_iphi}
\begin{align}
\label{seq:H_log_inTermsOf_irphi_iphi}
    H^\text{log} &= \frac{m \nu^2}{c^8} \Biggl\{\frac{1}{i_\phi^{\,7} i_{r\phi}^{\,3}}\left( -\frac{1291}{18}+\frac{170}{3}\gamma_\mathrm{E}\right)+\frac{170}{3
   i_\phi^{\,6} i_{r\phi}^{\,4}}+\frac{1}{i_\phi^{\,5} i_{r\phi}^{\,5}}\left(\frac{239}{5}-\frac{244}{5}\gamma_\mathrm{E}\right)-\frac{1346}{45
   i_\phi^{\,4} i_{r\phi}^{\,6}}+\frac{1}{i_\phi^{\,3} i_{r\phi}^{\,7}}\left(-\frac{17}{6}+\frac{74}{15} \gamma_\mathrm{E}\right) \nonumber\\
   &\qquad\qquad +  \ln \left(\frac{c \, i_{r\phi}^{\,2} (i_\phi+i_{r\phi})}{8
   i_\phi^{\,2}}\right)\left[-\frac{170}{3 i_\phi^{\,7} i_{r\phi}^{\,3}}+\frac{244}{5 i_\phi^{\,5}
   i_{r\phi}^{\,5}}-\frac{74}{15 i_\phi^{\,3} i_{r\phi}^{\,7}}\right] \Biggr\} \,,\\
\label{seq:H_hered_inTermsOf_irphi_iphi}
    H^\text{hered} &= \frac{m\nu ^2}{c^8}\Biggl\{ \ln
   \left(\frac{i_{r\phi}}{i_\phi}\right)\left[\frac{170}{\, i_\phi^{\,7} i_{r\phi}^{\,3}}-\frac{732}{5 \, i_\phi^{\,5}
   i_{r\phi}^{\,5}}+\frac{74}{5 \, i_\phi^{\,3} i_{r\phi}^{\,7}}\right]+\frac{96}{5}  \lambda_0\!
   \left(\sqrt{1-\frac{i_\phi^{\,2}}{i_{r\phi}^{\,2}}}\right)\left[\frac{1}{i_\phi^{\,5}
   i_{r\phi}^{\,5}}-\frac{1}{i_\phi^{\,7} i_{r\phi}^{\,3}}\right] \Biggr\} \,,
\end{align}\end{subequations}
where $\lambda_0$ is defined by~\eqref{eq:Lambda_0_def} and~\eqref{eq:Lambda_0_resummed} but is well approximated by~\eqref{eq:lambda_0_expansion} and~\eqref{eq:alpha_n_coeffs}.

The full Hamiltonian has now been obtained in terms of action variables 
\begin{align}\label{eq:H_inTermsOf_irphi_iphi_sum_loc_log_hered}
    H(i_{r\phi},i_\phi) = H^\mathrm{loc}(i_{r\phi},i_\phi)  + H^\mathrm{log}(i_{r\phi},i_\phi) + H^\mathrm{hered}(i_{r\phi},i_\phi)\,,
\end{align}
where  $H^\mathrm{loc}(i_\phi,i_{r\phi})$ was  given by Eq.~\eqref{eq:H_loc_inTermsOf_irphi_iphi}, but which is now to be understood as a relation in terms of the new, perturbed action variables $(i_\phi,i_{r\phi})$ [rather than in terms of the old, nonperturbed, local action variables $(i_{r\phi}^{\mathrm{loc}}, i_\phi^{\mathrm{loc}})$].

In the rest of this work, I will analogously split various quantities into a local part, a logarithmic part and a hereditary part. Let me now define precisely what is meant by this split. Consider a quantity $Q[\mathcal{H}]$ which is a functional of some Hamiltonian $\mathcal{H}$. The local piece of $Q$ is denoted by $Q^\text{loc}= Q[H^\text{loc}]$, and is defined as the quantity obtained if one only considers the local piece of the Hamiltonian. The logarithmic piece of $Q$ is then denoted by $Q^\text{log}= Q[H^\text{loc}+H^\text{log}] - Q^\text{loc}$, which is defined as the quantity one needs to add to $Q^\text{loc}$ if one now considers both the local and logarithmic pieces, but ignores the hereditary piece. Finally, one defines the hereditary piece $Q^\text{hered}= Q[H^\text{loc}+H^\text{log}+H^\text{hered}] - Q^\text{loc}-Q^\text{log}$, namely the missing piece one needs to add to $Q^\text{loc}+Q^\text{log}$ in order to account for the full physical Hamiltonian. Thus, of course, the only meaningful physical quantity is $Q=Q^\text{loc}+Q^\text{log}+Q^\text{hered}$, but this split is handy for presentation and computational purposes. One subtlety is that the quantity $Q$ can sometimes be expressed in terms of two different pairs of variables, say $(x_1,y_1)$ and $(x_2,y_2)$, which leads to the splits $Q(x_1,y_1)=Q^\text{loc}(x_1,y_1)+Q^\text{log}(x_1,y_1)+Q^\text{hered}(x_1,y_1)$ and $Q(x_2,y_2)=Q^\text{loc}(x_2,y_2)+Q^\text{log}(x_2,y_2)+Q^\text{hered}(x_2,y_2)$. Although the numerical equality $Q(x_1,y_1) = Q(x_2,y_2)$  always holds (up to neglected higher-order PN terms), this is not always the case for the individual pieces; e.g., one can have $Q^\text{hered}(x_1,y_1) \neq Q^\text{hered}(x_2,y_2)$. This is because there can be indirect logarithmic and hereditary corrections arising from the choice of variables.   

\subsection{Tail contributions to the radial action in terms of energy and angular momentum}
\label{subsec:tail_radial_action}

It is now straightforward to invert the map between energy and radial action. 
The local contribution to $i_{r\phi}$ in terms of $(\varepsilon,j)$ has already been  given in Eq.~\eqref{eq:irphi_loc_inTermsOf_varepsilon_j}. 
Then, recalling that $E = H(i_\phi,i_{r\phi})$ and $J = I_\phi$, I computed using Eq.~\eqref{seq:H_log_inTermsOf_irphi_iphi} the correction needed to account for the logarithmic piece in the Hamiltonian, which reads 
\begin{subequations}
\label{eq:irphi_log_hered_inTermsOf_varepsilon_j}
\begin{align}
\label{seq:irphi_log_inTermsOf_varepsilon_j}
    i_{r\phi}^\text{log} = i_{r}^\text{log} = \frac{\nu \,\varepsilon ^{7/2} }{c \,j^{3/2}}  \Biggl\{\frac{17}{6}+\frac{1346}{45
   \sqrt{j}}-\frac{239}{5 j}-\frac{170}{3 j^{3/2}}+\frac{1291}{18 j^2}+\left(\frac{74}{15}-\frac{244}{5 j}+\frac{170}{3 j^2}\right) \left[\ln
   \left(\frac{1+\sqrt{j}}{8 j \sqrt{\varepsilon }}\right) -\gamma_\text{E}\right]\Biggr\}\,.
\end{align}
Finally, the last correction, needed to account for the hereditary term in the Hamiltonian, follows from Eq.~\eqref{seq:H_hered_inTermsOf_irphi_iphi} and reads
\begin{align}
\label{seq:irphi_hered_inTermsOf_varepsilon_j}
    i_{r\phi}^\text{hered}  & = i_{r}^\text{hered}= \frac{\varepsilon ^{7/2} \nu }{c j^{3/2}}\Biggl\{ \ln j \left(\frac{37}{5}-\frac{366}{5
   j}+\frac{85}{j^2}\right)+\frac{96}{5}\lambda_0\!\left(\sqrt{1-j}\right)  \left(\frac{1}{j^2}-\frac{1}{j}\right)
    \Biggr\}\,.
\end{align} 
\end{subequations}
The final inverse maps are thus 
\begin{subequations}
\label{eq:irphi_ir_inTermsOf_varepsilon_j_sum_loc_log_hered}
\begin{align}
\label{seq:irphi_inTermsOf_varepsilon_j_sum_loc_log_hered}
    i_{r\phi}(\varepsilon,j) &= i_{r\phi}^\text{loc}(\varepsilon,j) + i_{r\phi}^\text{log}(\varepsilon,j) + i_{r\phi}^\text{hered}(\varepsilon,j) \,,\\
    \label{seq:ir_inTermsOf_varepsilon_j_sum_loc_log_hered}
    i_{r}(\varepsilon,j) &= i_{r}^\text{loc}(\varepsilon,j) + i_{r}^\text{log}(\varepsilon,j) + i_{r}^\text{hered}(\varepsilon,j) \,,
\end{align}
where the various components are given explicitly by Eqs.~\eqref{eq:irphi_loc_inTermsOf_varepsilon_j} and~\eqref{eq:irphi_log_hered_inTermsOf_varepsilon_j}.
\end{subequations}

\section{Fundamental frequencies in terms of energy and angular momentum}
\label{sec:frequencies}


\subsection{Radial and azimuthal frequencies $(n,\omega)$}

The fundamental frequencies are straightforward to obtain in the action-angle approach, since they are by construction conjugate to action variables; see Secs.~\ref{subsec:angles_loc} and~\ref{sec:tails}. Since the full Hamiltonian \mbox{$H = H^\text{loc} + H^\text{log} + H^\text{hered}$} has now been constructed in terms of action variables, one can immediately obtain the radial and azimuthal frequencies through partial differentiation~\cite{Goldstein}. Here, it will be instructive to follow the conventions of Sec.~\ref{subsec:tail_Hamiltonian} to split these frequencies into local, logarithmic and hereditary terms, which are defined as 
\begin{subequations}
\label{eq:n_omega_loc_log_hered_from_H}
    \begin{align}
        \label{seq:n_omega_loc_from_H}
        n^{\text{loc}}(i_{r\phi},i_\phi) &= \frac{1}{Gm^2 \nu} \frac{\partial H^\text{loc}}{\partial i_{r\phi}}\Bigg|_{i_\phi} \,, &  \omega^{\text{loc}}(i_{r\phi},i_\phi) &= n^\text{loc}(i_{r\phi},i_\phi) + \frac{1}{G m^2 \nu} \frac{\partial H^\text{loc}}{\partial i_\phi} \Bigg|_{i_{r\phi}} \,, \\
        \label{seq:n_omega_log_from_H}
        n^{\text{log}}(i_{r\phi},i_\phi) &= \frac{1}{G m^2 \nu} \frac{\partial H^\text{log}}{\partial i_{r\phi}} \Bigg|_{i_\phi} \,, &   \omega^{\text{log}}(i_{r\phi},i_\phi) &= n^\text{log}(i_{r\phi},i_\phi)+ \frac{1}{G m^2 \nu} \frac{\partial H^\text{log}}{\partial i_\phi} \Bigg|_{i_{r\phi}} \,, \\
        \label{seq:n_omega_hered_from_H}
        n^{\text{hered}}(i_{r\phi},i_\phi) &= \frac{1}{G m^2 \nu} \frac{\partial H^\text{hered}}{\partial i_{r\phi}} \Bigg|_{i_\phi} \,, & \omega^{\text{hered}}(i_{r\phi},i_\phi) &= n^\text{hered}(i_{r\phi},i_\phi)+ \frac{1}{G m^2 \nu} \frac{\partial H^\text{hered}}{\partial i_\phi} \Bigg|_{i_{r\phi}}\,.
    \end{align}
\end{subequations}
The total frequencies in terms of action variables   read 
\begin{subequations}
\label{eq:n_omega_inTermsOf_irphi_iphi_sum_loc_log_hered}
\begin{align}
\label{seq:n_inTermsOf_irphi_iphi_sum_loc_log_hered}
    n(i_{r\phi},i_\phi) &= n^\text{loc}(i_{r\phi},i_\phi) + n^\text{log}(i_{r\phi},i_\phi)+ n^\text{hered}(i_{r\phi},i_\phi) \,, \\*
\label{seq:omega_inTermsOf_irphi_iphi_sum_loc_log_hered}
    \omega(i_{r\phi},i_\phi) &= \omega^\text{loc}(i_{r\phi},i_\phi) + \omega^\text{log}(i_{r\phi},i_\phi) + \omega^\text{hered}(i_{r\phi},i_\phi)\,,
\end{align}
\end{subequations}
where different contributions to $n$ and $\omega$ are given, respectively, in Eqs.~\eqref{eq:n_loc_log_hered_inTermsOf_irphi_iphi} and~\eqref{eq:omega_loc_log_hered_inTermsOf_irphi_iphi}.
It is then immediate to replace the action variables $(i_{r\phi}, i_\phi)$ by their full expressions Eq.~\eqref{eq:irphi_ir_inTermsOf_varepsilon_j_sum_loc_log_hered} in terms of energy and angular momentum~$(\varepsilon,j)$, including both local and tail contributions. The map between fundamental frequencies $(n,\omega)$ and energy and angular momentum $(\varepsilon, j)$  is obtained schematically by 
\begin{subequations}\label{eq:n_omega_inTermsOf_varepsilon_j_procedure}
    \begin{align}\label{seq:n_inTermsOf_varepsilon_j_procedure}
    n(\varepsilon,j) &= \Biggl[n^\text{loc}(i_{r\phi},i_\phi)+n^\text{log}(i_{r\phi},i_\phi)+n^\text{hered}(i_{r\phi},i_\phi)\Biggr]_{i_{r\phi} \,\rightarrow\, i_{r\phi}^\text{loc}(\varepsilon,j)\,+\,i_{r\phi}^\text{log} \,(\varepsilon,j)\,+\,i_{r\phi}^\text{hered}(\varepsilon,j)} \quad  , \\
    \label{seq:omega_inTermsOf_varepsilon_j_procedure}
    \omega(\varepsilon,j) &= \Biggl[\omega^\text{loc}(i_{r\phi},i_\phi)+\omega^\text{log}(i_{r\phi},i_\phi)\,+\,\omega^\text{hered}(i_{r\phi},i_\phi)\Biggr]_{i_{r\phi} \,\rightarrow\, i_{r\phi}^\text{loc}(\varepsilon,j)\,+\,i_{r\phi}^\text{log}(\varepsilon,j)\,+\,i_{r\phi}^\text{hered}(\varepsilon,j)}  \quad , 
\end{align}
\end{subequations}
where $i_\phi$ is trivially replaced in terms of $(\varepsilon,j)$ using Eq.~\eqref{eq:iphi_inTermsOf_varepsilon_j}.
From Eq.~\eqref{eq:n_omega_inTermsOf_varepsilon_j_procedure}, it can be seen that the tail part appearing in Eq.~\eqref{eq:n_omega_inTermsOf_varepsilon_j_loc_log_hered} in fact arises in two distinct ways: there is (i)~a~direct contribution, arising from the logarithmic and hereditary pieces of the fundamental frequencies in terms of action variables provided in Eq.~\eqref{eq:n_omega_inTermsOf_irphi_iphi_sum_loc_log_hered}; and~(ii)~an indirect contribution, arising from the replacement of the action variable $i_{r\phi}$ appearing in the local, Newtonian piece of Eqs.~\eqref{eq:n_omega_inTermsOf_irphi_iphi_sum_loc_log_hered} by its 4PN expression~\eqref{eq:irphi_log_hered_inTermsOf_varepsilon_j}  in terms of $(\varepsilon,j)$, which contains tail contributions.
The direct contribution is in (numerical) agreement with Eqs.~(54)--(57) of Ref.~\cite{Cho:2021oai}, but the indirect contribution was overlooked in that work; the full map between $(n,\omega)$ and $(\varepsilon, j)$ presented hereafter  is thus novel.

The fundamental frequencies in terms of energy and angular momentum are then split as usual into 
\begin{subequations}
\label{eq:n_omega_inTermsOf_varepsilon_j_loc_log_hered}
    \begin{align}
    \label{seq:n_inTermsOf_varepsilon_j_sum_loc_log_hered}
    n(\varepsilon,j) &= n^\text{loc}(\varepsilon,j) + n^\text{log}(\varepsilon,j) + n^\text{hered}(\varepsilon,j) \,,\\*
\label{seq:omega_inTermsOf_varepsilon_j_sum_loc_log_hered}
    \omega(\varepsilon,j) &= \omega^\text{loc}(\varepsilon,j) + \omega^\text{log}(\varepsilon,j) + \omega^\text{hered}(\varepsilon,j) \,,
\end{align}
\end{subequations}
where the different contributions to $n$ and $\omega$ are given, respectively, in Eqs.~\eqref{eq:n_loc_log_hered_inTermsOf_varepsilon_j} and~\eqref{eq:omega_loc_log_hered_inTermsOf_varepsilon_j}. From the previous discussion, it is  known that this split differs from that of Eq.~\eqref{eq:n_omega_inTermsOf_irphi_iphi_sum_loc_log_hered}, such that \mbox{$n^\text{log}(\varepsilon,j) \neq n^\text{log}(i_{r\phi},i_\phi)$} and  \mbox{$n^\text{hered}(\varepsilon,j) \neq n^\text{hered}(i_{r\phi},i_\phi)$}. The conventions that were chosen for this split were described in general in Sec.~\ref{subsec:tail_Hamiltonian}, but for clarity, I will remind the reader what the definitions are in the particular case of $n(\varepsilon,j)$. The local piece is defined ignoring everything else, namely
\begin{subequations}
\begin{align}
    n^\text{loc}(\varepsilon,j) = \big[n^\text{loc}(i_{r\phi},i_\phi) \big]_{i_{r\phi} \rightarrow i_{r\phi}^\text{loc}(\varepsilon,j)}\,.
\end{align}
The logarithmic piece is then added as a perturbation to the local piece, namely
\begin{align}
    n^\text{log}(\varepsilon,j) =  \big[n^\text{loc}(i_{r\phi},i_\phi) +n^\text{log}(i_{r\phi},i_\phi) \big]_{i_{r\phi} \rightarrow i_{r\phi}^\text{loc}(\varepsilon,j)+i_{r\phi}^\text{log}(\varepsilon,j)} - n^\text{loc}(\varepsilon,j)  \,.
\end{align}
Finally, the hereditary piece is given as a perturbation to the sum of local and logarithmic contributions, namely 
\begin{align}
    n^\text{hered}(\varepsilon,j) = \big[n(i_{r\phi},i_\phi)\big]_{i_{r\phi} \rightarrow i_{r\phi}(\varepsilon,j)} - n^\text{loc}(\varepsilon,j) - n^\text{log}(\varepsilon,j) \,.
\end{align}  
\end{subequations}
I have checked that $n(\varepsilon,j)$ agrees at 3PN with Eq.~(4.22a)~of~Ref.~\cite{Akcay:2015pza}. Moreover, I have checked that $n^\text{loc}(\varepsilon,j)$ exactly reproduces  the local 4PN result provided in Eq.~(28b)~of~Ref.~\cite{Cho:2021oai}.


\subsection{Radial period and periastron advance $(P,K)$}

The radial period $P$ and the periastron advance $K$ can immediately be deduced from the radial and azimuthal frequencies $(n,\omega)$ through the relations $n = 2\pi/P$ and $\omega = Kn$. The radial period and periastron advance then read 
\begin{subequations}
\label{eq:P_K_inTermsOf_varepsilon_j_sum_loc_log_hered}
    \begin{align}
    \label{seq:P_inTermsOf_varepsilon_j_sum_loc_log_hered}
    P(\varepsilon,j) &= P^\text{loc}(\varepsilon,j) + P^\text{log}(\varepsilon,j) + P^\text{hered}(\varepsilon,j) \,,\\*
     \label{seq:K_inTermsOf_varepsilon_j_sum_loc_log_hered}
    K(\varepsilon,j) &= K^\text{loc}(\varepsilon,j) + K^\text{log}(\varepsilon,j) +K^\text{hered}(\varepsilon,j) \,,
\end{align}
\end{subequations}
where the different contributions to $P$ and $K$ are given, respectively, in Eqs.~\eqref{eq:P_loc_log_hered_inTermsOf_varepsilon_j} and~\eqref{eq:K_loc_log_hered_inTermsOf_varepsilon_j}.
I have checked that $K(\varepsilon,j)$ agrees at 3PN with Eq.~(4.22b)~of~Ref.~\cite{Akcay:2015pza} and that $K^\text{loc}(\varepsilon,j)$ agrees with Eq.~(9.9) of Ref.~\cite{Bini:2020nsb} after proper conversion from EOB variables (see Footnote~\ref{footnote:EOB_conversion}). I have also verified that $P^\text{loc}(\varepsilon,j)$ and $K^\text{loc}(\varepsilon,j)$ exactly reproduce the local 4PN results provided\footnote{Eq.~(28o) in that reference is also correct, up to a spurious closing square bracket in the expression.}   in the Supplemental Material of Ref.~\cite{Cho:2021oai}.

I also found it instructive to  perform the computation in two other ways.  The first alternative method, which is only relevant for the \textit{local} piece, does not involve action-angle variables. It consists in integrating Eq.~\eqref{eq:rDot_phiDot_loc_inTermsOf_r}. The solution reads~\cite{Blanchet:2013haa}
\begin{align}\label{eq:P_K_loc_from_contour_integral}
P^{\text{loc}} =   \oint \frac{\dd r}{\sqrt{\mathcal{R}(1/r)}} &&\text{and}&& K^{\text{loc}}= \frac{1}{2\pi} \oint \frac{\dd r \,  \mathcal{S}(1/r)}{\sqrt{\mathcal{R}(1/r)}}  \,,
\end{align}
which can be computed \emph{à la Sommerfeld}~\cite{Sommerfeld, Damour:1988mr} in terms of $\{A,B,C,D_n,F,I_n\}_{n\in [\![1,7]\!]}$ using the integration formula~\eqref{eq:Ipq_integral_formula}; see Eq.~\eqref{eq:P_K_loc_inTermsOf_ABCDFI} for explicit expressions of $P$ and $K$ in terms of these coefficients at 4PN.
One then replaces these coefficients by their expressions in terms of energy and angular momentum~\eqref{eq:ABCDFI_inTermsOf_varepsilon_j}, and find perfect agreement with Eqs.~\eqref{seq:P_loc_inTermsOf_varepsilon_j} and~\eqref{seq:K_loc_inTermsOf_varepsilon_j} in the local sector.
The second alternative method (valid for both the local and tail sectors) is by directly varying the radial action $I_r$ with respect to energy and angular momentum:
\begin{align} \label{eq:P_K_loc_from_ir_loc}
P = 2\pi \frac{\partial I_r}{\partial E} &&\text{and}& &
K =  - \frac{\partial I_r}{\partial J} \,,
\end{align}
see, e.g., Eqs.~(3.6)~and~(3.7)~of~Ref.~\cite{Damour:1988mr}. I checked that this second method recovers exactly the results of Eq.~\eqref{eq:P_K_inTermsOf_varepsilon_j_sum_loc_log_hered}.
%


\subsection{Dimensionless frequencies $(x,\iota)$}

Now that  these  frequencies are computed, it will prove very useful to compute the Blanchet parameters $(x,\iota)$, defined by~\cite{Blanchet:2013haa}
\begin{align}\label{eq:x_iota_inTermsOf_n_omega}
    x &= \left(\frac{G m \omega}{c^3}\right)^{2/3} &&\text{and}&
    \iota &= \frac{3 x}{K-1} = \frac{3 (Gm\omega)^{2/3}}{c^2(\omega/n-1)}\,.
\end{align}
These definitions are chosen such that $x = \mathcal{O}(1/c^2)$ and $\iota = \mathcal{O}(1)$; recall that $K-1 = \mathcal{O}(1/c^2)$. I thus find that the Blanchet parameters
read\footnote{\label{footnote:error_iota}Notice that $\iota(\varepsilon,j) = \iota^\text{loc}(\varepsilon,j) + \iota^\text{log}(\varepsilon,j) +\iota^\text{hered}(\varepsilon,j)$ disagrees at 3PN with Eq.~(8.23)~of~Ref.~\cite{Arun:2007sg}, although they agree at 2PN. The expression in that paper does not correspond to the 3PN local part $\iota^\text{loc}(\varepsilon,j)$ either. I believe that the discrepancy stems from an order-counting mistake in that work: indeed, the 3PN expression for $\iota$ has as much information as the 4PN periastron advance (which was  unknown at the time). 
To substantiate this claim, I have performed the following check. I computed $K(\varepsilon,j)$ using the relation $K = 1 + 3x/\iota$, along with $\iota(\varepsilon,j)$ as given by Eq.~(8.23)~of~Ref.~\cite{Arun:2007sg} and $x(\varepsilon,j)$ as given by Eq.~(7.10)~of~Ref.~\cite{Arun:2007sg}. I then obtained $K^\text{circ}(\varepsilon)$ using the relation $j^\text{circ}(\varepsilon)$ given by Eq.~(7.12)~of~Ref.~\cite{Arun:2007sg}. Finally, I obtained $K^\text{circ}(x)$ using the relation $\varepsilon^\text{circ}(x)$ given by Eq.~(4.11)~of~Ref.~\cite{Bernard:2016wrg}. This led to a 4PN-accurate expression for $K^\text{circ}(x)$, which disagrees with Eq.~(5.10)~of~Ref.~\cite{Bernard:2016wrg}. On the contrary, my result agrees in the circular limit with known results; see Sec.~\ref{sec:circular}. This mistake propagates into Eq.~(8.24)~of~Ref.~\cite{Arun:2007sg}, where the instantaneous piece of the 3PN energy flux is expressed in terms of the gauge invariant expressions $(x,\iota)$.}
\begin{subequations}
\label{eq:x_iota_inTermsOf_varepsilon_j_loc_log_hered}
    \begin{align}
    \label{seq:x_inTermsOf_varepsilon_j_sum_loc_log_hered}
    x(\varepsilon,j) &= x^\text{loc}(\varepsilon,j) + x^\text{log}(\varepsilon,j) + x^\text{hered}(\varepsilon,j) \,, \\*
     \label{seq:iota_inTermsOf_varepsilon_j_sum_loc_log_hered}
    \iota(\varepsilon,j) &= \iota^\text{loc}(\varepsilon,j) + \iota^\text{log}(\varepsilon,j) +\iota^\text{hered}(\varepsilon,j) \,, 
\end{align}
\end{subequations}
where the different contributions to $x$ and $\iota$ are given, respectively, in Eqs.~\eqref{eq:x_loc_log_hered_inTermsOf_varepsilon_j}~and~\eqref{eq:iota_loc_log_hered_inTermsOf_varepsilon_j}.
Finally, these expressions can be inverted to obtain $(\varepsilon,j)$ in terms of $(x, \iota)$, thus generalizing to eccentric orbits the result given in Eq.~(5.5) of~Ref.~\cite{Damour:2014jta}; this is the main result of this paper. I find that the energy and angular momentum are given in terms of the fundamental frequencies by
\begin{subequations}
\label{eq:varepsilon_j_inTermsOf_x_iota_sum_loc_log_hered}
\begin{align}
\label{seq:varepsilon_inTermsOf_x_iota_sum_loc_log_hered}
    \varepsilon(x,\iota) &=  \varepsilon^\text{loc}(x,\iota) + \varepsilon^\text{log}(x,\iota) + \varepsilon^\text{hered}(x,\iota) \,,\\*
\label{seq:j_inTermsOf_x_iota_sum_loc_log_hered}
    j(x,\iota) &=  j^\text{loc}(x,\iota) + j^\text{log}(x,\iota) + j^\text{hered}(x,\iota) \,, 
\end{align}
\end{subequations}
where the local, logarithmic and hereditary pieces are given explicitly by
\begin{subequations}
\label{eq:varepsilon_j_loc_log_hered_inTermsOf_x_iota}
\begin{align}
\label{seq:varepsilon_loc_inTermsOf_x_iota}
\varepsilon^{\text{loc}} &= x \Biggl\{1+x \Biggl[\frac{5}{4}-\frac{\nu}{12} -\frac{2}{\iota } \Biggr]+x^2 \Biggl[\frac{5}{8} -\frac{5}{8} \nu -\frac{\nu^2}{24} +\frac{5-2 \nu }{\sqrt{\iota }}+\frac{1}{\iota }\biggl(-5+\frac{\nu}{3}\biggr) +\frac{5}{\iota ^2} \Biggr] \nonumber\\
&  +x^3 \Biggl[-\frac{185}{192} -\frac{75}{64} \nu -\frac{25}{288} \nu ^2-\frac{35}{5184} \nu^3 +\frac{1}{\sqrt{\iota}}\biggl(\frac{105}{8}-\frac{35}{6} \nu -\frac{7}{6} \nu^2\biggr)+\frac{1}{\iota }\biggl(-\frac{15}{4}+\frac{15}{4} \nu +\frac{\nu^2}{4}\biggr) \nonumber\\
& \qquad +\frac{1}{\iota ^{3/2}}\biggl(-\frac{95}{8}+\nu \Bigl(-\frac{211}{9}+\frac{41}{96} \pi^2\Bigr) 
   +\frac{5}{2} \nu ^2\biggr)  +\frac{1}{\iota ^2}\biggl(\frac{35}{2}-\frac{7}{6} \nu\biggr)  -\frac{40}{3 \iota ^3} \Biggr] \nonumber\\
&  +x^4 \Biggl[-\frac{931}{384}+\frac{245}{1152} \nu +\frac{245}{1152} \nu^2+\frac{175}{5184} \nu^3+\frac{77}{31104} \nu^4 \nonumber\\
& \qquad   +\frac{1}{\sqrt{\iota }}\biggl(\frac{1575}{128}-\frac{865}{64} \nu -\frac{65}{32} \nu^2 -\frac{5}{16} \nu^3\biggr)  +\frac{1}{\iota}\biggl(\frac{935}{24}-\frac{125}{8} \nu +\frac{205}{36} \nu ^2+\frac{35}{648} \nu^3\biggr) \nonumber\\
& \qquad  +\frac{1}{\iota ^{3/2}}\biggl(-\frac{3255}{64}+\nu\Bigl(-\frac{26611}{864}-\frac{10969}{9216} \pi^2\Bigr) + \nu^2\Bigl(-\frac{6757}{432}+\frac{205}{576}\pi^2\Bigr) +\frac{65}{24}\nu^3\biggr) \nonumber\\
& \qquad +\frac{1}{\iota ^2}\biggl(\frac{135}{8}-\frac{135}{8} \nu -\frac{9}{8} \nu^2\biggr)  +\frac{1}{\iota ^{5/2}}\biggl(\frac{5723}{128}+\nu\Bigl(\frac{318371}{2880}+\frac{15359}{3072} \pi ^2\Bigr)  + \nu^2\Bigl(\frac{679}{32}-\frac{123}{64} \pi^2\Bigr)+\frac{55}{16}\nu ^3\biggr)  \nonumber\\
& \qquad   +\frac{1}{\iota ^3}\biggl(-\frac{175}{3}+\frac{35}{9} \nu \biggr) +\frac{110}{3 \iota ^4} \Biggr]\Biggr\} \,,\\
\label{seq:varepsilon_log_inTermsOf_x_iota}
    \varepsilon^\text{log} &= \frac{x^5 \nu}{ \iota  (1+\sqrt{\iota })}   \Biggl\{-\frac{44}{45} -\frac{2588}{45 \sqrt{\iota }}-\frac{2752}{45 \iota }+\frac{176}{45\iota ^{3/2}}+\frac{340}{3 \iota ^2}+\frac{340}{3 \iota ^{5/2}} \nonumber\\
   & \qquad \qquad + \left[\frac{296}{45}+\frac{296}{45 \sqrt{\iota }}-\frac{488}{15 \iota }-\frac{488}{15 \iota ^{3/2}}\right] \left(\ln (x)-2
   \ln \left(\frac{1+\sqrt{\iota }}{8 \iota }\right)+2 \gamma_\text{E} \right) \Biggr\} \,,\\
\label{seq:varepsilon_hered_inTermsOf_x_iota}
    \varepsilon^\text{hered} &= \frac{x^5 \nu}{\iota ^{3/2}}   \Biggl\{ -\frac{148}{15} + \frac{488}{5 \iota }-\frac{340}{3 \iota^2} + \ln \iota \left(-\frac{296}{15}+ \frac{488}{5 \iota }\right)  +\frac{128}{5 \iota}\lambda_0 \left(\sqrt{1-\iota }\right)+\frac{64}{5 \iota }\sqrt{1-\iota } \,\lambda_0' \!\left(\sqrt{1-\iota }\right) \Biggr\}  \,,\\
\label{seq:j_loc_inTermsOf_x_iota}
j^{\text{loc}} &= \iota \Biggl\{1+x \Biggl[\frac{5}{12} \nu +\frac{1}{\iota }\biggl(\frac{27}{4}-\frac{5}{2} \nu \biggr)\Biggr] \nonumber\\
&  +x^2 \Biggl[-\frac{35}{16} +\frac{25}{48} \nu +\frac{\nu ^2}{8} +\frac{5-2 \nu }{\sqrt{\iota }} +\frac{1}{\iota}\biggl(-\frac{35}{8}+\nu\Bigl(\frac{373}{16}-\frac{41}{128}\pi^2\Bigr)  -\frac{55}{24} \nu ^2\biggr) \nonumber\\
& \qquad  +\frac{1}{\iota ^2}\biggl(\frac{115}{16}+\nu\Bigl(-\frac{665}{12}+\frac{205}{128}\pi^2\Bigr)  -\frac{15}{8} \nu^2\biggr)\Biggr] \nonumber\\
&  +x^3\Biggl[-\frac{635}{192}+\frac{155}{192} \nu +\frac{5}{36} \nu^2+\frac{91}{5184} \nu^3  +\frac{1}{\sqrt{\iota }}\biggl(\frac{5}{8}+\frac{25}{6} \nu -\frac{19}{6} \nu^2\biggr) \nonumber\\
& \qquad  +\frac{1}{\iota}\biggl(-\frac{1085}{64}+\nu\Bigl(\frac{23689}{576}+\frac{6049}{24576} \pi^2\Bigr) 
   +\nu^2\Bigl(\frac{3173}{288}-\frac{287}{1536} \pi^2\Bigr) -\frac{65}{48}\nu^3\biggr) \nonumber\\
& \qquad  +\frac{1}{\iota ^{3/2}} \biggl(\frac{255}{8}+\nu\Bigl(-\frac{481}{9}+\frac{41}{96} \pi^2\Bigr)  +\frac{15}{2} \nu^2\biggr) \nonumber\\
& \qquad  +\frac{1}{\iota ^2}\biggl(-\frac{455}{32}+\nu \Bigl(\frac{1559}{576}-\frac{98771}{12288} \pi^2\Bigr) + \nu^2\Bigl(-\frac{5765}{72}+\frac{4715}{1536} \pi^2\Bigr)-\frac{85}{32} \nu ^3\biggr) \nonumber\\
& \qquad  +\frac{1}{\iota^3}\biggl(-\frac{4345}{192}+\nu \Bigl(\frac{351151}{864}-\frac{360025}{73728} \pi ^2\Bigr) + \nu^2\Bigl(-\frac{7505}{72}+\frac{1025}{384} \pi^2\Bigr) -5 \nu^3\biggr)   \Biggr]\Biggr\} \,, \\
\label{seq:j_log_inTermsOf_x_iota}
    j^\text{log} &= \frac{x^3 \sqrt{\iota } \,\nu}{1+\sqrt{\iota }}  \Biggl\{\frac{403}{90} + \frac{12421}{270 \sqrt{\iota }}-\frac{7567}{135 \iota }-\frac{3383}{15 \iota ^{3/2}}+\frac{3937}{54 \iota ^2}+\frac{11077}{54\iota ^{5/2}} \nonumber\\*
   &+ \left[-\frac{37}{15}-\frac{37}{15 \sqrt{\iota }}+\frac{122}{3 \iota }+\frac{122}{3 \iota ^{3/2}}-\frac{595}{9
   \iota ^2}-\frac{595}{9 \iota ^{5/2}}\right] \left(\ln (x)-2 \ln \left(\frac{1+\sqrt{\iota }}{8 \iota }\right)+2 \gamma_\mathrm{E}
   \right)\Biggr\} \,, \\
    \label{seq:j_hered_inTermsOf_x_iota}
    j^\text{hered} &= x^3 \nu  \Biggl\{-\frac{74}{15} +\frac{244}{5 \iota }-\frac{170}{3 \iota ^2}+ \ln \iota\,\left(\frac{37}{5}-\frac{122}{\iota }+\frac{595}{3 \iota ^2}\right)    +  \lambda_0 \! \left(\sqrt{1-\iota }\right)\left(-\frac{32}{\iota}+\frac{224}{5 \iota^2 }\right)+\frac{32}{5 \iota} \sqrt{1-\iota } \,\lambda_0'\!\left(\sqrt{1-\iota }\right)\Biggr\} \,.
\end{align}
\end{subequations}
The energy $\varepsilon$ is almost in agreement with the 3PN result in Eq.~(4.40a) of~\cite{Akcay:2015pza}, up to 3PN terms proportional to $\iota^{-2}$ and $\iota^{-3}$ which have probably been omitted in that reference. Moreover, I found that the angular momentum $j$ is in perfect agreement with the 2PN result in Eq.~(4.40b)~of~\cite{Akcay:2015pza}.

\section{Redshift invariant}
\label{sec:redshift}

The Detweiler redshift invariant was first introduced in the context of black-hole perturbation theory  as the ratio of the rates of change of Schwarzschild coordinate time and of proper time along
a particle’s geodesic in a perturbed and regularized
Schwarzschild metric~\cite{Detweiler:2008ft}. It was given a first interpretation as a constant of motion associated with the helical symmetry of spacetime, and a second interpretation as the redshift undergone by light emitted at the particle and reaching an asymptotic observer. It was then shown to be a very useful gauge-invariant quantity for comparison with post-Newtonian theory, where it is defined as~\cite{Blanchet:2009sd,Akcay:2015pza,LeTiec:2015kgg} 
\begin{align}
    z_1 &= \sqrt{- (g_{\alpha \beta})_1 \frac{v_1^\alpha v_1^\beta}{c^2}} \,,
\end{align}
where the 4-vector $v_1^\alpha=(c,v_1^i)$ is associated with the velocity 3-vector $v_1^i$ of the first particle, as well as the regularized value of the metric on the first particle $(g_{\alpha \beta})_1$ [the regularization procedure was historically Hadamard regularization, then was promoted to dimensional regularization]. 
Once orbit-averaged, the redshift invariant was shown to be linked, in a variational sense, to the fundamental frequencies $n$ and $\omega$, the masses $m_1$ and $m_2$, the energy $E$, the angular momentum $J$,  and the radial action $I_r$ through the \textit{first law of binary black hole mechanics}~\cite{LeTiec:2011ab, LeTiec:2015kgg}:
\begin{align}
    c^2\,\delta \dM = n\ \delta I_r + \omega \ \delta J  + c^2\,\langle z_1 \rangle \,\delta m_1 + c^2\,\langle z_2 \rangle \,\delta m_2 \,,
\end{align}
where I have introduced the ADM mass
\begin{align}\label{eq:ADM_mass_def}
\dM = m + E/c^2\,. 
\end{align}
This law was shown to hold even in the presence of hereditary effects~\cite{Blanchet:2017rcn}.
Using this law, one finds that the redshift invariant can also be obtained directly by differentiating the expression of the energy in terms of action-angle variables with respect to the masses of the two compact objects. However, one should now be careful with the scalings of various quantities with the masses $m_1$ and $m_2$. 
Using Eqs.~\eqref{eq:H_inTermsOf_irphi_iphi_sum_loc_log_hered} and~\eqref{eq:rescaled_actions_def},  one can express the ADM mass~\eqref{eq:ADM_mass_def} in terms of the dimensionful action variables $(I_r, I_\phi = J)$ and the masses $(m_1,m_2)$ of each particle.
The averaged redshift invariant (with respect to particle $1$) is then simply given by \cite{LeTiec:2011ab, Blanchet:2012at, Blanchet:2017rcn, LeTiec:2015kgg}
\begin{align}\label{eq:redshift_from_first_law}
z_1 &= \frac{\partial \dM}{\partial m_1}\Bigg|_{m_2, I_r, J} \,.
\end{align}
Of course, the initial result after differentiation is expressed in terms of the action variables, and one can then use Eq.~\eqref{seq:irphi_inTermsOf_varepsilon_j_sum_loc_log_hered} and Sec.~\ref{subsec:notations} to re-express it in terms of $(m,\nu, \delta, \varepsilon,j)$.

I finally find that the redshift invariant is given in terms of energy and angular momentum by
\begin{align}
\label{eq:redshift_inTermsOf_varepsilon_j_sum_loc_log_hered}
\langle z_1\rangle(\varepsilon, j) = \langle z_1^\text{loc}\rangle(\varepsilon, j) + \langle z_1^\text{log}\rangle(\varepsilon, j) + \langle z_1^\text{hered}\rangle(\varepsilon, j)\,,
\end{align}
where the local, logarithmic and hereditary pieces read 
\begin{subequations}
\label{eq:redshift_loc_log_hered_inTermsOf_varepsilon_j}
\begin{align}
\label{seq:redshift_loc_inTermsOf_varepsilon_j}
\langle z_1^\text{loc}\rangle &= 1+\varepsilon  \Biggl\{-\frac{3}{4}+\frac{\nu }{2} +\frac{3}{4} \delta  \Biggr\}
 + \varepsilon ^2  \Biggl\{\frac{15}{8} -\frac{3}{16} \nu +\frac{\nu^2}{4} +\delta  \left(-\frac{15}{8}+\frac{3}{16} \nu\right) +\frac{-3+3 \delta }{\sqrt{j}} \Biggr\} \nonumber\\*
& \qquad +\varepsilon ^3  \Biggl\{ -\frac{65}{16}-\frac{3}{32} \nu^2+\frac{\nu ^3}{8} +\delta \left(\frac{65}{16}+\frac{3}{32} \nu^2\right)  +\frac{1}{\sqrt{j}}\Biggl[\frac{105}{8}-\frac{33}{8}\nu+3 \nu ^2+\delta  \left(-\frac{105}{8}+\frac{33}{8} \nu \right)\Biggr] \nonumber\\
&\qquad\qquad +\frac{1}{j^{3/2}}\Biggl[-\frac{35}{2}+\frac{25}{4}\nu-5 \nu ^2 +\delta  \left(\frac{35}{2}-\frac{25}{4} \nu \right) \Biggr] \Biggr\}\nonumber\\
& \qquad  +\varepsilon ^4  \Biggl\{\frac{291}{32}+\frac{65}{64} \nu +\frac{15}{128} \nu ^2-\frac{3}{64} \nu ^3+\frac{\nu ^4}{16}+\delta  \left(-\frac{291}{32}-\frac{65}{64} \nu  -\frac{15}{128} \nu ^2+\frac{3}{64} \nu ^3\right) \nonumber\\
&\qquad\qquad +\frac{1}{\sqrt{j}}\Biggl[-\frac{5625}{128}+\frac{1125}{64}\nu -\frac{1749}{128} \nu^2+\frac{33}{8} \nu^3+\delta 
   \left(\frac{5625}{128}-\frac{1125}{64} \nu +\frac{549}{128} \nu^2\right) \Biggr] \nonumber\\
&\qquad\qquad  +\frac{1}{j}\Biggl[\frac{45}{2}-9 \nu +\delta  \left(-\frac{45}{2}+9 \nu \right) \Biggr] \nonumber\\
&\qquad\qquad +\frac{1}{j^{3/2}}\Biggl[\frac{1785}{16}+\nu\left(-\frac{13543}{96}+\frac{287}{256} \pi^2\right)  +  \nu ^2\left(\frac{9391}{96}-\frac{41}{64} \pi^2\right)-\frac{125}{8}\nu^3 \nonumber\\
&\qquad\qquad \qquad  +\delta \left[-\frac{1785}{16}+\nu\left(\frac{13543}{96}-\frac{287}{256} \pi^2\right)  -\frac{505}{32} \nu^2\right]\Biggr] \nonumber\\
&\qquad\qquad +\frac{1}{j^{5/2}}\Biggl[-\frac{693}{4}+\nu\left(\frac{875}{4}-\frac{861}{256} \pi^2\right)  +\nu ^2\left(-\frac{271}{2}+\frac{123}{64} \pi^2\right) +\frac{21}{2} \nu ^3\nonumber\\
&\qquad\qquad \qquad +\delta  \left[\frac{693}{4}+\nu\left(-\frac{875}{4}+\frac{861}{256}\pi ^2\right)  +\frac{21}{2} \nu^2\right] \Biggr] \Biggr\}\nonumber\\
& \qquad  +\varepsilon^5  \Biggl\{ -\frac{1293}{64}-\frac{291}{64} \nu -\frac{65}{256} \nu^2+\frac{15}{256} \nu^3 -\frac{3}{128} \nu^4+\frac{\nu ^5}{32}+\delta  \left(\frac{1293}{64}+\frac{291}{64} \nu +\frac{65}{256} \nu ^2-\frac{15}{256} \nu ^3+\frac{3}{128} \nu^4\right) \nonumber\\
&\qquad\qquad  +\frac{1}{\sqrt{j}}\Biggl[\frac{133485}{1024}-\frac{52215}{1024} \nu +\frac{39315}{1024} \nu ^2-\frac{14697}{1024} \nu ^3+\frac{525}{128} \nu ^4 \nonumber\\
& \qquad\qquad\qquad +\delta   \left(-\frac{133485}{1024}+\frac{52215}{1024} \nu -\frac{17835}{1024} \nu ^2 +\frac{3897}{1024} \nu ^3\right)\Biggr] \nonumber\\
&\qquad\qquad  +\frac{1}{j}\Biggl[-\frac{1215}{8}+\frac{405}{4} \nu -45 \nu ^2+9 \nu ^3+\delta  \left(\frac{1215}{8}-\frac{405}{4}\nu +\frac{45}{2} \nu^2\right)\Biggr] \nonumber\\
&\qquad\qquad +\frac{1}{j^{3/2}}\Biggl[-\frac{107625}{256}+\nu\left(\frac{376279}{512}-\frac{160827}{16384} \pi ^2\right)  +\nu ^2\left(-\frac{10703}{18}+\frac{78127}{12288}\pi ^2 \right)  \nonumber\\
& \qquad\qquad\qquad +\nu ^3\left(\frac{379681}{1536}-\frac{943}{512} \pi^2\right) -\frac{3415}{128} \nu ^4 \nonumber\\
& \qquad\qquad\qquad +\delta  \left[\frac{107625}{256}+\nu\left(-\frac{376279}{512}+\frac{160827}{16384} \pi ^2\right)  + \nu ^2\left(\frac{108451}{384}-\frac{4633}{2048} \pi ^2\right) -\frac{12515}{512}\nu ^3\right]\Biggr] \nonumber\\
&\qquad\qquad  +\frac{1}{j^2}\Biggl[\frac{1155}{4}+\nu\left(-\frac{2539}{8}+\frac{123}{64} \pi ^2\right)  +\frac{315}{4}\nu ^2-15 \nu ^3+\delta  \left[-\frac{1155}{4}+\nu\left(\frac{2539}{8}-\frac{123}{64} \pi ^2\right)  -\frac{165}{4}  \nu ^2\right] \Biggr] \nonumber\\
&\qquad\qquad  +\frac{1}{j^{5/2}}\Biggl[\frac{46431}{32}+\nu\left(-\frac{1018419}{320}+\frac{514611}{8192} \pi ^2\right) 
   +\nu ^2\left(\frac{311969}{120}-\frac{3095}{64} \pi ^2\right)  \nonumber\\
& \qquad\qquad\qquad + \nu ^3\left(-\frac{30379}{32}+\frac{7995}{512} \pi ^2\right) +\frac{651}{16} \nu ^4\nonumber\\
& \qquad\qquad\qquad +\delta  \left[-\frac{46431}{32}+\nu\left(\frac{1018419}{320}-\frac{514611}{8192} \pi ^2\right)  + \nu ^2\left(-\frac{18285}{16}+\frac{40221}{2048} \pi ^2\right) +\frac{1197}{32} \nu ^3\right]\Biggr] \nonumber\\
&\qquad\qquad +\frac{1}{j^{7/2}}\Biggl[-\frac{32175}{16}+ \nu\left(\frac{248057}{64}-\frac{1275315}{16384} \pi ^2\right)  + \nu ^2\left(-\frac{779989}{288}+\frac{671105}{12288} \pi ^2\right)  \nonumber\\
& \qquad\qquad\qquad + \nu ^3\left(\frac{77185}{96}-\frac{1025}{64} \pi ^2\right) -\frac{135}{8} \nu ^4 \nonumber\\
& \qquad\qquad\qquad +\delta  \left[\frac{32175}{16}+ \nu\left(-\frac{248057}{64}+\frac{1275315}{16384} \pi ^2\right) + \nu ^2\left(\frac{94625}{96}-\frac{5125}{256} \pi ^2\right) -\frac{495}{32} \nu ^3\right]\Biggr] \Biggr\} \,, \\
\label{seq:redshift_log_inTermsOf_varepsilon_j}
\langle z_1^\text{log}\rangle &=\frac{\varepsilon ^5 \nu}{j^{3/2}}  \Biggl\{-\frac{617}{60} +\frac{17}{3} \nu +\frac{617}{60} \delta    +\frac{1}{\sqrt{j}} \biggl[ -\frac{673}{5}+\frac{2692}{45} \nu +\frac{673}{5} \delta \biggr]  +\frac{1}{j}\biggl[\frac{1907}{10}-\frac{478}{5} \nu -\frac{1907}{10}\delta  \biggr] \nonumber\\
& \qquad\qquad +\frac{1}{j^{3/2}}\biggl[255 -\frac{340}{3} \nu  -255 \delta \biggr]   +\frac{1}{j^2}\biggl[-\frac{3533}{12} +\frac{1291}{9} \nu +\frac{3533}{12} \delta  \biggr]  \nonumber\\
& \qquad\qquad +\Biggl(\frac{111}{10}-\frac{74}{15} \nu  -\frac{111}{10} \delta +\frac{1}{j^2}\biggl[\frac{255}{2}-\frac{170}{3} \nu -\frac{255}{2} \delta  \biggr]+\frac{1}{j}\biggl[-\frac{549}{5}+\frac{244}{5} \nu +\frac{549}{5} \delta   \biggr]\Biggr)\nonumber\\
&\qquad\qquad\qquad \times \left(\ln \varepsilon  -2 \ln \left(\frac{1+\sqrt{j}}{8 j}\right) + 2 \gamma_\text{E} \right)\Biggr\} \,, \\
\label{seq:redshift_hered_inTermsOf_varepsilon_j}
\langle z_1^\text{hered}\rangle &= \frac{\varepsilon ^5 \nu}{j^{3/2}}  \Biggl\{ \ln j \left( -\frac{333}{10} +\frac{74}{5} \nu + \frac{333}{10}\delta   +\frac{1}{j}\bigg[\frac{1647}{5} -\frac{732}{5} \nu -\frac{1647}{5} \delta \bigg]  +\frac{1}{j^2}\bigg[-\frac{765}{2} +170 \nu  +\frac{765}{2}\delta   \bigg] \right) \nonumber\\
& \qquad\qquad+ \lambda_0 \left(\sqrt{1-j}\right)\Biggl(\frac{1}{j}\biggl[\frac{432}{5}-\frac{192}{5} \nu -\frac{432}{5} \delta  \biggr]+\frac{1}{j^2}\biggl[-\frac{432}{5}+\frac{192}{5} \nu +\frac{432}{5} \delta\biggr]\Biggr)\Biggr\} \,.
\end{align}
\end{subequations}
I have compared the previous expression with $\langle U_1 \rangle = 1/\langle z_1 \rangle$ given in Eq.~(4.39)~of~Ref.~\cite{Akcay:2015pza} in terms of $(\varepsilon,j)$ and found perfect agreement\footnote{\label{footnote:delta_def}Agreement is found after accounting for the fact that Refs.~\cite{Akcay:2015pza,Bini:2019lcd,Blanchet:2017rcn} use a  different convention for the relative mass than me: they assume that $m_1 \ge m_2$ and define $\Delta \equiv (m_2-m_1)/m = -\delta$.}
at 3PN. Moreover, I have explicitly checked that the `first integral' relation stemming from the first law of black hole mechanics, given by Eq.~(3.8)~of~\cite{LeTiec:2015kgg}, explicitly holds through 4PN order, including the tail term. This relation reads in my notations
\begin{align}
   m+ \frac{E}{c^2} =  m_1 \langle z_1 \rangle + m_2 \langle z_2 \rangle + \frac{2}{c^2}\Big(\omega \, J + n\, I_r \Big)\,.
\end{align}
To perform this check, I have written $(m_1,m_2)$ in terms of $(m,\nu,\delta)$ and expressed all quantities in terms of $(\varepsilon,j)$ using Eqs.~\eqref{seq:ir_inTermsOf_varepsilon_j_sum_loc_log_hered}, \eqref{eq:n_omega_inTermsOf_varepsilon_j_loc_log_hered}, \eqref{eq:redshift_inTermsOf_varepsilon_j_sum_loc_log_hered} and the definitions of Sec.~\ref{subsec:notations}; recall also that~$\langle z_2 \rangle$ is immediately obtained from~$\langle z_1 \rangle$ by performing the operation~$\delta\rightarrow - \delta$.

One now computes the redshift in terms of the Blanchet parameters $(x,\iota)$, which reads 
\begin{align}
\label{eq:redshift_inTermsOf_x_iota_sum_loc_log_hered}
\langle z_1\rangle(x, \iota) = \langle z_1^\text{loc}\rangle(x, \iota) + \langle z_1^\text{log}\rangle(x, \iota) + \langle z_1^\text{hered}\rangle(x, \iota) \,.
\end{align}
One should keep in mind the discussion in Sec.~\ref{subsec:split_log_hered} for the definition of this split; namely, recall that, e.g., $z_1^\text{hered}(x,\iota) \neq z_1^\text{hered}\Bigl(\varepsilon(x,\iota), j(x,\iota)\Bigr)$. I find that the local, logarithmic and hereditary pieces are given by 
\begin{subequations}
\label{eq:redshift_loc_log_hered_inTermsOf_x_iota}
\begin{align}
\label{seq:redshift_loc_inTermsOf_x_iota}
\langle z_1^\text{loc} \rangle &= 1+x \Biggl\{-\frac{3}{4} +\frac{\nu }{2}+\frac{3}{4}\delta\Biggr\} +x^2 \Biggl\{\frac{15}{16} +\frac{\nu }{2}+\frac{5}{24} \nu ^2 +\delta  \left(-\frac{15}{16}+\frac{\nu}{8}\right) +\frac{-3+3 \delta }{\sqrt{\iota }}  +\frac{1}{\iota }\Biggl[\frac{3}{2}-\frac{3}{2} \delta  -\nu \Biggr]\Biggr\} \nonumber\\
   &\qquad +x^3 \Biggl\{\frac{5}{32}+\frac{9}{32} \nu ^2+\frac{\nu ^3}{16}  +\delta 
   \left(-\frac{5}{32}+\frac{5}{16} \nu +\frac{\nu ^2}{32}\right) \nonumber\\
   &\qquad\qquad +\frac{1}{\sqrt{\iota }} \Biggl[\frac{15}{8}+\nu +2 \nu ^2+\delta  \left(-\frac{15}{8}+\frac{3}{2} \nu \right) \Biggr]  +\frac{1}{\iota }\Biggl[-\frac{15}{4}-2 \nu -\frac{5}{6} \nu ^2 +\delta  \left(\frac{15}{4}-\frac{\nu }{2}\right)\Biggr] \nonumber\\
   &\qquad\qquad +\frac{1}{\iota ^{3/2}}\Biggl[\frac{37}{8}+\frac{5}{2}\nu  -5 \nu ^2 +\delta \left(-\frac{37}{8}-\frac{5}{2} \nu \right) \Biggr]  +\frac{1}{\iota^2}\Biggl[-\frac{15}{4} +\frac{5}{2} \nu+\frac{15}{4} \delta \Biggr] \Biggr\} \nonumber\\
   &\qquad +x^4 \Biggl\{-\frac{37}{256}-\frac{5}{6} \nu +\frac{5}{64} \nu ^2+\frac{49}{864}\nu ^3+ \frac{91}{10368} \nu^4 +\delta  \left(\frac{37}{256}+\frac{45}{128} \nu+\frac{5}{128} \nu ^2+\frac{7}{1728} \nu^3 \right) \nonumber\\
   &\qquad\qquad+\frac{1}{\sqrt{\iota}}\Biggl[\frac{315}{128}+\frac{35}{8} \nu +\frac{77}{96} \nu ^2+\frac{7}{6} \nu ^3+\delta  \left(-\frac{315}{128}+\frac{35}{16} \nu +\frac{21}{32} \nu ^2 \right)\Biggr] \nonumber\\
   &\qquad\qquad +\frac{1}{\iota}\Biggl[-\frac{15}{16}-\frac{27}{16} \nu ^2 -\frac{3}{8} \nu ^3+\delta  \left(\frac{15}{16}-\frac{15}{8} \nu - \frac{3}{16} \nu^2 \right)\Biggr] \nonumber\\
   &\qquad\qquad +\frac{1}{\iota ^{3/2}}\Biggl[-\frac{285}{64}+ \nu\left(-\frac{1129}{48}+\frac{41}{128} \pi ^2\right)  + \nu ^2\left(\frac{3781}{144}-\frac{41}{96} \pi ^2\right) -\frac{25}{4} \nu ^3 \nonumber\\
   &\qquad\qquad\qquad\qquad +\delta 
   \left[\frac{285}{64}+ \nu\left(\frac{211}{12}-\frac{41}{128}\pi ^2\right)  -\frac{45}{16} \nu ^2\right]\Biggr]  \nonumber\\
   &\qquad\qquad +\frac{1}{\iota ^2} \Biggl[\frac{105}{8}+7 \nu +\frac{35}{12} \nu ^2+\delta  \left(-\frac{105}{8}+\frac{7}{4} \nu \right)\Biggr]\nonumber\\
   &\qquad\qquad +\frac{1}{\iota^{5/2}}\Biggl[-\frac{1797}{128}+\nu \left(\frac{355}{16}-\frac{123}{128} \pi ^2\right)  + \nu ^2\left(-\frac{1321}{32}+\frac{123}{64} \pi^2\right) -\frac{33}{4} \nu ^3  \nonumber\\
   &\qquad\qquad\qquad\qquad +\delta  \left[\frac{1797}{128}+ \nu \left(-\frac{355}{16}+\frac{123}{128} \pi ^2\right) -\frac{99}{32} \nu ^2\right]\Biggr] \nonumber\\
    & \qquad\qquad
   +\frac{1}{\iota ^3}\Biggl[10   -\frac{20}{3}  \nu -10 \delta \Biggr]   \Biggr\} \nonumber\\
   &\qquad +x^5 \Biggl\{-\frac{133}{512}-\frac{7}{6} \nu +\frac{385}{4608} \nu ^2-\frac{425}{6912} \nu ^3-\frac{1045}{41472} \nu ^4 -\frac{187}{62208} \nu ^5 \nonumber\\
    & \qquad\qquad\qquad  +\delta  \left(\frac{133}{512}-\frac{35}{768} \nu -\frac{35}{512} \nu ^2-\frac{25}{1728} \nu^3-\frac{55}{41472} \nu ^4\right) \nonumber\\
    & \qquad\qquad +\frac{1}{\sqrt{\iota}}\Biggl[\frac{1575}{1024}+\frac{355}{128} \nu -\frac{195}{256} \nu ^2+\frac{55}{64} \nu ^3+\frac{5}{16}\nu ^4+\delta  \left(-\frac{1575}{1024}+\frac{865}{256} \nu +\frac{195}{256} \nu ^2+\frac{5}{32} \nu^3\right)\Biggr] \nonumber\\
    & \qquad\qquad +\frac{1}{\iota }\Biggl[\frac{187}{32}+\frac{355}{24} \nu  +\frac{33}{8} \nu ^2-\frac{427}{108} \nu ^3 -\frac{91}{1296} \nu ^4 +\delta  \left(-\frac{187}{32}+\frac{75}{16} \nu  -\frac{41}{16} \nu ^2-\frac{7}{216} \nu^3\right)\Biggr] \nonumber\\
    & \qquad\qquad +\frac{1}{\iota ^{3/2}} \Biggl[-\frac{9765}{1024}+ \nu\left(-\frac{85201}{2304}-\frac{10969}{24576} \pi ^2\right)  \nonumber\\
   &\qquad\qquad\qquad\qquad + \nu ^2 \left(-\frac{7591}{6912}+\frac{18349}{36864} \pi ^2\right)  +\nu ^3\left(\frac{15269}{864}-\frac{205}{576}\pi ^2\right) -\frac{455}{96} \nu ^4 \nonumber\\
   &\qquad\qquad\qquad\qquad +\delta  \left[\frac{9765}{1024}+\nu\left(\frac{26611}{2304}+\frac{10969}{24576} \pi ^2\right)  + \nu ^2\left(\frac{6757}{768}-\frac{205}{1024} \pi ^2\right) -\frac{65}{32} \nu^3\right]\Biggr]\nonumber\\
    & \qquad\qquad +\frac{1}{\iota^2}\Biggl[\frac{135}{32}+\frac{243}{32} \nu ^2+\frac{27}{16} \nu ^3+\delta  \left(-\frac{135}{32}+\frac{135}{16} \nu +\frac{27}{32} \nu ^2\right)\Biggr] \nonumber\\
    & \qquad\qquad +\frac{1}{\iota^{5/2}}\Biggl[\frac{17169}{1024}+\nu \left(\frac{50527}{480}+\frac{15359}{4096} \pi ^2\right) \nonumber\\
   &\qquad\qquad\qquad\qquad +\nu ^2\left(-\frac{998489}{11520}-\frac{22001}{3072} \pi ^2\right)   + \nu ^3\left(-\frac{3065}{64}+\frac{615}{128}\pi^2 \right) -\frac{55}{4} \nu ^4 \nonumber\\
   &\qquad\qquad\qquad\qquad +\delta  \left[-\frac{17169}{1024}+ \nu\left(-\frac{318371}{3840}-\frac{15359}{4096} \pi^2\right)  + \nu ^2\left(-\frac{6111}{256}+\frac{1107}{512} \pi ^2\right) -\frac{165}{32}\nu^3\right]\Biggr] \nonumber\\
    & \qquad\qquad +\frac{1}{\iota ^3}\Biggl[-\frac{175}{4}-\frac{70}{3} \nu -\frac{175}{18} \nu ^2 +\delta  \left(\frac{175}{4}-\frac{35}{6} \nu \right)\Biggr] \nonumber\\
    & \qquad\qquad+\frac{1}{\iota ^{7/2}}\Biggl[\frac{46653}{1024}+ \nu \left(-\frac{662129}{2304}+\frac{162343}{24576} \pi ^2\right)  \nonumber\\
   &\qquad\qquad\qquad\qquad + \nu^2\left(\frac{1532149}{2304}-\frac{199243}{12288} \pi ^2\right) +\left(-\frac{11167}{48}+\frac{1025}{128} \pi ^2\right)
   \nu ^3-\frac{765}{32} \nu ^4 \nonumber\\
   &\qquad\qquad\qquad\qquad +\delta  \left[ -\frac{46653}{1024}+ \nu \left(\frac{662129}{2304}-\frac{162343}{24576} \pi ^2\right) + \nu ^2\left(-\frac{23099}{256}+\frac{3075}{1024} \pi ^2\right) -\frac{255}{32} \nu^3\right]\Biggr] \nonumber\\
    & \qquad\qquad +\frac{1}{\iota ^4}\Biggl[-\frac{55}{2} +\frac{55}{3} \nu +\frac{55}{2} \delta\Biggr] \Biggr\} \,, \\
\label{seq:redshift_log_inTermsOf_x_iota}
\langle z_1^\text{log} \rangle &= \frac{x^5 \nu}{\left(1+\sqrt{\iota }\right) \iota }  \Biggl\{-\frac{17}{6}  +\frac{233}{45} \nu  +\frac{17}{6} \delta   +\frac{1}{\sqrt{\iota }} \Biggl[-\frac{2947}{90}+\frac{551}{15} \nu+\frac{2947}{90} \delta  \Biggr]  +\frac{1}{\iota}\Biggl[\frac{161}{9}-\frac{2986}{45}\nu -\frac{161}{9}\delta  \Biggr]  \nonumber\\
    & \qquad\quad +\frac{1}{\iota^{3/2}}\Biggl[\frac{1567}{15}-\frac{9314}{45} \nu  -\frac{1567}{15} \delta \Biggr]  +\frac{1}{\iota ^2}\Biggl[-\frac{271}{18}+\frac{781}{9} \nu  +\frac{271}{18} \delta  \Biggr]  +\frac{1}{\iota ^{5/2}}\Biggl[-\frac{1291}{18}+\frac{1801}{9} \nu  +\frac{1291}{18}  \delta \Biggr] \nonumber\\
    & \qquad\quad  +\left(\ln x -2 \ln \left(\frac{1+\sqrt{\iota }}{8 \iota }\right) + 2 \gamma_\text{E} \right) \nonumber\\
    & \qquad\qquad\quad   \times \Biggl(\frac{37}{15} -\frac{74}{45} \nu  -\frac{37}{15} \delta    +\frac{1}{\sqrt{\iota }} \Biggl[\frac{37}{15}-\frac{74}{45}\nu-\frac{37}{15} \delta  \Biggr] +\frac{1}{\iota }\Biggl[-\frac{122}{5} +\frac{488}{15} \nu +\frac{122}{5}\delta\Biggr] \nonumber\\
    & \qquad\qquad\qquad\quad +\frac{1}{\iota ^{3/2}}\Biggl[-\frac{122}{5}+\frac{488}{15} \nu+\frac{122}{5} \delta  \Biggr]  +\frac{1}{\iota ^2} \Biggl[\frac{85}{3}-\frac{170}{3} \nu -\frac{85}{3}\delta \Biggr]  +\frac{1}{\iota ^{5/2}}\Biggl[\frac{85}{3} -\frac{170}{3} \nu  -\frac{85}{3} \delta \Biggr] \Biggr) \Biggr\} \,, \\
\label{seq:redshift_hered_inTermsOf_x_iota}
\langle z_1^\text{hered} \rangle &= \frac{x^5 \nu }{\iota ^{3/2}}   \Biggl\{ -\frac{74 \nu }{15}+\frac{244 \nu }{5 \iota } -\frac{170 \nu }{3 \iota ^2}   + \ln \iota\left(-\frac{37}{5}   + \frac{74 \nu }{15} +\frac{37}{5} \delta  +\frac{1}{\iota}\Biggl[\frac{366}{5}-\frac{488}{5} \nu  -\frac{366}{5} \delta \Biggr]+\frac{-85  +170  \nu + 85  \delta}{\iota^2} \right)  \nonumber\\
    & \qquad   +   \lambda_0\! \left(\sqrt{1-\iota }\right) \left( \frac{1}{\iota }\Biggl[\frac{96}{5} -\frac{128}{5} \nu  -\frac{96}{5} \delta\Biggr] +\frac{1}{\iota ^2}\Biggl[-\frac{96}{5} +\frac{192}{5} \nu  + \frac{96}{5} \delta  \Biggr] \right)+\frac{32  \nu }{5 \iota }  \sqrt{1-\iota } \ \lambda_0'\!\left(\sqrt{1-\iota }\right)  \Biggr\}  \,.
    \end{align}
 \end{subequations}
I have found that this expression is in perfect agreement${}^\text{\ref{footnote:delta_def}}$ with Eq.~(49)~of~Ref.~\cite{Bini:2019lcd} at 3PN.

Finally, I compared these results with the gravitational self-force literature at both geodesic and first self-force (1SF) orders. For that, I follow the notations of Ref.~\cite{Hopper:2015icj} and define new variables $(y,\lambda)$, which are similar to the Blanchet parameters $(x,\iota)$, except for the fact that they are normalized with respect to the `primary mass' $m_1$ rather than the total mass $m$. Namely, they are defined from the frequencies $(n,\omega)$ as
\begin{align}\label{eq:def_y_lambda}
    y = \left(\frac{G m_1 \omega}{c^3}\right)^{2/3} & & \text{and} & &\lambda = \frac{3y}{\omega/n-1}  \ . 
\end{align}
Introducing the `small' mass ratio $\epsilon = m_2/m_1 \le 1$, it is immediate to obtain the relations
\begin{align}
\label{eq:expansions_in_GSF_mass_ratio}
    m = m_1(1+\epsilon )\,, &&
    \nu = \frac{\epsilon}{(1+\epsilon)^2}\,, &&
    \delta = \frac{1-\epsilon}{1+ \epsilon}\,, &&  x = y(1+\epsilon)^{2/3}\,, &&\text{and}&&
    \iota = \lambda(1+\epsilon)^{2/3} \,.
\end{align}
One then performs the small mass ratio expansion in $\epsilon \ll 1$ of the redshift associated with the `secondary object' of mass~$m_2$ and replaces the variables $(y,\lambda)$ by their 4PN-accurate expressions in terms of the Darwin variables $(p,e)$; see~\eqref{seq:y_inTermsOf_p_e} and~\eqref{seq:lambda_inTermsOf_p_e}. The result then has the structure
\begin{align}\label{eq:redshift_GSF_expansion}
    \langle z_2\rangle(p,e) = \langle z^\text{geo}_2\rangle + \epsilon \, \langle z^\text{1SF}_2\rangle + \mathcal{O}(\epsilon^2) \,.
\end{align}
I find that the geodesic piece $\langle z^\text{geo}_2\rangle $  is in perfect agreement with the prediction from black hole perturbation theory; see~Eq.~\eqref{seq:z2_geo_inTermsOf_p_e}. I then compared the 1SF piece $\langle z^\text{1SF}_2\rangle = - \langle u_2^\text{1SF}\rangle \times\langle z_2^\text{geo}\rangle^2$  against Eq.~(5.5)~of~Ref.~\cite{Munna:2022gio}; see also Eq.~(5.1)~of~Ref.~\cite{Hopper:2015icj}. Once the enhancement function~\eqref{eq:Lambda_0_def} has been identified with Eq.~(5.10)~of~Ref.~\cite{Munna:2022gio}, I find perfect agreement at 4PN for arbitrary eccentricity. This latter agreement is probably the most stringent test of my results, because it is the only test that probes the effects of the tail term at 4PN on orbits with arbitrary eccentricity.

\section{Reductions to circular orbits}
\label{sec:circular}

In order to validate these results, it is instructive to take the limit of circular orbits. This is defined gauge-invariantly by the condition that the radial action vanishes, namely $i_r(\varepsilon,j)= 0$. This defines a `circular link', namely a relation between energy and angular momentum that must hold for circular orbits; this link is denoted by $j_\text{circ}(\varepsilon)$. Crucially, one needs to account for the fact that the radial action has both local and tail contributions, see Eq.~\eqref{seq:ir_inTermsOf_varepsilon_j_sum_loc_log_hered}. If one were to ignore the tail sector [namely setting $i_r^\text{loc}=0$ using Eq.~\eqref{eq:ir_loc_inTermsOf_varepsilon_j}], one would find the `local' relation
\begin{subequations}
\label{eq:jcirc_loc_log_hered_inTermsOf_varepsilon}
\begin{align}
\label{seq:jcirc_loc_inTermsOf_varepsilon}
    j^\text{loc}_\text{circ} &= 1+\varepsilon  \left(\frac{9}{4}+\frac{\nu }{4}\right)+\varepsilon ^2 \left(\frac{81}{16}-2 \nu +\frac{\nu ^2}{16}\right)+\varepsilon ^3
   \left(\frac{945}{64}+\nu\left[-\frac{7699}{192}+\frac{41}{32} \pi ^2\right]  +\frac{\nu ^2}{2}+\frac{\nu ^3}{64}\right) \nonumber\\*
   &\qquad +\varepsilon ^4
   \left(\frac{14337}{256}+ \nu\left[-\frac{218431}{1920}+\frac{677}{512} \pi^2\right] + \nu^2\left[\frac{229}{4}-\frac{123}{64} \pi ^2\right]
   +\frac{\nu ^4}{256}\right) \, .
\end{align}
Including the tail pieces then leads to the extra contributions
\begin{align}
\label{seq:jcirc_log_hered_inTermsOf_varepsilon}
    j^\text{log}_\text{circ} &= - \frac{64}{5} \varepsilon^4 \nu \Bigl[\ln(16\,\varepsilon) + 2 \gamma_\text{E}\Bigr]\,, &&&
     j^\text{hered}_\text{circ} &= 0 \,.
\end{align}
\end{subequations}
Thus, the full relation $j_\text{circ} = j^\text{loc}_\text{circ} + j^\text{log}_\text{circ} +j^\text{hered}_\text{circ}$ is given by
\begin{align}
\label{eq:jcirc_inTermsOf_varepsilon}
     j_\text{circ} &= 1+\varepsilon  \left(\frac{9}{4}+\frac{\nu }{4}\right)+\varepsilon ^2 \left(\frac{81}{16}-2 \nu +\frac{\nu ^2}{16}\right)+\varepsilon ^3
   \left(\frac{945}{64}+ \nu\left[-\frac{7699}{192}+\frac{41 \pi ^2}{32}\right] +\frac{\nu ^2}{2}+\frac{\nu ^3}{64}\right) \nonumber\\
   &\qquad+\varepsilon ^4
   \left(\frac{14337}{256}+\nu 
   \left[-\frac{218431}{1920}-\frac{128}{5} \gamma_\text{E} +\frac{677}{512} \pi ^2 -\frac{64}{5}\ln (16\,\varepsilon) \right]+ \nu ^2\left[\frac{229}{4}-\frac{123}{64} \pi ^2\right]+\frac{\nu ^4}{256} \right) \,.
\end{align}
This expression agrees with Eq.~(4.43) of Ref.~\cite{Akcay:2015pza} at 3PN. More importantly, I have checked that this result is in perfect agreement at 4PN with Eq.~(5.3) of~\cite{Damour:2014jta}, once the adequate map between variables is performed.

I then proceed to translate this into a circular link between $x$ and $\iota$. Recall that one actually loses a PN order due to the fact that the radial and azimuthal frequencies are degenerate at Newtonian order.$^\text{\ref{footnote:error_iota}}$ Thus, restricting to the local section, I find that the local, 3PN, circular link reads
\begin{subequations}
\label{eq:iota_K_circ_loc_log_hered_inTermsOf_x}
\begin{align}
\label{seq:iota_circ_loc_inTermsOf_x}
    \iota_\text{circ}^\text{loc}&=1+x \left(-\frac{9}{2}+\frac{7}{3} \nu \right)+x^2 \left(-\frac{9}{4}+\nu\left[\frac{397}{12}-\frac{41}{32} \pi ^2\right]  +\frac{28}{9}\nu^2\right) \nonumber\\
   & \qquad +x^3 \left(-\frac{27}{4}+ \nu\left[-\frac{36943}{216}+\frac{58265}{9216} \pi ^2\right] + \nu^2\left[\frac{445}{12}-\frac{41}{32}\pi
   ^2\right]+\frac{245}{81} \nu ^3\right) \,.
\end{align}
This expression immediately converts to the 4PN local periastron advance
\begin{align}
\label{seq:K_circ_loc_inTermsOf_x}
    K_\text{circ}^\text{loc}&= 1+3 x+x^2 \left(\frac{27}{2}-7 \nu \right)+x^3 \left(\frac{135}{2}+ \nu \left[-\frac{649}{4}+\frac{123}{32} \pi ^2\right]  +7 \nu ^2\right) \nonumber\\*
    & \quad +x^4
   \left(\frac{2835}{8}+ \nu \left[-\frac{60257}{72}+\frac{48007}{3072} \pi ^2\right]+ \nu ^2\left[\frac{5861}{12}-\frac{451}{32}\pi ^2\right]
  -\frac{98}{27} \nu^3\right) \,,
\end{align}
which perfectly agrees with the local 4PN periastron advance for circular orbits obtained in Eq.~(5.9)~of~Ref.~\cite{Bernard:2016wrg}.
One then computes the tail pieces, which read
\begin{align}
\label{seq:iota_circ_log_hered_inTermsOf_x}
    \iota_\text{circ}^\text{log}& = x^3 \nu \bigg\{\frac{1256}{45}\Bigl[\ln(16x)+2 \gamma_\text{E}\Bigr] - \frac{352}{15}\bigg\} \,,&&& \iota_\text{circ}^\text{hered}& = x^3 \nu \bigg\{- \frac{4432}{45}\ln 2 + \frac{486}{5} \ln3\bigg\} \,,\\*
\label{seq:K_circ_log_hered_inTermsOf_x}
    K_\text{circ}^\text{log}& = x^4 \nu \bigg\{-\frac{1256}{15}\Bigl[\ln(16x)+2 \gamma_\text{E}\Bigr] + \frac{352}{5}\bigg\} \,,&&& K_\text{circ}^\text{hered} & = x^4 \nu \bigg\{ \frac{4432}{15}\ln 2 - \frac{1458}{5} \ln 3\bigg\} \,.
\end{align}
\end{subequations}
One now adds everything together, namely $\iota_\text{circ} = \iota_\text{circ}^\text{loc}+\iota_\text{circ}^\text{log}+\iota_\text{circ}^\text{hered}$ and $K_\text{circ} = K_\text{circ}^\text{loc}+K_\text{circ}^\text{log}+K_\text{circ}^\text{hered}$, and finds
\begin{subequations}
\label{eq:iota_K_circ_inTermsOf_x}
\begin{align}
\label{seq:iota_circ_inTermsOf_x}
    \iota_\text{circ}&=1+x \left(-\frac{9}{2}+\frac{7}{3}\nu  \right)+x^2 \left(-\frac{9}{4}+ \nu \left[\frac{397}{12}-\frac{41}{32} \pi ^2\right]+\frac{28}{9}\nu^2\right) \nonumber\\
   &\qquad\! +x^3 \Biggl(-\frac{27}{4} +\nu \left[-\frac{210059}{1080}+\frac{2512}{45} \gamma_\text{E} +\frac{58265}{9216} \pi^2+\frac{592}{45} \ln 2 +\frac{486}{5} \ln 3 +\frac{1256}{45} \ln x\right] \nonumber\\*
   &\qquad\qquad\ \quad\qquad+\nu^2 \left[\frac{445}{12}-\frac{41}{32} \pi ^2\right] +\frac{245}{81} \nu ^3 \Biggr) \,,\\
\label{seq:K_circ_inTermsOf_x}
    K_\text{circ} &= 1+3 x+x^2 \left(\frac{27}{2}-7 \nu \right)+x^3 \left(\frac{135}{2}+\nu\left[-\frac{649}{4}+\frac{123}{32} \pi ^2\right]  +7 \nu ^2\right) \nonumber\\
    & \qquad\! +x^4
   \Biggl(\frac{2835}{8} +\nu 
   \left[-\frac{275941}{360}-\frac{2512}{15} \gamma_\text{E} +\frac{48007}{3072} \pi ^2-\frac{592}{15} \ln 2-\frac{1458}{5} \ln 3 -\frac{1256}{15}\ln x\right] \nonumber\\
   &\qquad\qquad\quad \qquad + \nu ^2 \left[\frac{5861}{12}-\frac{451}{32} \pi ^2\right]-\frac{98}{27}\nu^3\Biggr) \,.
\end{align}
\end{subequations}
Notably, the latter result is in perfect agreement with the 4PN periastron advance for circular orbits given in Eq.~(5.10)~of~Ref.~\cite{Bernard:2016wrg}. I have also found perfect agreement with Eq.~(5.4b)~of~Ref.~\cite{Damour:2014jta}, once the appropriate variable conversions have been performed.

Finally, the 4PN redshift can also be reduced to circular orbits; in this case, the angular brackets that denote orbit averaging can be dropped. It can first be obtained in terms of the energy $\varepsilon$ thanks to the relation $j_\text{circ}(\varepsilon)$ given in~Eq.~\eqref{eq:jcirc_inTermsOf_varepsilon}. The redshift on circular orbits in terms of $\varepsilon$ then reads
\begin{subequations}
\label{eq:redshift_circ_inTermsOf_varepsilon_x}
\begin{align}
\label{seq:redshift_circ_inTermsOf_varepsilon}
    z_1^\text{circ} = 1 &+\varepsilon  \Biggl\{-\frac{3}{4} +\frac{\nu }{2} +\frac{3}{4}\delta \Biggr\} +\varepsilon ^2 \Biggl\{-\frac{9}{8} -\frac{3}{16} \nu +\frac{\nu ^2}{4} +\delta  \left(\frac{9}{8}+\frac{3}{16} \nu \right)\Biggr\}\nonumber\\*
   & +\varepsilon ^3 \Biggl\{-\frac{81}{16}+\frac{5}{2} \nu -\frac{67}{32} \nu ^2 +\frac{\nu ^3}{8}+\delta 
   \left(\frac{81}{16}-\frac{5}{2}\nu +\frac{3}{32} \nu ^2\right)\Biggr\} \nonumber\\*
   & +\varepsilon ^4 \Biggl\{-\frac{891}{32}+ \nu \left(\frac{13723}{192}-\frac{287}{128} \pi
   ^2\right) +  \nu ^2\left(-\frac{15179}{384}+\frac{41}{32} \pi ^2\right)+\frac{29}{64} \nu ^3+\frac{\nu ^4}{16} \nonumber\\*
   &\qquad\qquad +\delta 
   \left[\frac{891}{32}+\nu \left(-\frac{13723}{192}+\frac{287}{128} \pi ^2 \right)  +\frac{89}{128} \nu ^2+\frac{3}{64} \nu ^3\right]\Biggr\} \nonumber\\*
   & +\varepsilon ^5
  \Biggl\{-\frac{10935}{64} +\nu  \left(\frac{118349}{320}+\frac{288}{5} \gamma_\text{E}-\frac{11997}{2048}\pi ^2+\frac{144}{5} \ln (16 \,\varepsilon )\right)   \nonumber\\*
   & \qquad\qquad  +\nu ^2  \left(-\frac{965807}{3840}+\frac{2891}{512} \pi ^2 -\frac{128}{5}\gamma_\text{E} -\frac{64}{5} \ln (16\,  \varepsilon )\right) + \nu ^3 \left(\frac{76189}{768}-\frac{205}{64} \pi ^2\right) -\frac{19}{128} \nu ^4+\frac{\nu ^5}{32}\nonumber\\*
   &\qquad\qquad  +\delta 
   \Biggl[\frac{10935}{64} +\nu \left(-\frac{118349}{320}+\frac{11997}{2048} \pi ^2 -\frac{288}{5} \gamma_\text{E} -\frac{144}{5} \ln (16\, \varepsilon )\right) \nonumber\\*
   &\qquad\qquad\qquad    + \nu ^2 \left(\frac{34561}{256}-\frac{1107}{256} \pi ^2\right) -\frac{31}{256} \nu ^3+\frac{3}{128}\nu ^4\Biggr]\Biggr\}\,.
\end{align}
It can alternatively be expressed in terms of $x$ thanks to the relation $\iota_\text{circ}(x)$ given in Eq.~\eqref{seq:iota_circ_inTermsOf_x}. It reads
\begin{align}
\label{seq:redshift_circ_inTermsOf_x}
     z_1^\text{circ} &= 1  +x \Biggl\{-\frac{3}{4}+\frac{\nu }{2} +\frac{3}{4} \delta  \Biggr\}+x^2 \Biggl\{-\frac{9}{16}-\frac{\nu }{2}+\frac{5}{24}\nu ^2 +\delta  \left(\frac{9}{16}+\frac{\nu }{8}\right) \Biggr\} \nonumber\\
   &  +x^3 \Biggl\{-\frac{27}{32}-\frac{\nu }{2}-\frac{39}{32} \nu ^2+\frac{\nu ^3}{16}+\delta  \left(\frac{27}{32}-\frac{19}{16} \nu +\frac{\nu^2}{32}\right)\Biggr\} \nonumber\\
   &  +x^4 \Biggl\{-\frac{405}{256}+ \nu\left(\frac{38}{3}-\frac{41 }{64}\pi ^2\right) +\nu^2\left(-\frac{3863}{576}+\frac{41}{192} \pi ^2\right) +\frac{973}{864} \nu ^3+\frac{91}{10368} \nu ^4  \nonumber\\
   & \qquad\qquad  +\delta  \Biggl[\frac{405}{256}+\nu\left(-\frac{6889}{384}+\frac{41}{64} \pi ^2\right)  +\frac{93}{128} \nu
   ^2+\frac{7}{1728} \nu ^3\Biggr]\Biggr\}\nonumber\\
   &  +x^5 \Biggl\{-\frac{1701}{512} +\nu  \left(-\frac{329}{15} +\frac{1291}{1024} \pi ^2+\frac{64}{5} \gamma_\text{E} +\frac{32}{5} \ln (16 x)\right) \nonumber\\
   &\qquad   +\nu ^2  \left(-\frac{1019179}{23040}+\frac{6703}{3072}\pi^2 +\frac{64}{15}\gamma_\text{E} +\frac{32}{15} \ln (16 x)\right) +\nu ^3 \left(\frac{356551}{6912}-\frac{2255}{1152} \pi ^2\right) -  \frac{5621}{41472}\nu^4 -\frac{187}{62208}\nu ^5 \nonumber\\*
   &\qquad  +\delta  \Biggl[\frac{1701}{512} +\nu  \left(\frac{24689}{3840} -\frac{1291}{1024} \pi ^2 -\frac{64}{5} \gamma_\text{E}-\frac{32}{5} \ln (16 x)\right) +\nu ^2 \left(\frac{71207}{1536}-\frac{451}{256} \pi ^2\right) -\frac{43 }{576}\nu ^3-\frac{55}{41472} \nu ^4 \Biggr] \Biggr\}\,.\nonumber\\*&
\end{align}
\end{subequations}
I find that the latter result is in perfect agreement${}^\text{\ref{footnote:delta_def}}$ at 4PN  with Eq.~(C3)~of~Ref.~\cite{Blanchet:2017rcn} and with~Eq.~(440)~of~Ref.~\cite{Blanchet:2013haa}.

\section{Application: gauge-invariant expressions for the 3PN fluxes}
\label{sec:flux}

In Refs.~\cite{Arun:2007rg, Arun:2009mc, Arun:2007sg}, the complete\footnote{In Ref.~\cite{Arun:2009mc}, it was claimed that the memory contribution to the angular momentum flux was divergent when integrated up to the infinite past. It was then argued from astrophysical arguments that physical systems only form at a finite time in the past; this divergent contribution was then conventionally taken to vanish. Later, I have shown rigorously  that this memory contribution is in fact always finite, even when integrating to the infinite past; see Appendix A of~Ref.~\cite{Trestini:2024mfs}.  The contribution is non-zero in general, but vanishes after orbit-averaging.} orbit-averaged fluxes of energy $\langle \mathcal{F}\rangle $ and angular momentum $\langle \mathcal{G} \rangle$ were obtained for nonspinning, structureless, eccentric systems at 3PN order in terms of the quasi-Keplerian variables $(x,e_t)$.  These variables are gauge-dependent, and results were obtained both in the harmonic and ADM gauges. I have not rederived these results, and in the rest of this section, I will trust that these expressions are correct.
Instead, I will use my new mapping between energy and angular momentum to express these fluxes, for the
first time,${}^\text{\ref{footnote:error_iota}}$
in terms of gauge-invariant frequencies $(x,\iota)$. 

One first introduces the variable $e_t$, which here denotes the time-eccentricity in 
ADM coordinates\footnote{ Note that in modified harmonic coordinates, one finds a different expression for $e_t$, given by Eq.~(25d) of Ref.~\cite{Memmesheimer:2004cv}. Differences between the two coordinates arise only at 2PN, whereas hereditary terms need to be controlled at a relative 1.5PN order with respect to the leading tail. For this reason, hereditary fluxes are given in the literature without specifying the coordinate system.} 
entering the 3PN quasi-Keplerian parametrization~\cite{Memmesheimer:2004cv}. It is given as an explicit function of $(\varepsilon,j)$ in Eq.~(20d) of Ref.~\cite{Memmesheimer:2004cv}. The starting point will be the 3PN expressions of the orbit-averaged fluxes in ADM coordinates, which are provided in terms of $(x,e_t)$. In the literature, these fluxes are further split into an instantaneous and a hereditary part. The energy flux is provided in Ref.~\cite{Arun:2007sg}, see Eqs.~(8.18)--(8.20) for the instantaneous part and Eqs.~(8.11)--(8.13) for the hereditary part; see also \cite{Arun:2007rg} for the hereditary part. The angular momentum flux is provided in~\cite{Arun:2009mc}, see  Eqs.~(4.10) and (4.11) for the instantaneous part and Eqs.~(5.29) and (4.12) for the hereditary part.

A few remarks about these results are in order. First, these expressions of these fluxes feature the enhancement functions $F(e)$ and $\tilde{F}(e)$, which have been replaced by their \textit{exact} expressions, given by Eq.~(8.13) of~Ref.~\cite{Arun:2007sg}  and Eq.~(5.24)~of~Ref.~\cite{Arun:2009mc}, respectively. This allows for the arbitrary constant $x_0$, which appears separately in the instantaneous and hereditary fluxes, to cancel out explicitly in the total flux. Secondly, the hereditary pieces of these fluxes feature a collection of enhancement functions which are not known in closed form. These are defined exactly as infinite sums over the Fourier coefficients\footnote{\label{footnote:FourierJij} The expressions for the Fourier coefficients of the current-type quadrupole moment $\mathrm{J}_{ij}$ provided in the literature all have typos. In Ref.~\cite{Loutrel:2016cdw}, Eq.~(A8) is lacking a total minus sign and should read ${}_{(p)}\widehat{\mathcal{J}}_{zy} = - \frac{\di}{2}(1-e^2) \frac{J_p(p e)}{p e}$. In the Appendix of Ref.~\cite{Arun:2007rg}, there appears to be a problem in the factorization of the $-1/4$ factor in Eq.~(A5a), which would instead read 
$${}_{(p)}\widehat{\mathcal{J}}_{xz}^{\text{(N)}} =  - \frac{1}{4}\sqrt{1 - e^2}\Biggl\{  3 e J_p(p e)  - (1+e^2)\Bigl[J_{p+1}(p e)  + J_{p-1}(p e)\Bigr] + \frac{e}{2} \Bigl[J_{p+2}(p e)  + J_{p-2}(p e)\Bigr]\Biggr\} \ .$$}
of multipolar moments, but are most practically approximated in closed form using small eccentricity expansions. Using asymptotic analysis, one can determine the $e \rightarrow 1$ behavior or the enhancement functions, which makes it possible to engineer sensible approximations valid for both small and large eccentricities~\cite{Forseth:2015oua, Loutrel:2016cdw}; see Sec.~\ref{subsec:resumming}. For the reader's convenience,  in Table~\ref{tab:enhancement}, I provide references  for the definitions of these various enhancement functions, as well as their small-eccentricity expansions and behavior as $e \rightarrow 1$.

\begin{table}[H]
    \centering
    \begin{tabular}{|c|c|c|c|}
	\hline
        Name & Definition & Small-$e$ expansions & Behavior as $e \rightarrow 1$\\ \hline \hline
         \multirow{2}{*}{$\varphi(e)$} & \multirow{2}{*}{(92) of~\cite{Loutrel:2016cdw}} & (B7a) of~\cite{Ebersold:2019kdc} & $\sim(1-e^2)^{-5}$\\  
         & & & (140) of~\cite{Loutrel:2016cdw}  \\ \hline
         \multirow{2}{*}{$\tilde{\varphi}(e)$} & \multirow{2}{*}{(92) of~\cite{Loutrel:2016cdw} } &  \multirow{2}{*}{(B7b) of~\cite{Ebersold:2019kdc}  } & $\sim(1-e^2)^{-7/2}$ \\  
         & & & (141) of~\cite{Loutrel:2016cdw}  \\ \hline
         \multirow{2}{*}{ $\psi(e)$ } & (6.1a) of~\cite{Arun:2007rg}  &  \multirow{2}{*}{ (B7c) of~\cite{Ebersold:2019kdc} } & $\sim (1-e^2)^{-6}$ \\ 
          &  (93), (97) \& (99) of~\cite{Loutrel:2016cdw} &  &  (169), (142) \& (143) of~\cite{Loutrel:2016cdw}  \\ \hline
         \multirow{2}{*}{ $\tilde{\psi}(e)$ } & (5.28a) of~\cite{Arun:2009mc} & \multirow{2}{*}{ (B7d) of~\cite{Ebersold:2019kdc} } &  $\sim(1-e^2)^{-9/2}$ \\ 
          &  (94), (98) \& (100) of~\cite{Loutrel:2016cdw} & &  (170), (144) \& (145) of~\cite{Loutrel:2016cdw}  \\ \hline
         \multirow{2}{*}{ $\zeta(e)$ }& (6.1b) of~\cite{Arun:2007rg}  & \multirow{2}{*}{ (B7g) of~\cite{Ebersold:2019kdc} } & $\sim (1-e^2)^{-6}$ \\ 
          &  (95), (97) \& (99) of~\cite{Loutrel:2016cdw} & &  (152), (142) \& (143) of~\cite{Loutrel:2016cdw}   \\  \hline
        \multirow{2}{*}{  $\tilde{\zeta}(e)$ } & (5.28b) of~\cite{Arun:2009mc}  &\multirow{2}{*}{  (B7h) of~\cite{Ebersold:2019kdc} }&  $\sim(1-e^2)^{-9/2}$   \\ 
          &  (96), (98) \& (100) of~\cite{Loutrel:2016cdw} & & (153), (144) \& (145) of~\cite{Loutrel:2016cdw}   \\  \hline
        \multirow{2}{*}{  $\kappa(e)$ } &  (6.1c) of~\cite{Arun:2007rg}   & \multirow{2}{*}{ (B7e) of~\cite{Ebersold:2019kdc} } & $\sim \ln (1-e^2)\ (1-e^2)^{-13/2}$  \\ 
          &   (8.13) of~\cite{Arun:2007sg}  \&  (101) of~\cite{Loutrel:2016cdw} & &  (8.13) of~\cite{Arun:2007sg} \& (171) of~\cite{Loutrel:2016cdw} \\ \hline 
        \multirow{2}{*}{ $\tilde{\kappa}(e)$ }&  (5.28c) of~\cite{Arun:2009mc} & \multirow{2}{*}{ (B7f) of~\cite{Ebersold:2019kdc} } &   $\sim \ln (1-e^2)\  (1-e^2)^{-5}$   \\ 
          & (5.24) of~\cite{Arun:2009mc} \&   (102) of~\cite{Loutrel:2016cdw} & &  (5.24) of~\cite{Arun:2009mc} \&  (172) of~\cite{Loutrel:2016cdw} \\ \hline 
        \multirow{2}{*}{ $\Lambda_0(e)$ }&    (4.1) of~\cite{Munna:2019fjz}  & \multirow{2}{*}{\eqref{eq:Lambda_0_small_e_expansion} of this paper} &   $\sim \ln (1-e^2)\  (1-e^2)^{-7/2}$   \\ 
          &~\eqref{eq:Lambda_0_def} of this paper  & &  (4.3) of~\cite{Munna:2019fjz} \&~\eqref{eq:Lambda_0_superasymptotic} of this paper   \\ \hline
    \end{tabular}
    \caption{\label{tab:enhancement} Enhancement functions entering the 3PN fluxes. The coefficients of the Fourier expansions entering the definitions are provided in Eqs. (49)-(52), (67)-(70) and (A1)-(A8) of Ref.~\cite{Loutrel:2016cdw}, but beware of the sign error${}^{\text{\ref{footnote:FourierJij}}}$ in Eq.~(A8) of that reference. }
\end{table}

In the aforementioned fluxes, $e_t$ and $x$ were replaced  by their expressions in terms of $(\varepsilon, j)$, which are given, respectively, by Eq.~(20d) of Ref.~\cite{Memmesheimer:2004cv} and Eq.~\eqref{seq:x_inTermsOf_varepsilon_j_sum_loc_log_hered} of this paper. After Taylor-expanding and keeping only terms up to 3PN order, I find that the gauge-invariant expressions for the energy and angular momentum fluxes in terms of $(\varepsilon, j)$ read
\begin{subequations}
\label{eq:fluxes_inTermsOf_varepsilon_j}
\begin{align}
\label{seq:energy_flux_inTermsOf_varepsilon_j}
    \langle \mathcal{F}\rangle &= \frac{32 c^5 \varepsilon ^5 \nu ^2}{5 G j^{3/2}} \Biggl\{\frac{37}{96} -\frac{61}{16 j} +\frac{425}{96 j^2}  \nonumber\\*
    & + \varepsilon 
   \Biggl[-\frac{3139}{2688} +\frac{481 \nu}{768} +\frac{1}{j}\Biggl(\frac{28349}{1792}-\frac{1805}{128} \nu \Biggr) +\frac{1}{j^2}\Biggl( -\frac{3355}{64}+\frac{32225}{768} \nu  \Biggr) +\frac{1}{j^3}\Biggl( \frac{13447}{256}-\frac{5635}{192} \nu  \Biggr) \Biggr] 
     \nonumber\\
    & +4 \pi \, \varepsilon ^{3/2}  j^{3/2}  \varphi \big(\sqrt{1-j}\big)  \nonumber\\
   & +\varepsilon^2 \Biggl[\frac{813485}{258048}-\frac{34529}{21504} \nu +\frac{8695}{12288} \nu ^2 +\frac{1}{\sqrt{j}} \Biggl(-\frac{185}{64}+\frac{37}{32} \nu\Biggr) +\frac{1}{j}\Biggl(-\frac{2235115}{64512}+\frac{1984163}{43008} \nu -\frac{55375}{2048} \nu ^2\Biggr) \nonumber\\*
   &\qquad  +\frac{1}{j^{3/2}} \Biggl(\frac{915}{32}-\frac{183}{16} \nu\Biggr) +\frac{1}{j^2}\Biggl(\frac{93623875}{774144}-\frac{759905}{2688} \nu +\frac{1766275}{12288} \nu ^2\Biggr) +\frac{1}{j^{5/2}}\Biggl(-\frac{2125}{64}+\frac{425}{32} \nu \Biggr)\nonumber\\
   &\qquad  +\frac{1}{j^3}\Biggl(-\frac{64129741}{165888}+\frac{12489631}{18432} \nu -\frac{349615}{1536} \nu^2\Biggr)+\frac{1}{j^4}\Biggl(\frac{29198255}{64512}-\frac{258051}{512} \nu +\frac{27405}{256} \nu ^2\Biggr)\Biggr] \nonumber\\
   &    +  \pi  \varepsilon ^{5/2} j^{3/2} \Biggl[ \left(-\frac{65}{2}+\frac{52}{j}+\frac{13}{6}\nu \right) \varphi \big(\sqrt{1-j}\big)-\frac{8191}{672} \psi \big(\sqrt{1-j}\big)  -\frac{583}{24}   \nu  \zeta  \big(\sqrt{1-j}\big) 
    \nonumber   \\
    &\qquad\qquad\qquad
    +\frac{j}{\sqrt{1-j}} \varphi'\big(\sqrt{1-j}\big) \left(\frac{17}{2}-\frac{7}{2} \nu +\frac{-4+4 \nu }{j}\right)  \Biggr]
   \nonumber\\
   & +\varepsilon^3 \Biggl[ -\frac{79616165}{11354112} +\frac{813485}{229376} \nu -\frac{191479}{114688} \nu ^2+\frac{22015}{32768} \nu^3
   +\frac{1}{\sqrt{j}}\Biggl(\frac{55995}{3584}-\frac{10909}{896} \nu +\frac{407}{128}  \nu^2\Biggr)
   \nonumber\\
   &\qquad 
   +\frac{1}{j}\Biggl(\frac{66618431179}{567705600} +\frac{99}{64}\pi^2+\nu\biggl(-\frac{9921055}{129024}-\frac{4059}{4096} \pi ^2\biggr)  +\frac{53371021 \nu ^2}{688128}-\frac{624585}{16384}\nu ^3 \Biggr)
   \nonumber\\
   &\qquad 
   +\frac{1}{j^{3/2}}\Biggl(-\frac{360071203}{2419200}+\nu\biggl(\frac{3556939}{16128}-\frac{1517 }{6144}\pi ^2\biggr)  -\frac{929}{16} \nu ^2\Biggr) 
   \nonumber\\
   &\qquad
   +\frac{1}{j^2}\Biggl(-\frac{24899642359}{15482880} -\frac{7895}{144}\pi^2+\nu\biggl(\frac{6222921445}{18579456}+\frac{259735}{12288} \pi ^2\biggr)  -\frac{6079515}{8192} \nu ^2+\frac{30490925}{98304} \nu^3 \Biggr)  \nonumber\\
   &\qquad
   +\frac{1}{j^{5/2}}\Biggl(\frac{2658103}{23040}+\nu\biggl(-\frac{305149}{384}+\frac{2501}{1024}\pi ^2\biggr)  +\frac{21685}{128}\nu^2\Biggr)
   \nonumber\\
   &\qquad 
   +\frac{1}{j^3}\Biggl(\frac{785110553611}{92897280} +\frac{86065}{288}\pi ^2+ \nu \biggl(-\frac{1529662327}{442368}-\frac{480725}{12288} \pi ^2\biggr) +\frac{903539077}{294912} \nu ^2  -\frac{20177465}{24576} \nu^3 \Biggr)  \nonumber\\
   &\qquad 
   +\frac{1}{j^{7/2}}\Biggl(\frac{15388747}{23040}+\nu \biggl(\frac{1611419}{2304}-\frac{17425}{6144} \pi ^2\biggr)  -\frac{485}{4} \nu^2\Biggr)
   \nonumber\\
   &\qquad
   +\frac{1}{j^4}\Biggl(-\frac{20625181571}{1228800} -\frac{8239}{16}\pi^2+\nu\biggl(\frac{871577351}{73728}-\frac{255717}{4096} \pi ^2\biggr)   -\frac{21152019}{4096}\nu ^2+\frac{1723365}{2048}\nu^3 \Biggr)   
   \nonumber\\
   &\qquad 
   -\frac{161249}{192 j^{9/2}}  +\frac{1}{j^5}\Biggl(\frac{25574348567}{2211840} +\frac{52745}{192} \pi ^2 + \nu \biggl(-\frac{3972009943}{387072}+\frac{208813}{2048} \pi ^2\biggr) +\frac{5787045}{2048} \nu ^2-\frac{148225}{512} \nu ^3 \Biggr)
   \nonumber\\*
   &\qquad 
   +\left(-\frac{10593}{4480 j} +\frac{168953}{2016 j^2} -\frac{263113}{576 j^3} +\frac{125939}{160 j^4} -\frac{161249}{384 j^5} \right) \ln ( \frac{ 64\, \de^{2\gamma_\text{E}} j^2 \,\varepsilon}{\big(1+\sqrt{j}\big)^2})
    -\frac{116761}{3675}  j^{3/2} \kappa \big(\sqrt{1-j}\big) 
     \Biggr] \Biggr\} \,,
     \end{align}
\begin{align}
\label{seq:angular_momentum_flux_inTermsOf_varepsilon_j}
    \langle \mathcal{G}\rangle &= \frac{32 c^2 m \varepsilon ^{7/2} \nu ^2}{5 j} \Biggl\{-\frac{7}{8}+\frac{15}{8 j}+\varepsilon  \left(\frac{8693}{2688} -\frac{305}{192} \nu +\frac{1}{j}\Biggl(-\frac{95}{8}+\frac{763}{64} \nu \Biggr)  +\frac{1}{j^2}\Biggl(\frac{1795}{128}-\frac{1175}{96} \nu \Biggr)\right) \nonumber\\*
& +4 j \pi  \varepsilon ^{3/2} \tilde{\varphi}\big(\sqrt{1-j}\big) \nonumber\\
& +\varepsilon ^2 \Biggl[-\frac{229361}{32256} +\frac{103349}{21504} \nu -\frac{6191}{3072} \nu ^2 +\frac{1}{\sqrt{j}}\Biggl(\frac{105}{16}-\frac{21}{8}\nu \Biggr) +\frac{1}{j}\Biggl(\frac{2986955}{193536}-\frac{126193}{2688} \nu+\frac{32109}{1024} \nu ^2\Biggr) \nonumber\\
& \qquad +\frac{1}{j^{3/2}}\Biggl(-\frac{225}{16}+\frac{45}{8}\nu \Biggr)+\frac{1}{j^2}\Biggl(-\frac{666995}{27648}+\frac{415765}{3072} \nu-\frac{60535}{768} \nu ^2\Biggr)+\frac{1}{j^3}\Biggl(\frac{1307683}{20736}-\frac{77287}{576} \nu+\frac{4655}{96} \nu ^2 \Biggr)\Biggr]  \nonumber\\
&  + \pi \varepsilon ^{5/2} j \Biggl[    \left(-25+\frac{40}{j}+\frac{5}{3} \nu \right) \tilde{\varphi}\big(\sqrt{1-j}\big)  -\frac{8191}{672}   \tilde{\psi }\big(\sqrt{1-j}\big)   -\frac{583}{24}  \nu  \tilde{\zeta }\big(\sqrt{1-j}\big)\nonumber\\
& \qquad \qquad +\frac{j}{\sqrt{1-j}}    \tilde{\varphi}'\big(\sqrt{1-j}\big)  \left(\frac{17}{2}-\frac{7}{2} \nu +\frac{-4+4 \nu }{j}\right)\Biggr]  \nonumber\\
&  +\varepsilon ^3 \Biggl[\frac{180026785}{11354112}-\frac{604753}{64512} \nu +\frac{1892143}{344064} \nu ^2-\frac{52549}{24576} \nu^3+\frac{1}{\sqrt{j}}\Biggl( -\frac{71395}{1792}+\frac{27453}{896} \nu-\frac{247}{32} \nu^2 \Biggr)  \nonumber\\
& \qquad  +\frac{1}{j}\Biggl(-\frac{926820527}{19353600}-\frac{23}{16} \pi^2+\nu\biggl(\frac{14170117}{221184}+\frac{615}{1024} \pi ^2\biggr)  -\frac{22787}{224} \nu ^2+\frac{456631}{8192} \nu ^3\Biggr) \nonumber\\
& \qquad    +\frac{1}{j^{3/2}}\Biggl(\frac{5104361}{40320}+ \nu \biggl(-\frac{4991}{24}+\frac{287}{512} \pi ^2 \biggr) +\frac{1557}{32} \nu ^2\Biggr)  \nonumber\\
&  \qquad  +\frac{1}{j^2}\Biggl(\frac{1315657259}{2211840}+\frac{505}{16} \pi^2+ \nu\biggl(-\frac{19790245}{73728}-\frac{4305}{2048} \pi ^2\biggr)  +\frac{178339885}{344064} \nu ^2 -\frac{1007455}{4096} \nu ^3\Biggr)   \nonumber\\
&   \qquad +\frac{1}{j^{5/2}}\Biggl(\frac{14825}{768}+ \nu \biggl(\frac{34575}{128}-\frac{615 \pi ^2}{512}\biggr) -\frac{1625}{32}\nu ^2\Biggr)  \nonumber\\
&   \qquad +\frac{1}{j^3}\Biggl(-\frac{458995577}{207360}-\frac{4655}{48} \pi^2+\nu \biggl(\frac{293771975}{165888}-\frac{1435}{64}\pi ^2\biggr) -\frac{1811929}{1536} \nu ^2+\frac{261835}{768} \nu ^3\Biggr)  \nonumber\\
&   \qquad  -\frac{3531}{16 j^{7/2}}  +\frac{1}{j^4}\Biggl(\frac{5567205457}{2580480}+\frac{1155}{16}\pi^2+ \nu  \biggl(-\frac{75465881}{32256}+\frac{74907}{2048} \pi ^2\biggr) +\frac{873067}{1024} \nu ^2 -\frac{36855}{256} \nu ^3\Biggr)  \nonumber\\*
&  \qquad  + \ln \left(\frac{64 \,\de^{2 \gamma_\text{E} } j^2 \varepsilon }{\left(1+\sqrt{j}\right)^2}\right) \left(\frac{2461}{1120 j}-\frac{10807}{224 j^2}+\frac{14231}{96 j^3}-\frac{3531}{32 j^4}\right) -\frac{116761}{3675} j \tilde{\kappa }\big(\sqrt{1-j}\big)\Biggr] \Biggr\} \,.
\end{align}
\end{subequations}

One can then replace $\varepsilon$ and $j$ in the previous fluxes by their newly-obtained expressions in terms of the Blanchet parameters $(x,\iota)$ provided in Eq.~\eqref{eq:varepsilon_j_inTermsOf_x_iota_sum_loc_log_hered}.
Note that even when keeping only 3PN terms in the fluxes, one will find traces of the 4PN equations of motion when working in these variables, e.g., the presence of the enhancement function $\lambda_0(e)$.  This is of course linked to the fact that at Newtonian order, the two frequencies $\omega$ and $n$ are degenerate --- one needs the 1PN equations of motion to find a finite value for $\iota = \frac{3x}{\omega/n -1}$ and control the pair of variables $(x,\iota)$ at leading order. Thus, I find that the gauge-invariant expressions for the fluxes in terms of $(x,\iota)$ read
\begin{subequations}
\label{eq:fluxes_inTermsOf_x_iota}
\begin{align}
\label{seq:energy_flux_inTermsOf_x_iota}
\langle \mathcal{F} \rangle &=\frac{32 c^5 x^5 \nu ^2 }{5 G \iota ^{3/2}}\Biggl\{\frac{37}{96}-\frac{61}{16 \iota }+\frac{425}{96 \iota ^2} \nonumber\\*
  & +x  \Biggl[\frac{139}{112}+\frac{259 }{1152} \nu +\frac{1}{\iota }\Biggl(-\frac{5297}{336}-\frac{2725}{384} \nu \Biggr) +\frac{1}{\iota ^2}\Biggl(\frac{1865}{24}+\frac{3775}{384}  \nu \Biggr) +\frac{1}{\iota   ^3}\Biggl(-\frac{289}{3}+\frac{3605}{384} \nu \Biggr)\Biggr] \nonumber\\
  & +4 \pi  x^{3/2} \iota ^{3/2}
   \varphi \big(\sqrt{1-\iota }\big) \nonumber\\
   &+x^2  \Biggl[\frac{744545}{258048}+\frac{19073}{32256} \nu+\frac{2849}{27648} \nu ^2+\frac{1}{\sqrt{\iota}}\Biggl(\frac{185}{48}-\frac{37}{24} \nu \Biggr)+\frac{1}{\iota}\Biggl(-\frac{2145781}{64512}+  \nu \biggl(-\frac{505639}{10752}+\frac{1517}{8192} \pi^2 \biggr)-\frac{105}{16}  \nu^2\Biggr)   \nonumber\\
   &\qquad +\frac{1}{\iota ^{3/2}}\Biggl(-\frac{305}{16}+\frac{61}{8} \nu \Biggr) +\frac{1}{\iota ^2} \Biggl(\frac{49183667}{387072}+\nu\biggl(\frac{14718145}{32256}-\frac{32595}{8192} \pi ^2\biggr) +\frac{37145 \nu ^2}{4608}\Biggr)  \nonumber\\
   &\qquad +\frac{1}{\iota ^3}\Biggl(-\frac{51894953}{82944}+ \nu\biggl(-\frac{583921}{512}+\frac{497125}{24576} \pi  ^2\biggr) +\frac{1625}{48} \nu ^2 \Biggr) \nonumber\\
   &\qquad +\frac{1}{\iota  ^4}\Biggl(\frac{267725837}{258048}+ \nu\biggl(\frac{1440583}{2304}-\frac{609875}{24576} \pi ^2\biggr)  +\frac{24395}{1024} \nu ^2\Biggr)\Biggr] \nonumber\\
  & + \pi x^{5/2}  \iota ^{3/2}  \Biggl[-\frac{8191}{672}  \psi \bigl(\sqrt{1-\iota}\bigr)  -\frac{583}{24} \nu  \zeta \bigl(\sqrt{1-\iota }\bigr)+ \frac{ \iota}{\sqrt{1-\iota }}  \varphi '\bigl(\sqrt{1-\iota }\bigr)  \Biggl(\frac{17}{2}-\frac{13}{3} \nu +\frac{1}{\iota }\left(-\frac{35}{2}+9 \nu \right)\Biggr) \Biggr]  \nonumber\\
  & +x^3 \Biggl[\frac{2635805}{405504} +\frac{14069}{55296} \nu ^2+\frac{34447}{995328} \nu ^3  +\frac{1}{\sqrt{\iota }}\Biggl(\frac{51335}{2688}-\frac{481}{192} \nu ^2 \Biggr) \nonumber\\
   &\qquad  +\frac{1}{\iota }\Biggl(-\frac{260378549}{141926400}+\frac{99}{64}\pi ^2 -\frac{5912035 }{11010048}\pi ^2\, \nu + \nu^2 \biggl(-\frac{1988257}{48384}+\frac{4551}{32768} \pi ^2 \biggr)-\frac{1454095}{331776}\nu ^3 \Biggr)\nonumber\\
   &\qquad  +\frac{1}{\iota^{3/2}}\Biggl(-\frac{3913177}{37800}+\frac{1517}{4608} \pi ^2 \, \nu +\frac{1153}{32} \nu ^2\Biggr)  \nonumber\\
   &\qquad  +\frac{1}{\iota^2}\Biggl(-\frac{2088085837}{1290240}-\frac{7895}{144} \pi^2+\frac{4733999}{344064} \pi ^2 \, \nu +\nu ^2 \biggl(\frac{22920665}{32256}-\frac{133045}{16384} \pi^2\biggr) +\frac{9515}{2048} \nu ^3\Biggr)  \nonumber\\
   &\qquad  +\frac{1}{\iota ^{5/2}}  \Biggl(-\frac{928043}{5760}-\frac{2501}{1536} \pi ^2 \, \nu -\frac{5605}{192} \nu^2 \Biggr) \nonumber\\
   &\qquad  +\frac{1}{\iota ^3}\Biggl(\frac{5979644669}{645120}+\frac{86065}{288} \pi ^2+\frac{249879115}{16515072}\pi ^2 \nu +\nu^2\biggl(-\frac{16437529}{7168}+\frac{283925}{4608} \pi ^2 \biggr) +\frac{812735}{18432}\nu ^3\Biggr)  \nonumber\\
   &\qquad   +\frac{3727559}{2880 \iota ^{7/2}}    +\frac{1}{\iota ^4}\Biggl(-\frac{898081441}{64800}-\frac{8239}{16} \pi^2-\frac{285258511}{589824} \pi ^2 \nu +\nu ^2\biggl(-\frac{11760667}{18432}-\frac{2622565}{49152} \pi ^2\biggr) +\frac{498325}{4096} \nu ^3 \Biggr)  \nonumber\\
   &\qquad   -\frac{161249}{192 \iota^{9/2}}  \!  + \frac{1}{\iota ^5}\!\Biggl(\! \frac{9189902317}{7741440} \! + \! \frac{52745}{192} \pi^2 \! + \! \frac{11338744391}{14155776} \! \pi ^2 \nu  + \nu^2 \biggl(\!\frac{101283605}{27648}\! -\! \frac{32151175}{294912} \pi ^2 \!\biggr)\! +\! \frac{833875}{12288} \nu ^3 \!\Biggr) \nonumber\\
   &\qquad  +\frac{ \nu  \sqrt{\iota }}{1+\sqrt{\iota }} \Biggl(\frac{891535}{3096576}-\frac{11724017}{3096576 \sqrt{\iota }}-\frac{741395903}{5160960 \iota } -\frac{2982728893}{15482880 \iota^{3/2}}  +\frac{13646721673}{7741440 \iota ^2} +\frac{2525941039}{1105920 \iota^{5/2}}   \nonumber\\
   &\qquad\qquad   -\frac{166284977507}{13934592 \iota ^3}   -\frac{198802962083}{13934592 \iota^{7/2}}+\frac{323444591125}{9289728 \iota ^4} +\frac{39949719725}{1032192 \iota^{9/2}}  -\frac{273905298995}{9289728 \iota ^5} -\frac{292937682995}{9289728 \iota^{11/2}} \Biggr) \nonumber\\
   &\qquad  +\frac{1}{\iota} \ln \left(\frac{64 \de^{2 \gamma_\text{E} } \, x \iota ^2}{\left(1+\sqrt{\iota }\right)^2}\right) \Biggl( -\frac{10593}{4480}+\frac{1369}{960} \nu  +\frac{1}{\iota}\biggl(\frac{168953}{2016}-\frac{2257}{48} \nu  \biggr)+\frac{1}{\iota^2}\biggl(-\frac{263113}{576}+\frac{133645}{288}\nu \biggr)     \nonumber\\
   &\qquad\qquad\qquad\qquad\qquad\qquad + \frac{1}{\iota^3}\biggl(\frac{125939}{160}-\frac{181475}{144}\nu\biggr) +\frac{1}{\iota ^4}\biggl(-\frac{161249}{384}+\frac{1770125 }{1728}\nu  \biggr)\Biggr)      \nonumber\\
   &\qquad  + \frac{\nu  \ln \iota}{\iota}   \left(-\frac{1369}{320} +\frac{2257}{16 \iota} -\frac{133645}{96 \iota ^2}+\frac{181475}{48 \iota ^3} -\frac{1770125}{576 \iota ^4} \right) 
-\frac{116761}{3675}\iota ^{3/2} \kappa \bigl(\sqrt{1-\iota }\bigr)     \nonumber\\
   &\qquad + \frac{ \nu}{\iota^2}  \lambda_0 \bigl(\sqrt{1-\iota }\bigr) \biggl(\frac{37}{2} -\frac{3309}{10 \iota } +\frac{5537}{6 \iota ^2}  -\frac{4165}{6 \iota^3} \biggr) +   \frac{\nu \sqrt{1-\iota }}{\iota ^2}  \lambda_0' \bigl(\sqrt{1-\iota  }\bigr)  \biggl(-\frac{37}{10}+\frac{61}{\iota } -\frac{595}{6 \iota^2} \biggr) \Biggr]\Biggr\} \,,
\end{align}
\begin{align}
\label{seq:angular_momentum_flux_inTermsOf_x_iota}
\langle \mathcal{G} \rangle &= \frac{32 c^2 m x^{7/2} \nu ^2}{5 \iota } \Biggl\{-\frac{7}{8}+\frac{15}{8 \iota }  +x \Biggl[-\frac{1597}{2688}-\frac{31}{32} \nu +\frac{1}{\iota ^2}\Biggl(-\frac{3125}{128}-\frac{275}{96} \nu \Biggr)+\frac{1}{\iota }\Biggl(\frac{535}{64}+\frac{61}{8} \nu \Biggr)\Biggr] \nonumber\\*
   & +4 \pi  x^{3/2} \iota \, \tilde{\varphi }\bigl(\sqrt{1-\iota }\bigr) \nonumber\\
   & +x^2 \Biggl[\frac{5857}{4608}-\frac{31085}{16128} \nu  - \frac{29}{36} \nu ^2  +\frac{1}{\sqrt{\iota }}\Biggl(-\frac{35}{8}+\frac{7}{4} \nu \Biggr)  +\frac{1}{\iota }\Biggl(-\frac{2004251}{96768}+ \nu\biggl(\frac{281177}{5376}-\frac{287}{1024} \pi ^2\biggr) +\frac{2723}{192} \nu ^2\Biggr)  \nonumber\\
   &\qquad +\frac{1}{\iota ^2}\Biggl(\frac{666785}{13824}+ \nu\biggl(-\frac{71445}{256}+\frac{2665}{1024} \pi ^2 \biggr)  -\frac{295}{24} \nu ^2\Biggr) +\frac{1}{\iota^3}\Biggl(\frac{5185091}{41472}+\nu\biggl(\frac{648227}{2304}-\frac{3075}{512} \pi ^2 \biggr)  -\frac{215}{192} \nu ^2\Biggr)\Biggr] \nonumber\\
   & +\pi  x^{5/2} \, \iota \Biggl[-\frac{8191}{672} \tilde{\psi}\bigl(\sqrt{1-\iota }\bigr) -\frac{583}{24} \nu  \tilde{\zeta }\bigl(\sqrt{1-\iota }\bigr) +\frac{\iota}{\sqrt{1-\iota }} \tilde{\varphi }'\bigl(\sqrt{1-\iota }\bigr)  \Biggl(\frac{17}{2}-\frac{13}{3} \nu +\frac{1}{\iota }\left(-\frac{35}{2}+9 \nu \right) \Biggr)  \Biggr] \nonumber\\
   & +x^3 \Biggl[\frac{9265825}{1892352} -\frac{106187}{48384} \nu ^2 -\frac{695}{1296} \nu ^3 +\frac{1}{\sqrt{\iota }}\Biggl(-\frac{545}{84}+\frac{83}{16} \nu ^2 \Biggr)  \nonumber\\
   &\qquad +\frac{1 }{\iota }  \Biggl(-\frac{702474313}{4838400}-\frac{23}{16} \pi ^2 +\frac{861053}{1376256} \pi ^2 \nu  +\nu ^2 \biggl( \frac{102461}{1152}-\frac{4387}{12288}\pi ^2 \biggr) +\frac{9763}{576} \nu ^3 \Biggr)\nonumber\\
   &\qquad +\frac{1 }{\iota ^{3/2}} \Biggl(\frac{4121}{2520}-\frac{287 }{768} \pi ^2 \nu -\frac{279}{16} \nu ^2 \Biggr)\nonumber\\
   &\qquad +\frac{1}{\iota ^2} \Biggl(\frac{751942171}{552960}+\frac{505}{16} \pi ^2 -\frac{655267}{114688}\pi ^2 \nu + \nu ^2 \biggl(-\frac{25632965}{32256}+\frac{35793}{4096} \pi ^2\biggr) -\frac{13585}{576} \nu ^3 \Biggr) \nonumber\\
   &\qquad +\frac{2675}{12 \iota^{5/2}}   +\frac{1}{\iota ^3}\Biggl(-\frac{560628743}{138240}-\frac{4655}{48} \pi^2 -\frac{19331455 }{589824} \pi ^2 \nu +  \nu ^2 \biggl(\frac{3126071}{2304}-\frac{415945}{12288} \pi ^2\biggr)-\frac{425}{32} \nu ^3\Biggr) \nonumber\\
   &\qquad -\frac{3531}{16 \iota ^{7/2}} +\frac{1}{\iota ^4}\Biggl(\frac{50018979133}{23224320}+\frac{1155}{16} \pi ^2 +\frac{14860511 \pi ^2 \nu }{98304}+ \nu ^2 \biggl(\frac{485659}{4608}+\frac{15375 \pi ^2}{4096}\biggr)+\frac{85}{48} \nu ^3\Biggr) \nonumber\\
   &\qquad +\frac{\nu \sqrt{\iota }  }{1+\sqrt{\iota }}\Biggl(-\frac{3695}{1536} -\frac{104473}{10752  \sqrt{\iota }} +\frac{17220089}{207360 \iota }  +\frac{249822191}{1451520 \iota ^{3/2}} -\frac{276365293}{362880 \iota ^2}  -\frac{403169593}{362880 \iota ^{5/2}}  \nonumber\\
   &\qquad \qquad  \qquad +\frac{22508185}{4608 \iota ^3} +\frac{25969945}{4608 \iota ^{7/2}} -\frac{2110187515}{290304 \iota ^4} -\frac{2254129915}{290304 \iota ^{9/2}} \Biggr)    \nonumber\\
   &\qquad +  \frac{1}{\iota}\ln\!  \left(\frac{64 \de^{2 \gamma_\text{E} } x \iota ^2}{\left(1+\sqrt{\iota }\right)^2}\right)  \!\!  \Biggl(\! \frac{2461}{1120}-\frac{259}{120} \nu    + \!\frac{1}{\iota }\biggl(\! -\frac{10807}{224}+\frac{269}{6}\nu \!  \biggr)  + \! \frac{1}{\iota ^2}\biggl( \! \frac{14231}{96}-\frac{15145}{72} \nu \!  \biggr)  +\! \frac{1}{\iota ^3}\biggl(\! -\frac{3531}{32}+\frac{2975}{12} \nu \!  \biggr) \! \Biggr) \nonumber\\
   &\qquad  + \frac{ \nu  \ln \iota}{\iota} \Biggl(\frac{259}{40 } -\frac{269}{2 \iota }+\frac{15145}{24 \iota ^2}-\frac{2975}{4 \iota ^3} \Biggr)   -\frac{116761}{3675} \iota \,  \tilde{\kappa }\left(\sqrt{1-\iota }\right)   \nonumber\\*
   &\qquad+\frac{\nu}{\iota ^2}   \lambda_0\bigl(\sqrt{1-\iota }\bigr) \biggl(-28+\frac{796}{5 \iota } -\frac{168}{\iota ^2} \biggr)+\frac{ \nu  \sqrt{1-\iota }}{\iota ^2} \lambda_0'\bigl(\sqrt{1-\iota }\bigr) \biggl(\frac{28}{5}-\frac{24}{\iota }\biggr) \Biggr]\Biggr\} \,.
\end{align}
\end{subequations}

Finally, one can take the circular limit of Eq.~\eqref{eq:fluxes_inTermsOf_x_iota} using the expression for $\iota^\text{circ}(x)$ that was obtained in Eq.~\eqref{seq:iota_circ_inTermsOf_x}. This operation requires Taylor-expanding the enhancement functions, so  the values of the enhancement functions and their derivatives are needed for $e=0$; these can be read off Eq.~(B7) of~Ref.~\cite{Ebersold:2019kdc}. The values of $\lambda_0(0)$ and $\lambda_0'(0)$ are also needed, and can be read off Eq.~\eqref{eq:lambda_0_expansion}. I  recover the usual expression for the 3PN circular energy flux $\langle \mathcal{F}^\text{circ} \rangle$, compare for example  against Eq.~(4) of Ref.~\cite{Blanchet:2023bwj}. I have also checked that the circular angular momentum flux thus obtained is related the circular energy flux by the usual relation $\langle \mathcal{G}^\text{circ} \rangle = \langle \mathcal{F}^\text{circ} \rangle / \omega$ at the required order. 

As a closing remark, note that I have ignored in this section the possible contribution of Schott terms, namely the difference between conservative and binding energies and angular momenta~\cite{Trestini:2025nzr}. This is licit because, after orbit-averaging, the Schott terms contribute only in the 4PN relations between $(x,\iota)$ and $(\varepsilon,j)$ [e.g., as a term $\propto \nu x^5$ in $\varepsilon(x,\iota)$], whereas only  the 3PN relations were used here. This is true even though the 3PN relation between $j$ and $\iota$ carries information about the 4PN equations of motion.

\section{Discussion}
\label{sec:discussion}

In this work, I have obtained the conservative energy and angular momentum in terms of the (orbital) radial and azimuthal frequencies at fourth post-Newtonian order. This result strongly relied on the action-angle approach, where the hereditary term was treated as a perturbation to the local Hamiltonian. It was then localized and Delaunay averaged using nonlocal shifts, along the lines of Refs.~\cite{Damour:2014jta, Damour:2015isa, Damour:2016abl}; it would be interesting to check whether one can recover these results by using directly the nonlocal Hamiltonian, as in Ref.~\cite{Bernard:2016wrg}. Thanks to the resummation of the enhancement function associated with the tails, I was able to ensure that this result is extremely accurate: the relative error remains $< 4 \cdot 10^{-6}$ for any value of the eccentricity. Using the first law of binary black hole mechanics, I have then  obtained the 4PN redshift for eccentric orbits, which I found to be in exact agreement with both the geodesic and postgeodesic redshifts obtained using black hole perturbation theory and  self-force techniques. As an application, I reexpressed in terms of the fundamental frequencies the 3PN orbit-averaged fluxes of energy and angular momentum, which were first computed by Refs.~\cite{Arun:2007rg, Arun:2007sg, Arun:2009mc} in terms of energy and angular momentum. 

This work is thus an important step towards the 4PN phasing for eccentric orbits, but several hurdles lie ahead. First, recall that it is the \textit{binding} energy and angular momentum, not the \textit{conservative} energy and angular momentum, which enter the flux balance law. The differences between the two are called Schott terms, and are yet to be computed for eccentric orbits. Most of these Schott terms are instantaneous, and have been obtained in harmonic gauge at 2.5PN and 3.5PN \cite{Iyer:1993xi,Iyer:1995rn,Pati:2002ux,Nissanke:2004er}
; these vanish upon orbit averaging, but it might be necessary to control their oscillatory contributions \cite{Boetzel:2019nfw}. Recently, it was discovered \cite{Trestini:2025nzr} that at 4PN, these Schott terms are non-vanishing, even after orbit-averaging (this is due to hereditary contributions). They have been computed in the case of circular orbits, but not eccentric orbits; this would require the 2.5PN decomposition of the quadrupole moment into a Fourier series. Moreover, it was noticed that the binding energy could be more straightforwardly obtained by introducing an effective, gravitational-wave frequency. It would be interesting to understand how to extend this procedure to the case of eccentric orbits, where there are two frequencies.

Furthermore, the fluxes will be required at 4PN order. If one wants to express them in terms of energy and angular momentum, several new ingredients will be needed. The first one is a complete quasi-Keplerian parametrization of the motion, extending the local parametrization of \cite{Cho:2021oai}. Indeed, I have here obtained the fundamental frequencies of the motion, but I do not control the precise evolution within an orbit. This is because I did not attempt to control the complete map between the angle variables and the relative position and velocity of the binary in harmonic coordinates; however, I have tried to control as many of the intermediate transformations as possible, see Eqs.~\eqref{seq:ell_prime_inTermsOf_loc} and \eqref{eq:ell_g_Irphi_Iphi_noprime_inTermsOf_prime}. The second ingredient is the Fourier decomposition of various multipolar moments; in particular, the 2PN Fourier decomposition was recently obtained in Ref.~\cite{Liu:2025tcj}; one also requires the 1PN decomposition of the mass-type octupole moment and current-type quadrupole moment as well as other moments at Newtonian order. These are necessary to treat the hereditary effects entering the flux, including the tails-of-memory~\cite{Trestini:2023wwg}. For the orbit-averaged fluxes, the enhancement functions should be resummed as in Ref.~\cite{Loutrel:2016cdw} in order to ensure that their approximate representation is valid for all eccentricities. Then, using the results of Refs.~\cite{Blanchet:2023sbv,Blanchet:2023bwj}, one should have all the ingredients to obtain the energy flux in terms of energy and angular momentum; the angular momentum flux should be similarly obtained, along the lines of the circular results of Ref.~\cite{Khairnar:2024rzs}.  However, in order to obtain the fluxes at 4PN in terms of the frequencies, it will be necessary to control the 5PN periastron advance, and thus the 5PN equations of motion; the requirement of this higher order is due to the degeneracy between frequencies at Newtonian order, as explained in Footnote \ref{footnote:error_iota}. The phasing should then be straightforward to obtain using the flux balance laws, modulo possible contributions of post-adiabatic effects~\cite{Boetzel:2019nfw}. Note that at 4PN, one should also include the contribution of the horizon fluxes \cite{Tagoshi:1997jy,Poisson:2005pi,Nagar:2011aa, Cunningham:2024dog}.

Finally, this work neglected the effects of spin, but they should be rather straightforward to include in the case of aligned or anti-aligned spins, even at 4PN. When the spins are not aligned, there is precession of the orbital plane, and three fundamental frequencies need to be computed, as well as the precession and nutation rates of each particle. This is a significantly  harder problem and should first be completed at lower orders.

\section*{Acknowledgments}
\label{sec:acknowledgments}

I thank Sashwat Tanay for discussions at a preliminary stage of this work, clarifications about Ref.~\cite{Cho:2021oai} and for convincing me of the usefulness of the action-angle approach. I also thank Luc Blanchet, Francisco Blanco, Thibault Damour, Jack Lewis and Adam Pound for discussions, suggestions, and clarifications about the localization of the tail Hamiltonian, as well as Quentin Henry for spotting a typo in Eq.~\eqref{seq:I7_inTermsOf_varepsilon_j} of a previous version of this work. I acknowledge support from the ERC Consolidator/UKRI Frontier Research Grant GWModels (selected by the ERC and funded by UKRI [Grant No. EP/Y008251/1]). This research was supported by the Munich Institute for Astro-, Particle and BioPhysics (MIAPbP) which is funded by the Deutsche Forschungsgemeinschaft (DFG, German Research Foundation) under Germany's Excellence Strategy -- EXC-2094 -- 390783311.

\appendix

\section{Redefinition of the phase-space variables for the localization of the hereditary Hamiltonian}
\label{app:shift_localization}

I first make contact with the notations of Ref.~\cite{Blanco:2024fte}. The phase-space variables are Delaunay variables and are denoted $\bm{Q} = Q^A = (\ell, g, I_{r\phi}, I_\phi)$. These variables are canonical so the symplectic form of the unperturbed problem is 
\begin{align}\label{eq:symplectic_form}
\Omega_{AB}^0  &= \begin{pmatrix} 0 &0 & -1 &0 \\ 0 &0 &0 & -1 \\ 1 & 0& 0 &0 \\ 0 & 1 & 0 &0 \end{pmatrix}\,.
\end{align}
The unperturbed Hamilton equations are then given by  
\begin{align}
\label{eq:Hamilton_equations_Q_A}
 \frac{\dd Q^A}{\dd t} = \Omega^{AB}_0 \partial_B (H^\text{loc}+H^\text{log}) \,,
\end{align}
where I have introduced the inverse of the unperturbed symplectic form (with indices upstairs) which reads
\begin{align}\label{eq:inverse_symplectic_form}
\Omega^{AB}_0  &= \begin{pmatrix} 0 &0 & 1 &0 \\ 0 &0 &0 & 1 \\ -1 & 0& 0 &0 \\ 0 & -1 & 0 &0 \end{pmatrix}\,.
\end{align}
In my case, the 2-point function defined in (8) of \cite{Blanco:2024fte} reads 
\begin{align}\label{eq:G2_of_Q1_Q2_sigma_def}
\mathcal{G}_2(\bm{Q}_1, \bm{Q}_2; \sigma) &= - \frac{2 G^2 m}{5 c^8} \frac{ \dI_{ij}^{(3)}(\bm{Q}_1) \dI_{ij}^{(3)}(\bm{Q}_2)}{|\sigma|}\,,
\end{align}
which corresponds to Eq.~(71) of Ref.~\cite{Blanco:2024fte} with $\cC = - {2 G^2 \dM}/{(5 c^8)} $. The third derivative of the quadrupole moment is explicitly expressed in terms of the phase-space variables using the Fourier decomposition
\begin{align}
 \dI_{ij}^{(3)}(\bm{Q}) &= \sum_{p\in\mathbb{Z}} (\di p n)^3 {}_p \widetilde{\dI}_{ij} \de^{\di p \ell}\,,
\end{align}
where one recalls that $n$ and ${}_p \widetilde{\dI}_{ij} = \mathcal{I}_2 \ {}_p \widehat{\dI}_{ij}$ are expressed in terms of the action variables $I_{r\phi}$ and $I_\phi$; see Eqs.~\eqref{eq:n_inTermsOf_Irphi_Iphi_Newtonian},~\eqref{eq:I2_expression} and~\eqref{eq:pIij_hat_expr}.   From Eq.~(8) of \cite{Blanco:2024fte} with $N=2$, one finds that the hereditary perturbation to the action reads
\begin{align}\label{eq:Snl_inTermsOf_G2}
S^\text{hered} &= - \frac{1}{2} \mathrm{Pf}_{2 \eta /c} \int \dd s_1 \dd s_2 \, \mathcal{G}_2(X_{s_1}(\bm{Q}), X_{s_2}(\bm{Q}) ; s_2-s_1) \,,
\end{align}
where $\eta$ is the scale defined in Eq.~\eqref{eq:def_eta_scale} to split the `logarithmic' and `hereditary' contributions to the tail action; I thus recover the hereditary piece of the tail action \eqref{seq:S_tail}.
Note that I have applied (in a relatively \textit{ad hoc} manner) a \textit{partie finie} regulator, which is compatible with the formalism of Ref.~\cite{Blanco:2024fte}; see the comment after Eq.~(70) in that reference. One thus applies the \textit{partie finie} regulator on Eq.~(10) of \cite{Blanco:2024fte} with $N=2$, and read off that
\begin{align}\label{eq:Phi_Q1_Q2_X}
\Phi(\bm{Q}_1,\bm{Q}_2, [X]) &\equiv \ \mathrm{Pf}_{2 \eta /c}\int \dd s_2  \ \mathcal{G}_2(\bm{Q}_1, X_{s_2}(\bm{Q}_2); s_2) \nonumber\\
&= - \frac{2 G^2 m}{5 c^8}\dI_{ij}^{(3)} (\bm{Q}_1) \mathrm{Pf}_{2\eta/c} \int \frac{\dd s}{|s|}  \dI_{ij}^{(3)}(X_{s}(\bm{Q}_2))  \\
&= - \frac{2 G^2 m}{5 c^8}\dI_{ij}^{(3)} (\bm{Q}_1)   \int_{0}^\infty \dd \tau  \ln\left(\frac{c \tau}{2\eta}\right) \Biggl\{\frac{\dd\dI_{ij}^{(3)}(X_{s}(\bm{Q}_2))}{\dd s}\Bigg|_{s=-\tau} - \frac{\dd\dI_{ij}^{(3)}(X_{s}(\bm{Q}_2))}{\dd s}\Bigg|_{s=\tau} \Biggr\} \,. \nonumber 
\end{align}
Following Eq.~(17) of Ref.~\cite{Blanco:2024fte} for $n=2$, I introduce the distribution
\begin{align}\label{eq:chi_of_s1_s2_def}
\chi(s_1, s_2) &\equiv \frac{\mathrm{sg}(s_1)-\mathrm{sg}(s_2)}{2} = \Theta(s_1) - \Theta(s_2) \,,
\end{align}
where $\mathrm{sg}(s)$ is the sign function and $\Theta(s)$ is the Heaviside function. Thus, one has the distributional identities \mbox{${\partial \chi}/{\partial s_1} = \delta(s_1)$} and \mbox{${\partial \chi}/{\partial s_2} = -\delta(s_2)$}.
Following Eq.~(20) of \cite{Blanco:2024fte} (with $n=2$ and $\mathcal{N}=1$), one now defines
\begin{align}\label{eq:K_Q1_Q2_def}
K_2^{(1)}(\bm{Q}_1, \bm{Q}_2) &\equiv \int \dd s_1 \dd s_2 \, \chi(s_1, s_2) \mathcal{G}_2(\bar{X}_{s_1}^{(0)}(\bm{Q}_1),\bar{X}_{s_2}^{(0)}(\bm{Q}_2); s_2-s_1) \nonumber\\
& \!\!\!\!  =  - \frac{2 G^2 m}{5 c^8} \,\mathrm{Pf}_{2\eta /c} \iint \dd s_1 \dd s_2 \frac{\Theta(s_1) - \Theta(s_2)}{|s_2-s_1|} \dI_{ij}^{(3)}(\bar{X}_{s_1}^{(0)}(\bm{Q}_1)) \dI_{ij}^{(3)}(\bar{X}_{s_2}^{(0)}(\bm{Q}_2)) \\
& \!\!\!\!   =  - \frac{2 G^2 m}{5 c^8} \!\!\!\!\sum_{(p,q)\in\mathbb{Z}^2}  \!\!  (\di p n_1)^3  (\di q n_2)^3 {}_p \widetilde{\dI}_{ij}(\bm{Q}_1) {}_q \widetilde{\dI}_{ij}(\bm{Q}_2) \de^{\di (p \ell_1 + q \ell_2)}  \, \mathrm{Pf}_{2 \eta/c} \iint \dd s_1 \dd s_2 \frac{\Theta(s_1) - \Theta(s_2)}{|s_2-s_1|}   \de^{\di (p n_1 s_1 + q n_2 s_2)} \,. \nonumber
\end{align}
I will need to take partial derivatives of the $K_2^{(1)}(\bm{Q}_1, \bm{Q}_2)$ with respect to $\bm{Q}_1$ and then evaluate at coincidence $\bm{Q}_1=\bm{Q}_2=\bm{Q}$. For $p+q \neq 0$, it will be enough to evaluate Eq.~\eqref{eq:K_Q1_Q2_def} explicitly using the  integration formula
\begin{subequations}
\label{eq:Pf_integral_formulas}
\begin{align}
\label{seq:Pf_integral_formula_AC_case}
\mathrm{Pf}_{2\eta/c} \iint \dd s_1 \dd s_2 \frac{\Theta(s_1) - \Theta(s_2)}{|s_2-s_1|}   \de^{\di (p n_1 s_1 + q n_2 s_2)} &= \frac{2\di}{pn_1+q n_2}\ln\left(\frac{|p| n_1}{|q|n_2}\right) \qquad\qquad\text{if } p n_1+q n_2\neq 0 \,,
\end{align}
then take a time-derivative with respect to $\bm{Q}_1$, and only at the end evaluate at coincidence. However, this procedure is somewhat pathological for $p+q=0$. In the latter case, it will be more useful to take derivatives with respect to $\bm{Q}_1$ before performing the integral. Any residual factor does not depend on $(s_1,s_2)$ and can be factored out of the integral. One then first evaluates at coincidence, uses $q= -p$ and only then performs the remaining integral, which is
\begin{align}
\label{seq:Pf_integral_formula_DC_case}
\mathrm{Pf}_{2 \eta/c} \iint \dd s_1 \dd s_2 \frac{\Theta(s_1) - \Theta(s_2)}{|s_2-s_1|}   \de^{\di p n( s_1 -  s_2)} = 0  \,.
\end{align}
\end{subequations}
Since I have now shown that the terms associated with $p+q=0$ are always vanishing in the end result, one can be more flexible and write
\begin{align}\label{eq:K_Q1_Q2_result}
K_2^{(1)}(\bm{Q}_1, \bm{Q}_2) & =  \frac{4 \di G^2 m}{5 c^8} \sum_{p + q \neq 0}     \frac{(p n_1)^3  (q n_2)^3}{pn_1+q n_2}\ln\left(\frac{|p| n_1}{|q|n_2}\right) {}_p \widetilde{\dI}_{ij}(\bm{Q}_1) {}_q \widetilde{\dI}_{ij}(\bm{Q}_2)   \de^{\di (p \ell_1 + q \ell_2)} \nonumber\\*[0.23cm]
& \qquad\qquad\quad+ \text{(terms that vanish at coincidence)} \,. 
\end{align}
Thus, I gladly find that the dependence on the scale $\eta$ has dropped out, which is what is expected: the transformation to localize the Hamiltonian should not depend on the arbitrary choice of splitting between local and hereditary Hamiltonian. One now has all the ingredients to compute the gauge transformation, which is given in Eq.~(37) of~Ref.~\cite{Blanco:2024fte} (for $\mathcal{N}=2$) by
\begin{align}\label{eq:xi_def}
\xi^A &= - \frac{1}{2} \Omega_0^{AB}\Biggl[\frac{\partial }{\partial Q_1^B} K_2^{(1)}(\bm{Q}_1, \bm{Q}_2)\Biggr]_{\bm{Q}_1=\bm{Q}_2=\bm{Q}} \,.
\end{align}
I find that this gauge vector is explicitly given by
\begin{subequations}
\label{eq:xi_value}
\begin{align}
\label{seq:xi_1_value}
\xi^1 = \ell' - \ell^\text{loc} &= -\frac{2\di G^2 m}{5 c^8} \sum_{p+q\neq 0} \frac{p^3 q^3}{p+q}  n^5 \Biggl\{   \frac{\partial ({}_p \widetilde{\dI}_{ij})}{\partial I_{r\phi}} \ln \left|\frac{p}{q}\right| + \frac{1}{n}\frac{\partial n}{\partial I_{r\phi}} \Bigg[1 + \left(2+ \frac{q}{p+q}\right)\ln \left|\frac{p}{q}\right| \Bigg] {}_p \widetilde{\dI}_{ij}  \Biggr\}\  {}_q \widetilde{\dI}_{ij} \de^{\di (p+q)\ell}  \,,  \\
\label{seq:xi_2_value}
\xi^{2} = g' - g^\text{loc} &=   -\frac{2\di G^2 m}{5 c^8} \sum_{p+q\neq 0} \frac{p^3 q^3}{p+q}  n^5   \ln \left|\frac{p}{q}\right| \frac{\partial ({}_p \widetilde{\dI}_{ij})}{\partial I_{\phi}}     {\ }_q \widetilde{\dI}_{ij} \ \de^{\di (p+q)\ell} \,, \\
\label{seq:xi_3_value}
\xi^3 = I_{r\phi}' - I_{r\phi}^\text{loc} &=  -\frac{2 G^2 m}{5 c^8} \sum_{p+q\neq 0} \frac{p^4 q^3}{p+q} \ln \left|\frac{p}{q}\right| n^5 {\ }_p \widetilde{\dI}_{ij} {\ }_q \widetilde{\dI}_{ij} \ \de^{\di (p+q)\ell} \,, \\
\label{seq:xi_4_value}
\xi^4 = I_{\phi}' - I_{\phi}^\text{loc} &= 0 \,,
\end{align}
\end{subequations}
where ${\partial n}/{\partial I_{r\phi}} = - {3G^2 m^5 \nu^3}/{I_{r\phi}^{4}}$. 

Finally, the associated localized Hamiltonian is given by Eq.~(62) of Ref.~\cite{Blanco:2024fte}, in which one neglects $\mathcal{O}(\epsilon^2)$ terms,  and reads 
\begin{align}\label{eq:H_hat_1_localized}
\hat{H}^{(1)}(\bm{Q}') = H_0(\bm{Q}') + \frac{1}{2}\Phi^{(1)}(\bm{Q}') \,,
\end{align}
where one defines
\begin{align}\label{eq:Phi_1_def_value}
\Phi^{(1)}(\bm{Q}') &= \Phi(\bm{Q}',\bm{Q}', [\bar{X}^0]) =   -\frac{4 G^2 m}{5 c^8} (n')^6 \left(\mathcal{I}_2'\right)^2 \sum_{(p,q)\in\mathbb{Z}^2}p^3 q^3 \ln\left(\frac{|p|}{2}\right)  {\ }_p\widehat{\dI}_{ij} {\ }_q\widehat{\dI}_{ij} \de^{\di(p+q)\ell'} 
\end{align}
analogously to Eq.~(50a) of Ref.~\cite{Blanco:2024fte}, in which  $\mathcal{O}(\epsilon^2)$ terms are neglected. Through this choice of notation, I have made it explicit that the new, localized Hamiltonian is to be expressed in terms of the new primed coordinates \mbox{$\bm{Q}'=(\ell',g',I_{r\phi}', I_\phi')$} and functions thereof, such as $n'$ and $\mathcal{I}_2'$.

Thus, Eqs.~\eqref{eq:H_hat_1_localized} and~\eqref{eq:Phi_1_def_value} exactly recover the result obtained in Eq.~\eqref{eq:H_hered_localized_expr}, which was obtained by naively replacing the equations of motion in the Hamiltonian, without controlling the associated transformation of the phase-space variables. Finally, as a caveat, note that the method exposed assumed that the \textit{partie finie} can be seamlessly applied; a more careful inspection would be necessary, for example, to rigorously control the absence of nonoscillatory terms in the gauge transformation \eqref{eq:xi_value}. I do not use this transformation in practice to derive the final results, so this derivation primarily serves the purpose of a proof of principle.

\section{Self force results}
\label{app:GSF}

In this section, I will consider the motion of a nonspinning test particle (of negligible mass $m_2$) following a geodesic around a Schwarzschild black hole of mass $m_1$; namely, I work at leading order in the mass ratio $\epsilon=m_2/m_1$.
It is then possible to analytically  express the frequencies $(n,\omega)$ and the small particle's averaged  redshift at geodesic order~$\langle z_2^\text{geo} \rangle$   in terms of the Darwin  semi-latus rectum and eccentricity $(p,e)$. The latter are directly related to the energy and angular momentum of the system by~\cite{Munna:2020civ}
\begin{align}\label{eq:cal_E_L_p_e}
    \mathcal{E} &= \sqrt{\frac{(p-2-2e)(p-2+2e)}{p(p-3-e^2)}} \,,&&&
    \mathcal{L} &= \frac{m_1\,p }{\sqrt{p-3-e^2}}\,,
\end{align}
where I have defined $\mathcal{E}=1-\varepsilon/2$ and $\mathcal{L} = m_1 \sqrt{j/\varepsilon}$.
The radial frequency $n$, the azimuthal frequency $\omega$ and the averaged  redshift variable $\langle z_2^\text{geo}\rangle$ can be straightforwardly related to the radial period in terms of coordinate time $P$, the radial period in terms of proper time $T$ and the accumulated azimuthal angle per radial period $\Phi$  through\footnote{In the gravitational self force literature~\cite{Akcay:2015pza},  the radial period is denoted $T_{r0}$ with respect to coordinate time and    $\mathcal{T}_{r0}$   with respect to  proper time.}
\begin{align}\label{eq:n_omega_z1_of_P_Phi_T}
n = \frac{2\pi}{P}\,, && \omega = \frac{\Phi}{P} \,,&& \langle z_2^\text{geo} \rangle = \frac{T}{P} \,.
\end{align}
The quantities $(P,\Phi,T)$ are then expressed~\cite{Cutler:1994pb, Barack:2010tm,Akcay:2015pza} as definite integrals in terms of $(p,e)$. These are given in Eq.~(B3)~of~Ref.~\cite{Blanchet:2017rcn} (see also Ref.~\cite{Barack:2010tm}), and read 
\begin{subequations}
\label{eq:P_Phi_T_integrals_p_e}
\begin{align}
\label{seq:P_integrals_p_e}
    P &= \frac{G m_1}{c^3} \int_0^{2\pi} \dd \chi \ \frac{p^2}{(p-2-2e \cos \chi)(1+e \cos\chi)^2} \sqrt{\frac{(p-2-2e)(p-2+2e)}{p-6 -2e \cos \chi}} \,,\\*
\label{seq:Phi_integrals_p_e}
    \Phi &= \int_0^{2\pi} \dd \chi \ \sqrt{\frac{p}{p-6-2e \cos\chi}} \,, \\*
\label{seq:T_integrals_p_e}
    T &= \frac{G m_1}{c^3} \int_0^{2\pi} \dd \chi \frac{p^{3/2}}{(1+e\cos\chi)^2}\sqrt{\frac{p-3-e^2}{p-6-2e\cos\chi}}\,.
\end{align}   
\end{subequations}
It is then relatively straightforward 
to perform the post-Newtonian (large $p$) expansion of these expressions. Recalling Eq.~\eqref{eq:n_omega_z1_of_P_Phi_T} as well as the definitions of $(y,\lambda)$ provided in Eq.~\eqref{eq:def_y_lambda}, one finds that the required relations read at 4PN:
\begin{subequations}
\label{eq:y_lambda_z2_geo_inTermsOf_p_e}
\begin{align}
\label{seq:y_inTermsOf_p_e}
    y &= \frac{1-e^2}{p} \Biggl\{1 +\frac{2 e^2}{p} +\frac{1}{p^2}\left[5-\frac{3}{2}e^2+5 e^4+\sqrt{1-e^2} \left(-5+5 e^2\right) \right] \nonumber\\
    & \qquad+\frac{1}{p^3} \left[10+34 e^2-13 e^4+\frac{40}{3} e^6+\sqrt{1-e^2} \left(-10-15 e^2+25
   e^4\right) \right] \nonumber\\
    & \qquad +\frac{1}{p^4}\left[\frac{469}{4}-\frac{101}{4} e^2+\frac{6969}{32} e^4 -\frac{355}{4} e^6 +\frac{110}{3} e^8+\sqrt{1-e^2}
   \Bigl(-\frac{469}{4}+\frac{693}{8} e^2-\frac{555}{8} e^4+100 e^6\Bigr)\right] \! \Biggr\} \,, \\
\label{seq:lambda_inTermsOf_p_e}
   \lambda &= \left(1-e^2\right) \Biggl\{1+\frac{1}{p}\left[-\frac{9}{2}+\frac{7}{4} e^2 \right] +\frac{1 }{p^2}\left[\frac{11}{4}-12 e^2+\frac{73}{16}e^4+\sqrt{1-e^2} \left(-5+5 e^2\right) \right] \nonumber\\*
   & \quad +\frac{1}{p^3}\left[-\frac{77}{4}+\frac{407}{16} e^2-\frac{2409}{64} e^4 +\frac{2341}{192}e^6 +\sqrt{1-e^2} \Bigl(\frac{25}{2}-\frac{145}{4} e^2 +\frac{95}{4} e^4 \Bigr) \right]  \nonumber\\*
   & \quad +\frac{1}{p^4}   \left[\frac{5719}{16}  \!-\frac{195}{4} e^2  \! +\frac{11827}{16} e^4  \!-\frac{33585}{128} e^6  \!+ \frac{93211}{768} e^8  \!+  \! \sqrt{1-e^2}
   \Bigl(-\frac{2495}{8}+\frac{4165}{16} e^2-\frac{1025}{4} e^4+\frac{4925}{16} e^6\Bigr) \! \right] \! \Biggr\} \,,\\
\label{seq:z2_geo_inTermsOf_p_e}
   \langle z_2^\text{geo}\rangle & =1-\frac{3 \left(1-e^2\right)}{2 p}
   +\frac{\left(1-e^2\right)^{3/2}}{p^2}\bigg[ -6+\frac{39}{8} \sqrt{1-e^2}\bigg]
   +\frac{\left(1-e^2\right)^{3/2}}{p^3} \bigg[7-30 e^2+\sqrt{1-e^2}
   \left(-\frac{139}{16}+\frac{235}{16}e^2\right)\bigg] \nonumber\\*
   &
   +\frac{\left(1-e^2\right)^{3/2}}{p^4} \bigg[-\frac{309}{4}+\frac{159}{2} e^2-\frac{507}{4}
   e^4 + \sqrt{1-e^2} \left(\frac{9483}{128}-\frac{6027}{64} e^2+\frac{5643}{128} e^4\right)\bigg] \nonumber\\*
   &
   +\frac{\left(1-e^2\right)^{3/2}}{p^5} \bigg[-\frac{33}{8}-\frac{1629}{2} e^2+\frac{2265}{4}e^4-\frac{981}{2} e^6+\sqrt{1-e^2} \left(-\frac{645}{256}+\frac{139983}{256}e^2-\frac{148623}{256} e^4+\frac{33861}{256} e^6\right)\bigg] \,.
\end{align}
\end{subequations}

Finally, in Sec.~\ref{sec:redshift}, I compare the redshift at postgeodesic order, namely at subleading order in the mass ratio; this is also called the first self-force (1SF) order. In that case, the relations that have just been derived cannot all hold; the maps \eqref{eq:cal_E_L_p_e} and \eqref{eq:P_Phi_T_integrals_p_e} can only be simultaneously valid in the test-particle limit. One will thus need to make a choice for the definition of the Darwin parameters $(p,e)$. One possible choice (which I will \textit{not} make here) is to define $(p,e)$ from the energy and angular momentum map~\eqref{eq:cal_E_L_p_e}; this would induce $\mathcal{O}(\epsilon)$ corrections to the frequency map~\eqref{eq:P_Phi_T_integrals_p_e}. In order to be consistent with the conventions\footnote{Indeed, Ref.~\cite{Munna:2022gio} performs a fixed-frequency expansion: ``\textit{To achieve a gauge-invariant result, we make the assumption that the (observable) radial libration frequency is held fixed in going from the background geodesic to the first-order perturbed orbit.}''}  used in the literature for $\langle z_2^\text{1SF}\rangle(p,e)$, the choice that I will make is instead to define $(p,e)$ from the frequency map~\eqref{eq:P_Phi_T_integrals_p_e}; the energy and angular momentum map~\eqref{eq:cal_E_L_p_e} will thus acquire corrections of order~$\mathcal{O}(\epsilon)$. Note that isofrequency pairing~\cite{Warburton:2013yj} restrict such a definition to large enough $p$, which is not a problem in the post-Newtonian limit. 

\section{Explicit expressions for the small eccentricity expansions of the enhancement functions}
\label{app:general_term}

In this Appendix, I derive a closed form expression for $\alpha_n$ for any $n\in\mathbb{N}$. Consider the definition \eqref{eq:Lambda_0_def} of $\Lambda_0(e)$, and replace the Fourier modes ${}_{p}\widehat\dI_{ij}(e)$ by their explicit expression \eqref{eq:pIij_hat_expr} in terms of Bessel functions. Expanding, I find that the most general type of term one needs to consider is
\begin{align}
    Q_{\alpha\beta\gamma} = \sum_{p=1}^{\infty} p^{4-\gamma} \ln\left(\frac{p}{2}\right) J_{p-\alpha}(pe) J_{p-\beta}(pe) \,,
\end{align}
where $\alpha\in\{0,1\}$, $\beta\in\{0,1\}$, and $\gamma\in\{0,1,2\}$. The enhancement function can then be written as
\begin{align}\label{eq:Lambda_0_Q_abc}
    \Lambda_0(e) = \sum_{\{\mathcal{C},\alpha,\beta,\gamma,\delta\}\in \mathcal{S}} \mathcal{C} \,e^{\delta} Q_{\alpha\beta\gamma}\,,
\end{align}
where $\mathcal{S}$ is the set of allowed parameters entering the sum, $\mathcal{C}$ is the numerical prefactor of each term entering the sum and $\delta$ is an integer. In $Q_{\alpha\beta\gamma}$, I replaced the Bessel functions by their Taylor expansion 
\begin{align}
    J_n(x) = \sum_{m=0}^{\infty} \frac{(-1)^m}{m! \,(m+n)!} \left(\frac{x}{2}\right)^{2m+n} \,,
\end{align}
expanded and reordered the sum. I then removed one of the nested sums using Vandermonde's identity:
\begin{align}
    \sum_{n=0}^{k} \begin{pmatrix} p-\beta\\n \end{pmatrix}  \begin{pmatrix} p-\alpha\\k-n \end{pmatrix} =  \begin{pmatrix} 2p-\alpha-\beta\\k \end{pmatrix} \,.
\end{align}
I find that $Q_{\alpha\beta\gamma} = \sum_{p=1}^\infty e^{2p-\alpha-\beta}K_{p\alpha\beta\gamma}$ where
\begin{align}
    K_{p\alpha\beta\gamma}&= \frac{(-1)^p}{2^{2p-\alpha-\beta}} \begin{pmatrix}
        2p-\alpha-\beta\\ p-\alpha
    \end{pmatrix}\sum_{k=1}^{p} \frac{(-1)^k k^{2p+4-\alpha-\beta-\gamma}\ln(k/2)}{(p-k)!(p+k-\alpha-\beta)!}\,.
\end{align}
Using \texttt{Wolfram Mathematica}, I then inject this expression into Eq.~\eqref{eq:Lambda_0_Q_abc}, summing over all contributions labeled by elements of $\mathcal{S}$. I obtain the structure
\begin{align}
    \Lambda_0(e) = \sum_{p=-1}^{\infty} e^{2p} \mathcal{A}_{p} \,.
\end{align}
Upon simplification, I then find as expected that $\mathcal{A}_{\,-1}=\mathcal{A}_0=0$ and that, for $p\ge 1$, the coefficients read
\begin{align}\label{eq:coeff_calA_p}
   \mathcal{A}_p&= \frac{1}{2^{2p+4}\,(p!)^2}\Bigg[(p+2)^{2 p+4} \ln \left(\frac{p+2}{2}\right)-2p\left(p^2+6 p+3\right) (p+1)^{2 p+2} \ln \left(\frac{p+1}{2}\right)\Bigg]\nn\\
   &+\frac{(2 p)!(-1)^p}{2^{2 p-1} (p!)^2 }\sum_{k=1}^p\frac{ (-1)^{k} k^{2 p+2} \ln \left(\frac{k}{2}\right)}{(p-k+2)! (p+k+2)!}\Bigg\{\frac{k^4}{3}+k^2 \left(\frac{9}{4} p^3+\frac{113}{24} p^2+\frac{17}{8} p-\frac{2}{3}\right)\nn\\
   & \qquad\qquad\qquad\qquad\qquad\qquad\qquad\qquad\qquad\quad+ \frac{p^6}{2}+\frac{9}{4}p^5+\frac{7}{3} p^4-2 p^3-\frac{47
   }{12}p^2-p+\frac{1}{3}\Bigg\}\,. 
\end{align}
Performing the small eccentricity expansion of the relation \eqref{eq:Lambda_0_resummed} and recalling the definition~\eqref{eq:lambda_0_expansion} of the $\alpha_n$ coefficients entering the small eccentricity expansion of $\lambda_0(e)$, I finally find that for $n\ge 2$:
\begin{align}\label{eq:alpha_n_general}
    \alpha_n= \frac{1}{n+1}+ \frac{73}{24n}+\frac{37}{96(n-1)} - \frac{2}{3}\sum_{p=0}^{n} (-1)^{n-p} \begin{pmatrix} 7/2\\n-p\end{pmatrix} \,\mathcal{A}_{p+1} \,.
\end{align}
The $n\in\{0,1\}$ contributions can be treated as special cases; see Eqs.~\eqref{eq:alpha_n_coeffs}.

\section{Expressions for the fundamental frequencies}
\label{app:frequencies_lengthy}

\subsection{$n$ and $\omega$ in terms of $i_{r\phi}$ and $i_\phi$}
\label{subapp:n_omega_of_irphi_iphi}

The local, logarithmic, and hereditary components of the radial frequency in terms of the action variables read
\begin{subequations}
\label{eq:n_loc_log_hered_inTermsOf_irphi_iphi}
\begin{align}
\label{seq:n_loc_inTermsOf_irphi_iphi}
n^\text{loc} &= \frac{1}{G m}\Bigg\{\frac{1}{i_{r\phi}^3}+\frac{1}{c^2}\Bigg[-\frac{15}{2 i_{r\phi}^5}+\frac{9}{i_{\phi} i_{r\phi}^4}+\frac{\nu }{2 i_{r\phi}^5}\Bigg] \nonumber\\
& \qquad +\frac{1}{c^4}\Bigg[\frac{435}{8 i_{r\phi}^7}-\frac{525}{4 i_{\phi} i_{r\phi}^6}+\frac{54}{i_{\phi}^2
   i_{r\phi}^5}+\frac{105}{4 i_{\phi}^3 i_{r\phi}^4}+ \nu\left(-\frac{45}{8 i_{r\phi}^7}+\frac{15}{i_{\phi} i_{r\phi}^6}-\frac{15}{2 i_{\phi}^3 i_{r\phi}^4}\right)  +\frac{3 \nu ^2}{8 i_{r\phi}^7}\Bigg] \nonumber\\
& \qquad+\frac{1}{c^6}\Bigg[-\frac{6363}{16
   i_{r\phi}^9}+\frac{5775}{4 i_{\phi} i_{r\phi}^8}-\frac{1350}{i_{\phi}^2 i_{r\phi}^7}-\frac{1515}{8 i_{\phi}^3 i_{r\phi}^6}+\frac{315}{i_{\phi}^4 i_{r\phi}^5}+\frac{693}{4 i_{\phi}^5 i_{r\phi}^4} \nonumber\\
& \qquad\qquad + \nu\Bigg(\frac{805}{16 i_{r\phi}^9}-\frac{525}{2 i_{\phi} i_{r\phi}^8}+\frac{405}{2 i_{\phi}^2 i_{r\phi}^7}-\frac{90}{i_{\phi}^4 i_{r\phi}^5}+\frac{\frac{7135}{24}-\frac{205 }{128}\pi^2}{i_{\phi}^3 i_{r\phi}^6} +\frac{-\frac{375}{2}+\frac{369}{128} \pi ^2}{i_{\phi}^5 i_{r\phi}^4} \Bigg)  \nonumber\\
& \qquad\qquad +  \nu^2\Bigg(-\frac{45}{8 i_{r\phi}^9}+\frac{21}{i_{\phi} i_{r\phi}^8}-\frac{25}{i_{\phi}^3 i_{r\phi}^6}+\frac{63}{8 i_{\phi}^5
   i_{r\phi}^4}\Bigg) +\frac{5 \nu ^3}{16 i_{r\phi}^9}\Bigg] \nonumber\\
& \qquad +\frac{1}{c^8}\Bigg[\frac{376515}{128 i_{r\phi}^{11}}-\frac{456327}{32 i_{\phi} i_{r\phi}^{10}}+\frac{85365}{4 i_{\phi}^2 i_{r\phi}^9}-\frac{323925}{64 i_{\phi}^3
   i_{r\phi}^8}-\frac{64305}{8 i_{\phi}^4 i_{r\phi}^7}-\frac{38745}{32 i_{\phi}^5 i_{r\phi}^6}+\frac{20307}{8 i_{\phi}^6 i_{r\phi}^5}+\frac{96525}{64 i_{\phi}^7 i_{r\phi}^4} \nonumber\\
& \qquad\qquad +  \nu\Bigg(-\frac{53565}{128 i_{r\phi}^{11}}+\frac{51705}{16 i_{\phi} i_{r\phi}^{10}}-\frac{10395}{2 i_{\phi}^2 i_{r\phi}^9} +\frac{\frac{467473}{96}-\frac{404795}{4096} \pi ^2}{i_{\phi}^5 i_{r\phi}^6} +\frac{\frac{24105}{4}-\frac{1845}{64} \pi ^2}{i_{\phi}^4 i_{r\phi}^7}  \nonumber\\
& \qquad\qquad\qquad +\frac{-\frac{5025}{2}+\frac{1107}{32} \pi ^2}{i_{\phi}^6 i_{r\phi}^5} +\frac{-\frac{417473}{144}+\frac{868903}{24576} \pi ^2 }{i_{\phi}^3 i_{r\phi}^8} +\frac{-\frac{248057}{96}+\frac{425105}{8192} \pi ^2}{i_{\phi}^7 i_{r\phi}^4} \Bigg) \nonumber\\
& \qquad\qquad  +  \nu^2\Bigg(\frac{7725}{128 i_{r\phi}^{11}}-\frac{1755}{4 i_{\phi} i_{r\phi}^{10}}+\frac{945}{2 i_{\phi}^2
   i_{r\phi}^9}-\frac{1125}{2 i_{\phi}^4 i_{r\phi}^7}+\frac{132}{i_{\phi}^6 i_{r\phi}^5}+\frac{\frac{18925}{32}-\frac{3075}{256}\pi ^2}{i_{\phi}^7 i_{r\phi}^4}  \nonumber\\
& \qquad\qquad\qquad\quad   +\frac{\frac{10829}{12}-\frac{287}{64} \pi ^2}{i_{\phi}^3  i_{r\phi}^8} +\frac{-\frac{43215}{32}+\frac{5535 \pi ^2}{256}}{i_{\phi}^5 i_{r\phi}^6}  \Bigg) \nonumber\\
& \qquad\qquad  + \nu ^3\Bigg(-\frac{375}{64 i_{r\phi}^{11}}+\frac{27}{i_{\phi} i_{r\phi}^{10}}-\frac{105}{2 i_{\phi}^3 i_{r\phi}^8}+\frac{315}{8 i_{\phi}^5 i_{r\phi}^6}-\frac{135}{16 i_{\phi}^7 i_{r\phi}^4}\Bigg) +\frac{35 \nu
   ^4}{128 i_{r\phi}^{11}}\Bigg]\Bigg\} \,, \\
\label{seq:n_log_inTermsOf_irphi_iphi}
n^{\text{log}} &= \frac{\nu}{c^8 G m}  \Biggl\{\frac{119}{6 i_\phi^3 i_{r\phi}^8}+\frac{2692}{15 i_\phi^4
   i_{r\phi}^7}-\frac{239}{i_\phi^5 i_{r\phi}^6}-\frac{680}{3 i_\phi^6
   i_{r\phi}^5}+\frac{1291}{6 i_\phi^7 i_{r\phi}^4} \nonumber\\*
    &\qquad\qquad\quad +\frac{1}{i_\phi+i_{r\phi} }\Biggl[-\frac{148}{15 i_\phi^2
   i_{r\phi}^8}-\frac{74}{5 i_\phi^3 i_{r\phi}^7}+\frac{488}{5 i_\phi^4
   i_{r\phi}^6}+\frac{732}{5 i_\phi^5 i_{r\phi}^5}-\frac{340}{3 i_\phi^6
   i_{r\phi}^4}-\frac{170}{i_\phi^7 i_{r\phi}^3}\Biggr] \nonumber\\*
   & \qquad\qquad\quad +\Biggl(\frac{518}{15
   i_\phi^3 i_{r\phi}^8}-\frac{244}{i_\phi^5 i_{r\phi}^6}+\frac{170}{i_\phi^7
   i_{r\phi}^4}\Biggr) \Biggl[\ln \left(\frac{c\, i_{r\phi}^2 (i_\phi+i_{r\phi})}{8
   i_\phi^2}\right) -\gamma_\mathrm{E} \Biggr]\Biggr\}\,,\\
\label{seq:n_hered_inTermsOf_irphi_iphi}
n^{\text{hered}} &= \frac{\nu}{c^8 G m} \Biggl\{\frac{74}{5 i_\phi^3 i_{r\phi}^8}-\frac{732}{5 i_\phi^5
   i_{r\phi}^6}+\frac{170}{i_\phi^7 i_{r\phi}^4}+\left[-\frac{518}{5 i_\phi^3
   i_{r\phi}^8}+\frac{732}{i_\phi^5 i_{r\phi}^6}-\frac{510}{i_\phi^7
   i_{r\phi}^4}\right] \ln \left(\frac{i_{r\phi}}{i_\phi}\right) \nonumber\\*
   & \qquad\qquad\quad +\left[-\frac{96}{i_\phi^5 i_{r\phi}^6}+\frac{288}{5 i_\phi^7 i_{r\phi}^4}\right] \lambda_0
   \left(\sqrt{1-\frac{i_\phi^2}{i_{r\phi}^2}}\right)- \frac{96}{5 i_\phi^5
   i_{r\phi}^6}  
   \sqrt{1-\frac{i_\phi^2}{i_{r\phi}^2}} \lambda_0'\left(\sqrt{1-\frac{i_\phi^2}{i_{r\phi}^2}}\right) \Biggr\} \,.
\end{align}
\end{subequations}

The local, logarithmic, and hereditary components of the azimuthal frequency in terms of the action variables read 
\begin{subequations}
\label{eq:omega_loc_log_hered_inTermsOf_irphi_iphi}
\begin{align}
\label{seq:omega_loc_inTermsOf_irphi_iphi}
\omega^\text{loc} &= \frac{1}{G m}\Bigg\{\frac{1}{i_{r\phi}^3}+\frac{1}{c^2}\Bigg[-\frac{15}{2 i_{r\phi}^5}+\frac{9}{i_{\phi} i_{r\phi}^4}+\frac{3}{i_{\phi}^2 i_{r\phi}^3}+\frac{\nu }{2 i_{r\phi}^5}\Bigg] \nonumber\\
& \qquad +\frac{1}{c^4}\Bigg[\frac{435}{8 i_{r\phi}^7}-\frac{525}{4 i_{\phi}
   i_{r\phi}^6}+\frac{111}{4 i_{\phi}^2 i_{r\phi}^5}+\frac{213}{4 i_{\phi}^3 i_{r\phi}^4}+\frac{105}{4 i_{\phi}^4 i_{r\phi}^3} \nonumber\\
& \qquad\qquad\!\!\! + \nu \left(-\frac{45}{8 i_{r\phi}^7}+\frac{15}{i_{\phi} i_{r\phi}^6}+\frac{3}{i_{\phi}^2
   i_{r\phi}^5}-\frac{15}{2 i_{\phi}^3 i_{r\phi}^4}-\frac{15}{2 i_{\phi}^4 i_{r\phi}^3}\right)+\frac{3 \nu ^2}{8 i_{r\phi}^7}\Bigg] \nonumber\\
& \qquad +\frac{1}{c^6}\Bigg[-\frac{6363}{16 i_{r\phi}^9}+\frac{5775}{4 i_{\phi} i_{r\phi}^8}-\frac{4575}{4
   i_{\phi}^2 i_{r\phi}^7}-\frac{5115}{8 i_{\phi}^3 i_{r\phi}^6}+\frac{1611}{8 i_{\phi}^4 i_{r\phi}^5}+\frac{1953}{4 i_{\phi}^5 i_{r\phi}^4}+\frac{1155}{4 i_{\phi}^6 i_{r\phi}^3} \nonumber\\
& \qquad\qquad\!\!\! + \nu\Bigg(\frac{805}{16 i_{r\phi}^9}-\frac{525}{2 i_{\phi} i_{r\phi}^8}+\frac{165}{i_{\phi}^2 i_{r\phi}^7}+\frac{\frac{8755}{24}-\frac{205}{128} \pi ^2}{i_{\phi}^3 i_{r\phi}^6}   +\frac{\frac{707}{8}-\frac{123}{128} \pi ^2}{i_{\phi}^4 i _{r\phi}^5}   +\frac{-\frac{555}{2}+\frac{369 \pi ^2}{128}}{i_{\phi}^5 i_{r\phi}^4}  +\frac{-\frac{625}{2}+\frac{615}{128} \pi ^2}{i_{\phi}^6 i_{r\phi}^3}  \Bigg)  \nonumber\\
& \qquad\qquad\!\!\! 
   +  \nu^2 \Bigg(-\frac{45}{8 i_{r\phi}^9}+\frac{21}{i_{\phi} i_{r\phi}^8}+\frac{3}{i_{\phi}^2 i_{r\phi}^7}-\frac{25}{i_{\phi}^3 i_{r\phi}^6}-\frac{15}{i_{\phi}^4 i_{r\phi}^5}+\frac{63}{8 i_{\phi}^5 i_{r\phi}^4}+\frac{105}{8
   i_{\phi}^6 i_{r\phi}^3}\Bigg)+\frac{5 \nu ^3}{16 i_{r\phi}^9} \Bigg] \nonumber\\
& \qquad +\frac{1}{c^8}\Bigg[\frac{376515}{128 i_{r\phi}^{11}}-\frac{456327}{32 i_{\phi} i_{r\phi}^{10}}+\frac{632217}{32 i_{\phi}^2 i_{r\phi}^9}+\frac{17535}{64
   i_{\phi}^3 i_{r\phi}^8}-\frac{653265}{64 i_{\phi}^4 i_{r\phi}^7}-\frac{210225}{32 i_{\phi}^5 i_{r\phi}^6}+\frac{42483}{32 i_{\phi}^6 i_{r\phi}^5}+\frac{340209}{64 i_{\phi}^7 i_{r\phi}^4}+\frac{225225}{64 i_{\phi}^8
   i_{r\phi}^3}\nonumber\\
& \qquad\qquad\!\!\!  + \nu\Bigg(-\frac{53565}{128 i_{r\phi}^{11}}+\frac{51705}{16 i_{\phi} i_{r\phi}^{10}}-\frac{77415}{16 i_{\phi}^2 i_{r\phi}^9}+\frac{\frac{853153}{96}-\frac{483515}{4096} \pi ^2}{i_{\phi}^5 i_{r\phi}^6} +\frac{\frac{226273}{96}-\frac{263099}{4096} \pi ^2}{i_{\phi}^6 i_{r\phi}^5} 
\nonumber\\
& \qquad\qquad\qquad   +\frac{\frac{229621}{48}-\frac{112031 \pi ^2}{8192}}{i_{\phi}^4 i_{r\phi}^7}+\frac{-\frac{604583}{144}+\frac{868903}{24576} \pi^2}{i_{\phi}^3 i_{r\phi}^8} +\frac{-\frac{609857}{96}+\frac{850193}{8192} \pi ^2}{i_{\phi}^7 i_{r\phi}^4}  +\frac{-\frac{1736399}{288}+\frac{2975735}{24576} \pi ^2}{i_{\phi}^8 i_{r\phi}^3} \Bigg) \nonumber\\
& \qquad\qquad\!\!\!  + \nu ^2 \Bigg(\frac{7725}{128 i_{r\phi}^{11}}-\frac{1755}{4 i_{\phi} i_{r\phi}^{10}}+\frac{1695}{4 i_{\phi}^2 i_{r\phi}^9}+\frac{\frac{132475}{96}-\frac{7175}{256} \pi ^2}{i_{\phi}^8 i_{r\phi}^3}   +\frac{\frac{25261}{32}-\frac{3075}{256} \pi^2 }{i_{\phi}^7 i_{r\phi}^4} \nonumber\\
& \qquad\qquad\qquad   +\frac{\frac{24493}{24}-\frac{287}{64} \pi ^2}{i_{\phi}^3 i_{r\phi}^8}  +\frac{-\frac{703}{4}-\frac{123}{64} \pi ^2}{i_{\phi}^4 i_{r\phi}^7} +\frac{-\frac{55215}{32}+\frac{5535}{256} \pi^2}{i_{\phi}^5 i_{r\phi}^6} +\frac{-\frac{38991}{32}+\frac{5535}{256} \pi ^2}{i_{\phi}^6 i_{r\phi}^5} \Bigg) \nonumber\\
& \qquad\qquad\!\!\! + \nu^3  \Biggl( \! -\frac{375}{64 i_{r\phi}^{11}}\!  + \! \frac{27}{i_{\phi} i_{r\phi}^{10}}\! +\! \frac{3}{i_{\phi}^2 i_{r\phi}^9}\! -\! \frac{105}{2 i_{\phi}^3
   i_{r\phi}^8}\! -\! \frac{45}{2 i_{\phi}^4 i_{r\phi}^7}\! +\! \frac{315}{8 i_{\phi}^5 i_{r\phi}^6}\! +\! \frac{315}{8 i_{\phi}^6 i_{r\phi}^5}\! -\! \frac{135}{16 i_{\phi}^7 i_{r\phi}^4}\! -\! \frac{315}{16 i_{\phi}^8 i_{r\phi}^3}\Biggr)   +\frac{35 \nu ^4 }{128 i_{r\phi}^{11}}  \Bigg]\Bigg\} \,, \\
\label{seq:omega_log_inTermsOf_irphi_iphi}
\omega^{\text{log}} &= \frac{\nu}{c^8 G m}  \Biggl\{\frac{119}{6 i_\phi^3 i_{r\phi}^8}+\frac{5639}{30 i_\phi^4
   i_{r\phi}^7}-\frac{5371}{45 i_\phi^5 i_{r\phi}^6}-\frac{1397}{3 i_\phi^6
   i_{r\phi}^5}-\frac{749}{6 i_\phi^7 i_{r\phi}^4}+\frac{9037}{18 i_\phi^8
   i_{r\phi}^3} \nonumber\\*
   & \qquad\qquad\quad  +\frac{1 }{i_\phi+i_{r\phi}}\Biggl[-\frac{148}{15 i_\phi^2 i_{r\phi}^8}-\frac{148}{15 i_\phi^3
   i_{r\phi}^7}+\frac{1612}{15 i_\phi^4 i_{r\phi}^6}+\frac{488}{5 i_\phi^5
   i_{r\phi}^5}-\frac{3164}{15 i_\phi^6 i_{r\phi}^4}-\frac{340}{3 i_\phi^7
   i_{r\phi}^3}+\frac{340}{3 i_\phi^8 i_{r\phi}^2} \Biggr] \nonumber\\*
   & \qquad\qquad\quad  +\left(\frac{518}{15
   i_\phi^3 i_{r\phi}^8}+\frac{74}{5 i_\phi^4 i_{r\phi}^7}-\frac{244}{i_\phi^5
   i_{r\phi}^6}-\frac{244}{i_\phi^6 i_{r\phi}^5}+\frac{170}{i_\phi^7
   i_{r\phi}^4}+\frac{1190}{3 i_\phi^8 i_{r\phi}^3}\right) \Biggl[\ln \left(\frac{c\,
   i_{r\phi}^2 (i_\phi+i_{r\phi})}{8 i_\phi^2}\right) -\gamma_\mathrm{E}\Biggr]\Biggr\} \,,\\
\label{seq:omega_hered_inTermsOf_irphi_iphi}
\omega^{\text{hered}} &= \frac{\nu}{c^8 G m} \Biggl\{\frac{74}{5 i_\phi^3 i_{r\phi}^8}-\frac{74}{5 i_\phi^4
   i_{r\phi}^7}-\frac{732}{5 i_\phi^5 i_{r\phi}^6}+\frac{732}{5 i_\phi^6
   i_{r\phi}^5}+\frac{170}{i_\phi^7 i_{r\phi}^4}-\frac{170}{i_\phi^8
   i_{r\phi}^3}  \nonumber\\*
   & \qquad\qquad\quad  +\left[-\frac{518}{5 i_\phi^3 i_{r\phi}^8}-\frac{222}{5 i_\phi^4
   i_{r\phi}^7}+\frac{732}{i_\phi^5 i_{r\phi}^6}+\frac{732}{i_\phi^6
   i_{r\phi}^5}-\frac{510}{i_\phi^7 i_{r\phi}^4}-\frac{1190}{i_\phi^8
   i_{r\phi}^3}\right] \ln \left(\frac{i_{r\phi}}{i_\phi}\right) \nonumber\\*
   & \qquad\qquad\quad  +\left[-\frac{96}{i_\phi^5 i_{r\phi}^6}-\frac{96}{i_\phi^6 i_{r\phi}^5}+\frac{288}{5 i_\phi^7
   i_{r\phi}^4}+\frac{672}{5 i_\phi^8 i_{r\phi}^3}\right] \lambda_0
   \left(\sqrt{1-\frac{i_\phi^2}{i_{r\phi}^2}}\right)  \nonumber\\*
   & \qquad\qquad\quad  + \frac{96}{5}\left[-\frac{1}{i_\phi^5
   i_{r\phi}^6}+\frac{1}{i_\phi^6 i_{r\phi}^5}\right]
   \sqrt{1-\frac{i_\phi^2}{i_{r\phi}^2}} \lambda_0
   '\left(\sqrt{1-\frac{i_\phi^2}{i_{r\phi}^2}}\right)\Biggr\} \,.
\end{align}
\end{subequations}

\subsection{$n$ and $\omega$ in terms of $\varepsilon$ and $j$}
\label{subapp:n_omega_of_epsilon_j}

The local, logarithmic, and hereditary components of the radial frequency in terms of the conserved energy and angular momentum read 
\begin{subequations}
\label{eq:n_loc_log_hered_inTermsOf_varepsilon_j}
\begin{align}
\label{seq:n_loc_inTermsOf_varepsilon_j}
n^{\text{loc}} &= \frac{c^3 \varepsilon^{3/2}}{G m} \Biggl\{1+\varepsilon  \Biggl[-\frac{15}{8}+\frac{\nu }{8}\Biggr]+\varepsilon^2 \Biggl[\frac{555}{128}+\frac{15}{64} \nu +\frac{11}{128} \nu^2+\frac{1}{\sqrt{j}}\biggl(-\frac{15}{2}+3 \nu \biggr)\Biggr] \nonumber\\
& +\varepsilon^3 \Biggl[-\frac{9795}{1024}-\frac{1665}{1024} \nu -\frac{105}{1024} \nu^2 + \frac{45}{1024} \nu^3 +\frac{1}{\sqrt{j}}\biggl(\frac{255}{8}-\frac{135}{8} \nu +\frac{15}{4} \nu^2\biggr) \nonumber\\*
& \qquad +\frac{1}{j^{3/2}}\biggl(-\frac{105}{2}+\nu\Bigl(\frac{218}{3}-\frac{41}{64} \pi^2\Bigr)  -\frac{15}{2} \nu^2 \biggr)\Biggr]  \nonumber\\
& +\varepsilon ^4 \Biggl[\frac{698643}{32768}+\frac{48975}{8192} \nu +\frac{6105}{16384}\nu^2 -\frac{345}{8192} \nu^3+\frac{723}{32768} \nu^4  +\frac{1}{\sqrt{j}}\biggl(-\frac{3375}{32}+\frac{825}{16} \nu -\frac{525}{32}\nu^2+\frac{57}{16} \nu^3\biggr) \nonumber\\*
& \qquad
+\frac{1}{j} \biggl(\frac{225}{4}-45 \nu +9 \nu^2 \biggr)  +\frac{1}{j^{3/2}}\biggl(\frac{4725}{16}+ \nu\Bigl(-\frac{5111}{9}+\frac{50329}{6144}\pi^2\Bigr)  + \nu ^2\Bigl(\frac{1289}{6}-\frac{451}{256}\pi^2\Bigr) -\frac{135}{8} \nu^3 \biggr)   \nonumber\\*
& \qquad +\frac{1}{j^{5/2}}\biggl(-\frac{9009}{16}+ \nu\Bigl(\frac{293413}{240}-\frac{51439}{2048} \pi^2\Bigr) + \nu ^2\Bigl(-\frac{7013}{16}+\frac{123}{16} \pi ^2\Bigr) +\frac{105}{8}\nu^3\biggr)\Biggr]\Biggr\} \,,  \\
\label{seq:n_log_inTermsOf_varepsilon_j}
    n^{\text{log}} &=  \frac{c^3 \varepsilon^{11/2} \nu}{G m \, j 
   \left(1+\sqrt{j}\right) }  \Biggl\{\frac{22}{15} +\frac{1294}{15 \sqrt{j}} +\frac{1376}{15 j} -\frac{88}{15 j^{3/2}}-\frac{170}{j^2}  -\frac{170}{j^{5/2}}\nonumber\\
   &\qquad\qquad\quad\qquad+\left(-\frac{148}{15} -\frac{148}{15 \sqrt{j}} +\frac{244}{5 j} +\frac{244}{5
   j^{3/2}} \right) \left[\ln \varepsilon  -2
   \ln \left(\frac{1+\sqrt{j}}{8 j}\right) + 2 \gamma_\mathrm{E}  \right]\Biggr\} \,, \\
   \label{seq:n_hered_inTermsOf_varepsilon_j}
   n^{\text{hered}} &= \frac{c^3 \varepsilon^{11/2} \nu}{G m j^{3/2}}  \Biggl\{\frac{74}{5} -\frac{732}{5 j} + \frac{170}{j^{2}}+ \ln j \left(\frac{148}{5} -\frac{732}{5 j}\right)
    -\frac{192}{5 j} \lambda_0 \left(\sqrt{1-j}\right) -\frac{96}{5j} \sqrt{1-j} \, \lambda_0'\!\left(\sqrt{1-j}\right)  \Biggr\} \,.
\end{align}
\end{subequations}
 
The local, logarithmic, and hereditary components of the azimuthal frequency in terms of the conserved energy and angular momentum read 
\begin{subequations}
\label{eq:omega_loc_log_hered_inTermsOf_varepsilon_j}
\begin{align}
\label{seq:omega_loc_inTermsOf_varepsilon_j}
    \omega^{\text{loc}} &= \frac{c^3 \varepsilon ^{3/2} }{G m} \Biggl\{1+\varepsilon \Biggl[-\frac{15}{8}+\frac{3}{j}+\frac{\nu }{8}\Biggr] \nonumber\\*
& +\varepsilon ^2 \Biggl[\frac{555}{128}+\frac{15}{64} \nu +\frac{11}{128} \nu^2 +\frac{1}{\sqrt{j}}\biggl(-\frac{15}{2}+3 \nu\biggr) +\frac{1}{j}\biggl(-\frac{75}{8}+\frac{15}{8} \nu \biggr) +\frac{1}{j^2}\biggl(\frac{105}{4}-\frac{15}{2} \nu \biggr) \Biggr]  \nonumber\\
& +\varepsilon ^3 \Biggl[-\frac{9795}{1024}-\frac{1665}{1024} \nu -\frac{105}{1024} \nu^2 + \frac{45}{1024} \nu^3  +\frac{1}{\sqrt{j}}\biggl(\frac{255}{8}-\frac{135}{8} \nu +\frac{15}{4} \nu ^2\biggr)  +\frac{1}{j}\biggl(\frac{2685}{128}-\frac{225}{64} \nu +\frac{153}{128} \nu ^2\biggr) \nonumber\\*
& \qquad +\frac{1}{j^2}\biggl(-\frac{4095}{32}+\nu\left(\frac{4043}{32}-\frac{123}{128} \pi^2\right)  -\frac{195}{16} \nu^2\biggr)  +\frac{1}{j^{3/2}}\biggl(-75+  \nu\left(\frac{245}{3}-\frac{41}{64} \pi^2\right) -\frac{15}{2}\nu^2\biggr) \nonumber\\*
& \qquad +\frac{1}{j^3}\biggl(\frac{1155}{4}+\nu\left(-\frac{625}{2}+\frac{615}{128}\pi^2\right)  +\frac{105}{8} \nu^2\biggr)\Biggr] \nonumber\\
& +\varepsilon ^4 \Biggl[\frac{698643}{32768}+\frac{48975}{8192} \nu +\frac{6105}{16384} \nu^2 - \frac{345}{8192} \nu^3 +\frac{723}{32768} \nu^4 \nonumber\\*
& \qquad +\frac{1}{\sqrt{j}}\biggl(-\frac{3375}{32}+\frac{825}{16} \nu -\frac{525}{32} \nu^2 + \frac{57}{16} \nu^3\biggr)  +\frac{1}{j} \biggl(\frac{9765}{1024}-\frac{43395}{1024} \nu +\frac{6891}{1024} \nu ^2+\frac{747}{1024} \nu^3\biggr) \nonumber\\*
& \qquad +\frac{1}{j^{3/2}}\biggl(\frac{6705}{16}+\Bigl(-\frac{46153}{72}+\frac{50329}{6144}\pi^2\Bigr) \nu + \nu^2\Bigl(\frac{2767}{12}-\frac{451}{256} \pi^2\Bigr) -\frac{135}{8} \nu^3\biggr) \nonumber\\*
& \qquad +\frac{1}{j^2} \biggl(\frac{171675}{512}+\nu\Bigl(-\frac{173677}{384}+\frac{50329}{8192}\pi^2\Bigr)  + \nu ^2 \Bigl(\frac{82751}{512}-\frac{1353}{1024}\pi^2\Bigr) -\frac{3405}{256} \nu^3\biggr)\nonumber\\*
& \qquad +\frac{1}{j^{5/2}}\biggl(-\frac{14679}{16}+\nu\Bigl(\frac{378133}{240}-\frac{55375}{2048}\pi ^2\Bigr)  + \nu^2\Bigl(-\frac{7733}{16}+\frac{123}{16}\pi^2\Bigr) +\frac{105}{8}\nu ^3\biggr)\nonumber\\*
& \qquad +\frac{1}{j^3}\biggl(-\frac{31185}{16}+ \nu\Bigl(\frac{44141}{12}-\frac{294095}{4096}\pi^2\Bigr) 
   +\nu^2\Bigl(-\frac{74205}{64}+\frac{20295}{1024}\pi^2\Bigr) +\frac{2205}{64}\nu^3\biggr) \nonumber\\*
& \qquad +\frac{1}{j^4}\biggl(\frac{225225}{64}+\nu \Bigl(-\frac{1736399}{288}+\frac{2975735}{24576}\pi ^2\Bigr) +\nu ^2\Bigl(\frac{132475}{96}-\frac{7175}{256}\pi^2\Bigr) -\frac{315}{16}\nu^3\biggr) \Biggr]\Biggr\} \,, \\
\label{seq:omega_log_inTermsOf_varepsilon_j}
    \omega^{\text{log}} &= \frac{c^3 \varepsilon ^{11/2} \nu}{G j \left(1+\sqrt{j}\right) m}  \Biggl\{\frac{22}{15} +\frac{997}{10
   \sqrt{j}} +\frac{20677}{90 j} -\frac{7831}{45 j^{3/2}} -\frac{4233}{5 j^2} +\frac{877}{18 j^{5/2}} +\frac{11077}{18 j^3}\nonumber\\
   & \quad\qquad+\left(-\frac{148}{15} -\frac{259}{15 \sqrt{j}} +\frac{207}{5 j} +\frac{854}{5 j^{3/2}} +\frac{122}{j^2} -\frac{595}{3 j^{5/2}}  -\frac{595}{3 j^3}\right) \left[\ln \varepsilon  -2 \ln \left(\frac{1+\sqrt{j}}{8
   j}\right) + 2 \gamma_\mathrm{E} \right]\Biggr\} \,, \\
   \label{seq:omega_hered_inTermsOf_varepsilon_j}
    \omega^{\text{hered}} &= \frac{c^3 \varepsilon^{11/2} \nu}{G m j^{3/2}} 
    \Biggl\{\frac{74}{5} -\frac{74}{5 \sqrt{j}} -\frac{732}{5 j} +\frac{732}{5 j^{3/2}} +\frac{170}{j^{2}}  -\frac{170}{j^{5/2}}   + \ln j\,\left(\frac{148}{5} +\frac{111}{5 \sqrt{j}} -\frac{732}{5 j} -\frac{366}{j^{3/2}} + \frac{595}{j^{5/2}} \right) \nonumber\\
   & \quad\qquad +\lambda_0\! \left(\sqrt{1-j}\right) \left(-\frac{192}{5 j} -\frac{96}{j^{3/2}}+ \frac{672}{5 j^{5/2}}\right) + \frac{96}{5} \sqrt{1-j}  \ \lambda_0'\!\left(\sqrt{1-j}\right) \left( -\frac{1}{j}+\frac{1}{j^{3/2}}\right) \Biggr\}\,.
\end{align}
\end{subequations}

\subsection{$P$ and $K$ in terms of $\varepsilon$ and $j$}
\label{subapp:P_K_of_epsilon_j}

The local, logarithmic, and hereditary components of the radial period in terms of the conserved energy and angular momentum read %
\begin{subequations}
\label{eq:P_loc_log_hered_inTermsOf_varepsilon_j}
\begin{align}
\label{seq:P_loc_inTermsOf_varepsilon_j}
   P^\text{loc} &= \frac{2\pi G m}{c^3 \varepsilon
   ^{3/2}} \Biggl\{1+\varepsilon  \Biggl[\frac{15}{8}-\frac{\nu
   }{8}\Biggr]+\varepsilon ^2 \Biggl[-\frac{105}{128}-\frac{45}{64} \nu -\frac{9}{128} \nu^2 +\frac{1}{\sqrt{j}}\biggl(\frac{15}{2}-3 \nu
   \biggr) \Biggr] \nonumber\\
   & +\varepsilon ^3
   \Biggl[-\frac{105}{1024}+\frac{525}{1024} \nu -\frac{75}{1024} \nu^2 - \frac{25}{1024} \nu^3+\frac{1}{\sqrt{j}}\biggl(-\frac{15}{4}+\frac{15}{4} \nu -3 \nu^2\biggr)  \nonumber\\
   &\qquad   +\frac{1}{j^{3/2}}\biggl(\frac{105}{2}+ \nu \Bigl(-\frac{218}{3}+\frac{41}{64} \pi^2\Bigr)  +\frac{15}{2} \nu^2\biggr) \Biggr] \nonumber\\
   &  +\varepsilon ^4
   \Biggl[-\frac{693}{32768}+\frac{735}{8192} \nu -\frac{735}{16384} \nu^2-\frac{105}{8192} \nu^3-\frac{245}{32768} \nu^4 +\frac{1}{\sqrt{j}} \biggl(\frac{45}{16} \nu^2 - \frac{9}{4} \nu^3\biggr) \nonumber\\
   &\qquad   +\frac{1}{j^{3/2}}\biggl(-\frac{1575}{16}+\nu\Bigl(\frac{20323}{72}-\frac{35569
   }{6144}\pi^2\Bigr)  + \nu^2 \Bigl(-\frac{4045}{24}+\frac{205}{128}\pi^2\Bigr) +15 \nu ^3\biggr)
   \nonumber\\
   &\qquad   +\frac{1}{j^{5/2}}\biggl(\frac{9009}{16}+ \nu \Bigl(-\frac{293413}{240}+\frac{51439}{2048}\pi^2\Bigr)  + \nu^2 \Bigl(\frac{7013}{16}-\frac{123}{16}\pi^2\Bigr) -\frac{105}{8} \nu^3\biggr) \Biggr] \Biggr\} \,, \\
   \label{seq:P_log_inTermsOf_varepsilon_j}
    P^\text{log} &= \frac{2\pi \, G m \,  \varepsilon^4  \nu}{c^3 j
   \left(1+\sqrt{j}\right)}  \Biggl\{-\frac{44}{15} -\frac{2588}{15 \sqrt{j}} -\frac{2752}{15 j} +\frac{176}{15 j^{3/2}} +\frac{340}{j^2} + \frac{340}{j^{5/2}}\nonumber\\
   & \qquad+\left(\frac{296}{15} +\frac{296}{15 \sqrt{j}} -\frac{488}{5 j} -\frac{488}{5 j^{3/2}}\right) \left[\ln \varepsilon -2
   \ln \left(\frac{1+\sqrt{j}}{8 j}\right) + 2 \gamma_\mathrm{E}\right]\Biggr\} \,, \\
   \label{seq:P_hered_inTermsOf_varepsilon_j}
    P^\text{hered} &= \frac{2\pi \, G m \,  \varepsilon^4  \nu}{c^3 j \left(1+\sqrt{j}\right)}  \Biggl\{-\frac{266}{15} -\frac{562}{3 \sqrt{j}} -\frac{556}{15 j} +\frac{2372}{15 j^{3/2}}+\frac{170}{j^2} +\frac{170}{j^{5/2}}\nonumber\\
   & \qquad +\left(\frac{296}{15}+\frac{296}{15 \sqrt{j}} -\frac{488}{5 j}  -\frac{488}{5 j^{3/2}} \right) \left[\ln \varepsilon  -2 \ln \left(\frac{1+\sqrt{j}}{8}\right)+\frac{1}{2} \ln j+ 2 \gamma_\text{E} \right] \nonumber\\
   & \qquad +\frac{192}{5}  \lambda_0\!\left(\sqrt{1-j}\right) \left(\frac{1}{j} + \frac{1}{j^{3/2}}\right)
   +\frac{96}{5} \sqrt{1-j}\,  \lambda_0'\!\left(\sqrt{1-j}\right) \left(\frac{1}{j}+\frac{1}{j^{3/2}}\right) \Biggr\}\,.
\end{align}
\end{subequations}

The local, logarithmic, and hereditary components of the periastron advance in terms of the conserved energy and angular momentum read %
\begin{subequations}
\label{eq:K_loc_log_hered_inTermsOf_varepsilon_j}
\begin{align}
\label{seq:K_loc_inTermsOf_varepsilon_j}
   K^\text{loc} &= 1+\frac{3 \varepsilon }{j}+ \frac{\varepsilon^2}{j} \Biggl[  -\frac{15}{4}+\frac{3}{2} \nu +\frac{1}{j}\biggl( \frac{105}{4}-\frac{15}{2} \nu  \biggr)\Biggr]\nonumber\\
   &  + \frac{\varepsilon^3}{j} \Biggl[\frac{15}{16}-\frac{15}{16} \nu +\frac{3}{4}\nu^2   + \frac{1}{j} \biggl(\! - \frac{315}{4}+ \nu\Bigl(109-\frac{123}{128} \pi^2\Bigr) 
  \!  - \!\frac{45}{4} \nu^2\biggr)  + \frac{1}{j^2}\biggl( \frac{1155}{4}+ \nu\Bigl(\! - \frac{625}{2}+\frac{615}{128}\pi^2 \Bigr)  +\frac{105}{8}\nu^2 \biggr)     \Biggr]\nonumber\\
   &  + \frac{\varepsilon^4}{j}
   \Biggl[ -\frac{15}{32} \nu^2+\frac{3}{8}\nu^3+ \frac{1}{j} \biggl(\frac{4725}{64}+  \nu \Bigl(-\frac{20323}{96}+\frac{35569}{8192}\pi^2 \Bigr) + \nu^2 \Bigl(\frac{4045}{32}-\frac{615}{512} \pi^2 \Bigr) -\frac{45}{4}\nu^3\biggr)   \nonumber\\
   &\qquad   + \frac{1}{j^2} \biggl(-\frac{45045}{32}+  \nu \Bigl(\frac{293413}{96}-\frac{257195}{4096} \pi^2 \Bigr) + \nu^2 \Bigl(-\frac{35065}{32}+\frac{615}{32}\pi^2 \Bigr) +\frac{525}{16} \nu^3  \biggr) \nonumber\\
   &\qquad  + \frac{1}{j^3}\biggl(\frac{225225}{64}+ \nu \Bigl(-\frac{1736399}{288}+\frac{2975735}{24576} \pi^2 \Bigr)  + \nu^2\Bigl(\frac{132475}{96}-\frac{7175}{256}\pi^2 \Bigr) -\frac{315}{16} \nu^3 \biggr)    \Biggr] \,, \\
    \label{seq:K_log_inTermsOf_varepsilon_j}
    K^\text{log} &=\frac{\varepsilon ^4 \nu}{\left(1+\sqrt{j}\right) j^{3/2}}  \Biggl\{\frac{403}{30}+\frac{12421}{90
   \sqrt{j}} -\frac{7567}{45 j} -\frac{3383}{5 j^{3/2}}  +\frac{3937}{18
   j^2} +\frac{11077}{18 j^{5/2}} \nonumber\\
   & \qquad+\left[-\frac{37}{5}-\frac{37}{5 \sqrt{j}} +\frac{122}{j}  -\frac{595}{3
   j^2}+\frac{122}{j^{3/2}}-\frac{595}{3 j^{5/2}} \right] \left(\ln \varepsilon  -2 \ln
   \left(\frac{1+\sqrt{j}}{8 j}\right) + 2 \gamma_\mathrm{E} \right)\Biggr\} \,, \\
\label{seq:K_hered_inTermsOf_varepsilon_j}
    K^\text{hered} &= \frac{\varepsilon ^4 \nu}{j^2}  \Biggl\{-\frac{74}{5} +\frac{732}{5 j} -\frac{170}{j^2}+\ln j \left(\frac{111}{5} -\frac{366}{j} +\frac{595}{j^2}\right)  \nonumber\\*
   & \qquad\qquad +\lambda_0 \!\left(\sqrt{1-j}\right) \left(-\frac{96}{ j}+\frac{672}{5 j^2}\right) +\frac{96}{5 j} \sqrt{1-j} \,\lambda_0'\!\left(\sqrt{1-j}\right) \Biggr\} \,.
\end{align}
\end{subequations}
 
\subsection{$x$ and $\iota$ in terms of $\varepsilon$ and $j$}
\label{subapp:x_iota_of_epsilon_j}

The local, logarithmic, and hereditary components of the Blanchet parameter $x$ in terms of the conserved energy and angular momentum read 
\begin{subequations}
\label{eq:x_loc_log_hered_inTermsOf_varepsilon_j}
\begin{align}
\label{seq:x_loc_inTermsOf_varepsilon_j}
x^{\text{loc}} &= \varepsilon\Biggl\{1 +\varepsilon \Biggl[-\frac{5}{4}+\frac{2}{j}+\frac{\nu }{12}\Biggr] +\varepsilon^2 \Biggl[\frac{5}{2} +\frac{5 \nu }{24}+\frac{\nu ^2}{18} +\frac{-5+2 \nu }{\sqrt{j}} +\frac{1}{j}\biggl(-5+\frac{7}{6} \nu \biggr) +\frac{1}{j^2}\biggl(\frac{33}{2}-5 \nu \biggr) \Biggr] \nonumber\\
&  +\varepsilon^3 \Biggl[-\frac{235}{48}-\frac{25}{24}\nu-\frac{25}{576} \nu^2+\frac{35}{1296} \nu^3 +\frac{1}{\sqrt{j}}\biggl(\frac{145}{8}-\frac{235}{24} \nu +\frac{29}{12} \nu^2\biggr)   +\frac{1}{j}\biggl(\frac{35}{4}-\frac{5}{3} \nu +\frac{25}{36} \nu^2\biggr) \nonumber\\*
& \qquad +\frac{1}{j^{3/2}}\biggl(-45+\nu\Bigl(\frac{472}{9}-\frac{41}{96} \pi^2\Bigr)  -5 \nu ^2\biggr) +\frac{1}{j^2}\biggl(-\frac{565}{8}+\nu\Bigl(\frac{1903}{24}-\frac{41}{64}\pi^2\Bigr)  -\frac{95}{12}\nu^2\biggr) \nonumber\\*
& \qquad +\frac{1}{j^3}\biggl(\frac{529}{3}+\nu \Bigl(-\frac{610}{3}+\frac{205}{64}\pi^2\Bigr) +\frac{35}{4} \nu^2\biggr)\Biggr]\nonumber\\
&  +\varepsilon^4 \Biggl[\frac{241}{24}+\frac{235}{72} \nu +\frac{55}{288}\nu^2-\frac{85}{5184}\nu^3+\frac{25}{1944} \nu^4 \nonumber\\*
& \qquad +\frac{1}{\sqrt{j}} \biggl(-\frac{6875}{128}+\frac{625}{24} \nu -\frac{10535}{1152} \nu^2+\frac{1279}{576} \nu^3\biggr)  +\frac{1}{j}\biggl(\frac{1385}{96}-\frac{1165}{48} \nu +\frac{575}{144} \nu^2+\frac{65}{162} \nu^3\biggr) \nonumber\\*
& \qquad +\frac{1}{j^{3/2}}\biggl(\frac{895}{4}+\nu\Bigl(-\frac{81311}{216}+\frac{47869}{9216} \pi^2\Bigr) 
   +\nu^2\Bigl(\frac{31301}{216}-\frac{2665}{2304} \pi^2\Bigr) -\frac{265}{24}\nu^3\biggr) \nonumber\\*
& \qquad +\frac{1}{j^2}\biggl(\frac{4825}{32}+\Bigl(-\frac{17429}{72}+\frac{45409}{12288} \pi^2\Bigr) \nu +\Bigl(\frac{14219}{144}-\frac{41}{48}\pi^2\Bigr) \nu^2-\frac{605}{72}\nu^3\biggr) \nonumber\\*
& \qquad +\frac{1}{j^{5/2}}\biggl(-\frac{4223}{8}+\nu\Bigl(\frac{38797}{40}-\frac{18021}{1024}\pi^2\Bigr) 
   +\nu^2\Bigl(-\frac{7493}{24}+\frac{41}{8}\pi^2\Bigr) +\frac{35}{4} \nu ^3\biggr) \nonumber\\*
& \qquad +\frac{1}{j^3}\biggl(-\frac{52295}{48}+\nu\Bigl(\frac{319793}{144}-\frac{277859}{6144} \pi ^2\Bigr)  + \nu^2\Bigl(-\frac{53885}{72}+\frac{10045}{768} \pi^2\Bigr) +\frac{1085}{48} \nu^3\biggr)  \nonumber\\*
& \qquad +\frac{1}{j^4}\biggl(\frac{202531}{96}+ \nu\Bigl(-\frac{1631819}{432}+\frac{2857655}{36864} \pi^2\Bigr)  +\nu^2\Bigl(\frac{130315}{144}-\frac{7175}{384} \pi^2\Bigr) -\frac{105}{8}\nu^3\biggr) \Biggr]  \Biggr\} \,,\\
\label{seq:x_log_inTermsOf_varepsilon_j}
    x^\text{log} &= \frac{\varepsilon ^5 \nu}{\left(1+\sqrt{j}\right) j}  \Biggl\{\frac{44}{45} +\frac{997}{15
   \sqrt{j}}+\frac{20677}{135 j}  -\frac{15662}{135 j^{3/2}} -\frac{2822}{5 j^2} +\frac{877}{27 j^{5/2}}  +\frac{11077}{27 j^3} \nonumber\\
   &\qquad\quad +\left[-\frac{296}{45} -\frac{518}{45 \sqrt{j}} +\frac{138}{5 j} +\frac{1708}{15 j^{3/2}} +\frac{244}{3 j^2} -\frac{1190}{9 j^{5/2}} -\frac{1190}{9 j^3} \right] \left(\ln \varepsilon  -2 \ln \left(\frac{1+\sqrt{j}}{8 j}\right) + 2 \gamma_\mathrm{E} \right) \Biggr\} \,, \\
   \label{seq:x_hered_inTermsOf_varepsilon_j}
    x^\text{hered} &=  \frac{\varepsilon ^5 \nu}{j^{3/2}} 
    \Biggl\{\frac{148}{15} -\frac{148}{15 \sqrt{j}}-\frac{488}{5 j}   +\frac{488}{5 j^{3/2}}+\frac{340}{3j^2} -\frac{340}{3 j^{5/2}}    + \ln j  \, \left(\frac{296}{15} +\frac{74}{5 \sqrt{j}}  -\frac{488}{5 j} -\frac{244}{j^{3/2}} +\frac{1190}{3 j^{5/2}}\right) \nonumber\\
   &\qquad\qquad + \lambda_0\! \left(\sqrt{1-j}\right) \left(-\frac{128}{5 j} -\frac{64}{j^{3/2}} +\frac{448}{5 j^{5/2}} \right) + \frac{64}{5}  \sqrt{1-j} \, \lambda_0'\!\left(\sqrt{1-j}\right) \left(-\frac{1}{j}+\frac{1}{j^{3/2}}\right)\Biggr\} \,.
\end{align}
\end{subequations}

The local, logarithmic, and hereditary components of the Blanchet parameter $\iota$ in terms of the conserved energy and angular momentum read${}^\text{\ref{footnote:error_iota}}$
\begin{subequations}
\label{eq:iota_loc_log_hered_inTermsOf_varepsilon_j}
\begin{align}
\label{seq:iota_loc_inTermsOf_varepsilon_j}
\iota^{\text{loc}} &= j \Biggl\{1+\varepsilon  \Biggl[-\frac{5}{12} \nu +\frac{1}{j}\biggl(-\frac{27}{4}+\frac{5}{2} \nu \biggr)\Biggr] \nonumber\\
&  +\varepsilon ^2 \Biggl[\frac{35}{16}+\frac{\nu ^2}{72}+\frac{-5+2 \nu}{\sqrt{j}}+\frac{1}{j}\biggl(\frac{205}{16}+\nu\Bigl(-\frac{1201}{48}+\frac{41}{128} \pi^2\Bigr)  +\frac{35}{24}\nu^2\biggr)+\frac{1}{j^2}\biggl(-\frac{331}{16}+\nu\Bigl(\frac{725}{12}-\frac{205}{128} \pi^2\Bigr)  +\frac{15}{8} \nu^2\biggr)\Biggr] \nonumber\\
&  +\varepsilon ^3 \Biggl[-\frac{415}{192}-\frac{385}{192} \nu -\frac{\nu ^3}{1296}\nonumber +\frac{1}{\sqrt{j}}\biggl(\frac{95}{8}-\frac{115}{24} \nu +\frac{17}{12} \nu^2\biggr) \nonumber\\*
& \qquad +\frac{1}{j}\biggl(-\frac{135}{8}+\nu\Bigl(\frac{6583}{288}-\frac{25729}{24576}\pi^2\Bigr)  +\nu^2\Bigl(-\frac{2771}{288}+\frac{41}{384}\pi^2\Bigr) +\frac{125}{144} \nu^3\biggr)   +\frac{1}{j^{3/2}}\biggl(-\frac{5}{4}+\nu\Bigl(\frac{202}{9}-\frac{41}{96} \pi^2\Bigr)  \biggr) \nonumber\\*
& \qquad +\frac{1}{j^2}\biggl(\frac{1335}{16}+\nu\Bigl(-\frac{144977}{576}+\frac{163715}{12288} \pi ^2\Bigr) 
   + \nu^2\Bigl(\frac{13625}{144}-\frac{5125}{1536} \pi^2\Bigr) +\frac{95}{32} \nu^3\biggr)  \nonumber\\*
& \qquad +\frac{1}{j^3} \biggl(-\frac{15937}{96}+ \nu\Bigl(\frac{60971}{216}-\frac{909335}{73728} \pi ^2\Bigr) + \nu^2\Bigl(-\frac{7675}{288}+\frac{1025}{768}\pi^2\Bigr) +\frac{5}{16}\nu^3\biggr) \Biggr]\Biggr\} \,,\\
\label{seq:iota_log_inTermsOf_varepsilon_j}
    \iota^\text{log} &= \frac{\varepsilon ^3 \nu \sqrt{j} }{1+\sqrt{j}}  \Biggl\{-\frac{403}{90}  -\frac{12421}{270
   \sqrt{j}}+\frac{7567}{135 j}+\frac{3383}{15 j^{3/2}}-\frac{3937}{54 j^2}-\frac{11077}{54
   j^{5/2}}\nonumber\\
   &\qquad + \left(\frac{37}{15}+\frac{37}{15 \sqrt{j}}-\frac{122}{3
   j}-\frac{122}{3 j^{3/2}}+\frac{595}{9 j^2}+\frac{595}{9 j^{5/2}}\right) \left(\ln \varepsilon  -2 \ln \left(\frac{1+\sqrt{j}}{8 j}\right) + 2 \gamma_\mathrm{E} \right) \Biggr\} \,,\\
\label{seq:iota_hered_inTermsOf_varepsilon_j}
    \iota^\text{hered} &= \varepsilon ^3 \nu  \Biggl\{\frac{74}{15}-\frac{244}{5 j}+\frac{170}{3 j^2}+\ln j  \left(-\frac{37}{5}+\frac{122}{j}-\frac{595}{3 j^2}\right)
     + \lambda_0
   \left(\sqrt{1-j}\right) \left(\frac{32}{j}-\frac{224}{5 j^2}\right) -\frac{32}{5  j} \sqrt{1-j} \,\lambda_0' \! \left(\sqrt{1-j}\right) \Biggr\} \,.
\end{align}
\end{subequations}

\section{Results in terms of $(A, B, C, D_n, F, I_n)$ and $(\mathcal{A},\mathcal{B},\mathcal{C},\mathcal{D}_n)$}
\label{app:ABC}

\subsection{Expressions for $\cA$, $\cB$, $\cC$, and $\cD_n$}
\label{subapp:ABCD_calligraphic}

The coefficients entering the local expression for $p_r^2 = \mathcal{I}(1/r)$  are given here explicitly in terms of energy and angular momentum at 4PN order; see Eqs.~\eqref{eq:cal_I_def} and \eqref{eq:cal_I_of_s_poly} for their definition.

\begin{subequations}
\label{eq:cal_ABCD_inTermsOf_varepsilon_j}
\begin{align}
\label{seq:cal_A_inTermsOf_varepsilon_j}
 \cA &= -c^2 m^2 \varepsilon  \nu ^2 \Biggl\{1+\varepsilon  \left(-\frac{1}{4}+\frac{3}{4} \nu \right)+\varepsilon ^2 \left(-\frac{\nu }{8}+\frac{\nu ^2}{2}\right)+\varepsilon ^3 \left(-\frac{5}{64} \nu ^2+\frac{5}{16} \nu ^3\right)+\varepsilon ^4 \left(-\frac{3}{64} \nu ^3+\frac{3}{16} \nu^4\right)\Biggr\} \\
\label{seq:cal_B_inTermsOf_varepsilon_j}
 \cB &= G m^3 \nu ^2 \Biggl\{1+\varepsilon  \left(-2+\frac{\nu }{2}\right)+\varepsilon ^2 \left(\frac{1}{2}-\frac{\nu }{2}+\frac{\nu ^2}{4}\right)+\varepsilon ^3 \left(-\frac{\nu ^2}{4}+\frac{\nu ^3}{8}\right)+\varepsilon ^4 \left(\frac{\nu ^2}{32}-\frac{\nu ^3}{8}+\frac{\nu ^4}{16}\right)\Biggr\} \\
 \label{seq:cal_C_inTermsOf_varepsilon_j}
 \cC &= -\frac{G^2 j m^4 \nu ^2}{c^2 \varepsilon } \Biggl\{1+\frac{\varepsilon}{j}  (-6-\nu ) +\frac{15 \varepsilon ^2}{2 j} +\frac{\varepsilon ^3}{j} \left(-\frac{15}{8}+\frac{19 \nu }{12}+\frac{\nu ^2}{6}\right)+\frac{\varepsilon ^4}{j} \left(-\frac{103 \nu }{60}+\frac{20 \nu ^2}{3}+\frac{53 \nu
   ^3}{60}\right)\Biggr\} \\
 \label{seq:cal_D1_inTermsOf_varepsilon_j}
 \cD_1 &= \frac{G^3 j m^5 \nu ^2}{c^4 \varepsilon } \Biggl\{-\nu +\varepsilon  \left(\frac{\nu }{2}+\frac{\nu ^2}{2}+\frac{1}{j}\left[\frac{17}{2}+\frac{5 \nu }{2}+2 \nu ^2\right]\right)+\varepsilon ^2 \left(-\frac{\nu ^2}{8}+\frac{1}{j}\left[-9+\left(\frac{53}{4}-\frac{\pi
   ^2}{16}\right) \nu +\frac{\nu ^2}{6}-\nu ^3\right]\right) \nonumber\\
   & \qquad\qquad+\varepsilon ^3 \left(\frac{5}{16} \nu^3 +\frac{1}{j}\left[\frac{9}{4}+ \nu \left(-\frac{29653}{480}+\frac{1385}{1024} \pi ^2 \right) + \nu ^2 \left(\frac{23011}{3600}+\frac{12823}{2048}\pi^2\right) +\frac{209}{80} \nu ^3\right]\right)\Biggr\} \\
   \label{seq:cal_D2_inTermsOf_varepsilon_j}
 \cD_2 &= \frac{G^4 j m^6 \nu ^2}{c^6 \varepsilon } \Biggl\{-\nu -3 \nu ^2+\varepsilon  \left(-\frac{79}{24} \nu -\frac{19}{12} \nu ^2 +3 \nu ^3+\frac{1}{j}\left[8+\left(-\frac{415}{12}+\frac{23}{16} \pi ^2 \right) \nu +\frac{25}{12} \nu ^2 +5 \nu
   ^3\right]\right) \nonumber\\
   & \qquad\qquad\qquad+\varepsilon^2 \left(\frac{2057}{480} \nu  -\frac{923}{48} \nu^2 +\frac{509}{240} \nu ^3 -\frac{3}{4} \nu^4 \right. \nonumber\\
   & \left. \qquad \qquad\qquad\qquad\qquad+\frac{1}{j}\left[-\frac{129}{16}+\nu \left(\frac{4966451}{14400}-\frac{86677}{6144} \pi^2\right) 
   + \nu ^2\left(\frac{368869}{7200}-\frac{103557}{4096} \pi^2 \right) -\frac{11}{10} \nu ^3 - 5 \nu ^4\right]\right)\Biggr\}  \\
   \label{seq:cal_D3_inTermsOf_varepsilon_j}
 \cD_3 &= \frac{G^5 j^2 m^7 \nu ^2}{c^8 \varepsilon ^2} \Biggl\{\frac{3}{4} \nu ^2+\varepsilon  \left(-\frac{9}{8} \nu ^3+\frac{1}{j}\left[\nu\left(\frac{35}{24}-\frac{3}{32} \pi ^2\right)  +\frac{17}{3}\nu ^2 -10 \nu ^3\right]\right) \nonumber\\
 &\qquad\qquad +\varepsilon ^2 \left(\frac{9 \nu^4}{16} +\frac{1}{j}\left[\nu \left(-\frac{203867}{4800}+\frac{3957}{2048} \pi^2\right)  +\nu^2 \left(-\frac{31133}{480}+\frac{54219}{4096}\pi^2\right) -\frac{2197}{80} \nu^3 + 15 \nu ^4\right] \right.\nonumber\\
 & \left.\qquad\qquad\qquad\qquad  +\frac{1}{j^2}\left[6+ \nu\left(-\frac{2568557}{7200}+\frac{44539}{3072} \pi ^2\right)  + \nu^2 \left(-\frac{65051}{1200}+\frac{143311}{6144}\pi ^2\right) -\frac{41}{60} \nu ^3 +14 \nu^4\right]\right)\Biggr\}  \\
   \label{seq:cal_D4_inTermsOf_varepsilon_j}
 \cD_4 &= \frac{G^6 j^2 m^8 \nu ^3}{c^{10} \varepsilon ^2} \Biggl\{-\frac{5}{6}-\frac{14}{3} \nu  +5 \nu ^2 +\varepsilon  \left(\frac{799}{160}-\frac{3353}{128} \nu  +\frac{10193}{480} \nu^2 - 10 \nu ^3 \right. \nonumber\\
 & \left. \qquad\qquad\qquad\qquad\qquad\qquad\qquad\quad  +\frac{1}{j}\left[\frac{4586503}{28800}-\frac{79631}{12288}\pi^2 + \nu\left(\frac{328037}{1920}-\frac{546673}{24576} \pi ^2\right) +\frac{5993}{120} \nu ^2 -35 \nu ^3\right]\right)\Biggr\}  \\
   \label{seq:cal_D5_inTermsOf_varepsilon_j}
 \cD_5 &= \frac{G^7 j^3 m^9 \nu ^3}{c^{12} \varepsilon ^3} \Biggl\{ \! -\frac{5}{8}\nu ^2 \! +\varepsilon  \left(\! -\frac{5}{8}\nu ^2+\frac{25}{16} \nu ^3+\frac{1}{j}\!\left[\frac{15541}{640} \!- \frac{375}{4096} \pi^2 \!+ \nu\left(-\frac{7687}{240} \!+\frac{38655}{8192} \pi ^2\right) 
   -\frac{18899}{320} \nu ^2+\frac{105}{4}\nu ^3\right]\right)\Biggr\}  \\
   \label{seq:cal_D6_inTermsOf_varepsilon_j}
 \cD_6 &= \frac{G^8 j^3 m^{10} \nu ^3}{c^{14} \varepsilon ^3} \Biggl\{\frac{369}{80}-\frac{1151}{64} \nu +\frac{9953}{640} \nu^2 -7 \nu ^3\Biggr\} \\
   \label{seq:cal_D7_inTermsOf_varepsilon_j}
 \cD_7 &= \frac{G^9 j^4 m^{11}\nu ^5}{c^{16} \varepsilon ^4} \left(-\frac{35}{128}+\frac{35}{64} \nu \right)  
\end{align}
\end{subequations} 

\subsection{Expressions for $A$, $B$, $C$, $D_n$, $F$, and $I_n$}

The coefficients entering the local expressions for $\dot{r}^2 = \mathcal{R}(1/r)$ and $\dot{\phi} = \mathcal{S}(1/r)$ are given here explicitly in terms of energy and angular momentum at 4PN order; see Eqs.~ \eqref{eq:rDot_phiDot_loc_inTermsOf_r}  for their definition.

\label{subapp:ABCDFI}
\begin{subequations}
\label{eq:ABCDFI_inTermsOf_varepsilon_j}
\begin{align}
\label{seq:A_inTermsOf_varepsilon_j}
    A &= -c^2 \varepsilon  \Biggl\{1+\varepsilon  \left(\frac{3}{4}-\frac{9}{4}\nu\right)+\varepsilon ^2 \left(\frac{1}{2}-\frac{19}{8} \nu+2 \nu
   ^2\right)+\varepsilon ^3 \left(\frac{5}{16}-2 \nu +\frac{211}{64} \nu^2-\frac{7}{8} \nu ^3\right) \nonumber\\
   &\qquad\qquad\quad + \varepsilon ^4 \left(\frac{3}{16}-\frac{3
   \nu }{2}+\frac{117}{32} \nu ^2-\frac{21}{8} \nu ^3+\frac{3}{16} \nu^4\right)\Biggr\} \\
\label{seq:B_inTermsOf_varepsilon_j}
   B &= G m \Biggl\{1+\varepsilon  \left(3-\frac{7}{2} \nu \right)+\varepsilon ^2
   \left(\frac{9}{4}-12 \nu +\frac{21}{4} \nu ^2\right)+\varepsilon ^3
   \left(\frac{3}{2}-\frac{93}{8} \nu +\frac{171}{8} \nu ^2-\frac{35}{8} \nu^3\right) \nonumber\\
   & \qquad\,\,\qquad +\varepsilon ^4 \left(\frac{15}{16}-\frac{75}{8} \nu
   +\frac{1755}{64} \nu ^2-\frac{1431}{64} \nu^3+\frac{35}{16} \nu^4\right)\Biggr\} \\
\label{seq:C_inTermsOf_varepsilon_j}
   C &=  -\frac{G^2 j m^2}{c^2 \varepsilon } \Biggl\{1+\varepsilon  \left(1+\frac{10-5 \nu }{j}-3 \nu
   \right)+\varepsilon ^2 \left(\frac{3}{4}-\frac{15}{4} \nu +\frac{15}{4} \nu^2+\frac{1}{j}\left[\frac{37}{2}-61 \nu +18 \nu ^2\right]\right) \nonumber\\
   & \qquad\qquad\quad +\varepsilon^3
   \left(\frac{1}{2}-\frac{27}{8}\nu +\frac{25}{4} \nu^2-\frac{5}{2} \nu^3+\frac{1}{j}\left[\frac{111}{8}-\frac{517}{4} \nu +\frac{317}{2} \nu ^2 -\frac{57}{2}\nu ^3\right]\right) \nonumber\\
   & \qquad\qquad\quad +\varepsilon ^4 \left(\frac{5}{16}-\frac{21}{8}\nu+\frac{447}{64}\nu^2-\frac{195}{32} \nu ^3+\frac{15}{16} \nu^4+\frac{1}{j}\left[\frac{37}{4}-\frac{883}{8} \nu +\frac{2857}{8} \nu^2-\frac{5821}{24}\nu ^3+26 \nu ^4\right]\right) 
   \Biggr\} \\
\label{seq:D1_inTermsOf_varepsilon_j}
   D_1 &= \frac{G^3 j m^3}{c^4 \varepsilon } \Biggl\{8-3 \nu +\varepsilon  \left(8-\frac{73}{2} \nu +\frac{19}{2}\nu^2 +\frac{1}{j}\left[\frac{53}{2}-\frac{83}{2} \nu +10 \nu^2\right]\right) \nonumber\\
   & \qquad\qquad+\varepsilon ^2 \left(6-\frac{87}{2} \nu +\frac{603}{8} \nu^2-12 \nu ^3+\frac{1}{j}\left[\frac{79}{2}+ \nu\left(-\frac{3563}{12}+\frac{\pi
   ^2}{16}\right) +226 \nu ^2-37 \nu ^3\right]\right)\nonumber\\
   & \qquad\qquad +\varepsilon ^3
   \left(4-\frac{153 \nu }{4}+110 \nu ^2-\frac{1497}{16} \nu ^3 +7 \nu
   ^4 \right.\nonumber\\
   & \qquad\qquad \qquad\ \left.+\frac{1}{j}\left[\frac{237}{8}+  \nu\left(-\frac{47391}{160}-\frac{4059}{1024} \pi^2\right) +  \nu^2 \left(\frac{1364939}{1200}-\frac{39045}{2048}\pi^2\right)-\frac{9201}{16} \nu ^3+60 \nu
   ^4\right]\right)\Biggr\} \\
\label{seq:D2_inTermsOf_varepsilon_j}
   D_2 &= \frac{G^4 j m^4}{c^6 \varepsilon } \Biggl\{-33+33 \nu -6 \nu ^2 \nonumber\\
   &\quad +\varepsilon  \left(-33+\frac{1857}{8}\nu-\frac{603}{4} \nu ^2+15 \nu ^3+\frac{1}{j}\left[-50+ \nu\left(\frac{2381}{12}+\frac{19 \pi ^2}{16}\right)
   -\frac{441}{4} \nu ^2 +16 \nu^3\right]\right)\nonumber\\
   &\quad  +\varepsilon ^2
   \left(-\frac{99}{4}+\frac{7845}{32} \nu -\frac{2547}{4} \nu ^2 + \frac{4789}{16} \nu ^3-\frac{15}{4} \nu^4 \right. \nonumber\\
   & \quad\qquad \left.+\frac{1}{j}\left[-\frac{1059}{16}+ \nu\left(\frac{3604429}{14400}+\frac{156613}{6144} \pi^2\right)   + \nu ^2\left(-\frac{3910231}{2400}+\frac{296181}{4096}  \pi^2 \right) +\frac{2437}{4} \nu ^3 -53 \nu^4\right]\right)\Biggr\} \\
\label{seq:D3_inTermsOf_varepsilon_j}
   D_3 &= \frac{G^5 j^2 m^5}{c^8 \varepsilon ^2} \Biggl\{-\nu -\frac{\nu ^2}{4}+\varepsilon  \left(-\frac{3}{2} \nu +\frac{17}{4} \nu ^2+\frac{27}{8} \nu^3 +\frac{1}{j}\left[94+ \nu\left(-\frac{5477}{24}+\frac{\pi ^2}{32}\right)  +82 \nu
   ^2-2 \nu ^3\right]\right) \nonumber\\
   &\qquad\qquad\quad +\varepsilon ^2 \left(-\frac{3}{2} \nu +\frac{123}{16}\nu^2-\frac{51}{8} \nu ^3-\frac{207}{16} \nu^4 \right.\nonumber\\
   &\qquad\qquad\qquad \left. +\frac{1}{j}\left[94+ \nu\left(-\frac{3612053}{4800}-\frac{14845}{2048} \pi^2\right)  + \nu^2\left(\frac{9266407}{7200}-\frac{157955}{4096} \pi^2\right) -\frac{4531}{16} \nu ^3 -45 \nu^4\right] \right.\nonumber\\
   &\qquad\qquad\qquad \left. +\frac{1}{j^2}\left[\frac{303}{4}+\nu \left(\frac{398869}{1440}-\frac{157231 \pi
   ^2}{3072}\right) + \nu ^2\left(\frac{658717}{720}-\frac{465743 \pi
   ^2}{6144}\right) -\frac{2725}{12} \nu ^3 +10 \nu ^4\right]\right)\Biggr\}\\
\label{seq:D4_inTermsOf_varepsilon_j}
   D_4 &= \frac{G^6 j^2 m^6}{c^{10} \varepsilon ^2} \Biggl\{\frac{71}{4} \nu  + \frac{\nu ^2}{2}-8 \nu ^3 \nonumber\\*
    &\qquad +\varepsilon
    \left(\frac{157}{32} \nu  +\frac{1065}{128}\nu ^2-\frac{2431}{32} \nu ^3 + 66 \nu ^4 \right. \nonumber  \\*
   & \qquad\quad\left.  +\frac{1}{j}\left[-\frac{1669}{8}+\nu \left(\frac{13907891}{28800}+\frac{352829}{12288} \pi^2\right)   + \nu ^2\left(-\frac{27087917}{28800}+\frac{2004947}{24576} \pi^2\right) +\frac{37}{8} \nu ^3 +64 \nu
   ^4\right]\right)\Biggr\}\\
\label{seq:D5_inTermsOf_varepsilon_j}
   D_5 &= \frac{G^7 j^3 m^7 \nu}{c^{12}
   \varepsilon ^3}  \Biggl\{\frac{13}{8}\nu ^2 +\varepsilon  \left(\frac{27}{8}\nu ^2 - \frac{193}{16} \nu ^3 \right. \nonumber\\
   & \qquad\qquad\qquad\qquad\qquad \left. +\frac{1}{j}\left[-\frac{53219}{3200}-\frac{16203}{4096} \pi^2 +\nu \left(\frac{13757}{480}-\frac{178221}{8192}\pi ^2\right) 
   +\frac{12073}{64} \nu^2 - \frac{327}{4}\nu^3 \right]\right)\Biggr\} \\
\label{seq:D6_inTermsOf_varepsilon_j}
   D_6 &= \frac{G^8 j^3 m^8 \nu}{c^{14} \varepsilon ^3} \left\{-\frac{43}{8}+\frac{1077}{32} \nu -\frac{20033}{384} \nu^2 +27 \nu ^3\right\}\\
\label{seq:D7_inTermsOf_varepsilon_j}
   D_7 &= \frac{G^9 j^4 m^9\nu
   ^3}{c^{16} \varepsilon ^4} \left\{\frac{125}{128}-\frac{165}{64} \nu \right\}  \\
\label{seq:F_inTermsOf_varepsilon_j}
   F &= \frac{G m}{c}\sqrt{\frac{j}{\varepsilon}} \Biggl\{1+\varepsilon  \left(\frac{1}{2}-\frac{3}{2}\nu\right)+\varepsilon ^2 \left(\frac{1}{4}-\frac{9}{8} \nu +\frac{3}{4} \nu^2\right) +\varepsilon ^3 \left(\frac{1}{8}-\frac{3}{4} \nu +\frac{17}{16} \nu ^2 -\frac{\nu^3}{8}\right) \nonumber\\
   &\qquad\qquad\qquad\  +\varepsilon ^4 \left(\frac{1}{16}-\frac{15}{32} \nu +\frac{65}{64} \nu ^2-\frac{35}{64} \nu ^3\right) \Biggr\} \\
\label{seq:I1_inTermsOf_varepsilon_j}
   I_1 &= \frac{G^2  m^2}{c^3 }\sqrt{\frac{j}{\varepsilon}} \Biggl\{-4+2 \nu +\varepsilon  \left(-2+11 \nu -4 \nu
   ^2\right)+\varepsilon ^2 \left(-1+\frac{31}{4} \nu -\frac{55}{4} \nu ^2+3
   \nu ^3\right) \nonumber\\
   & \qquad\qquad\qquad +\varepsilon ^3 \left(-\frac{1}{2}+5 \nu -\frac{113}{8} \nu^2+10 \nu ^3-\nu ^4\right)\Biggr\} \\
\label{seq:I2_inTermsOf_varepsilon_j}
   I_2 &= \frac{G^3 \sqrt{j} m^3}{c^5
   \sqrt{\varepsilon }} \Biggl\{\frac{17}{2}-11 \nu +5 \nu ^2+\varepsilon 
   \left(\frac{17}{4}-\frac{327}{8} \nu +\frac{157}{4} \nu ^2 -\frac{25}{2} \nu^3\right)\nonumber\\
   & \qquad\qquad\qquad +\varepsilon ^2 \left(\frac{17}{8}-\frac{881}{32} \nu +99 \nu^2-\frac{3061}{48} \nu^3 +\frac{25}{2} \nu^4\right)\Biggr\} \\
\label{seq:I3_inTermsOf_varepsilon_j}
   I_3 &= \frac{G^4 j^{3/2} m^4}{c^7 \varepsilon ^{3/2}} \Biggl\{\frac{\nu }{2}-\nu ^2+\varepsilon  \left(\frac{\nu
   }{2}-2 \nu ^2+3 \nu ^3+\frac{1}{j}\left[-13+ \nu \left(\frac{1183}{24}+\frac{\pi^2}{32}\right) -\frac{73}{2}\nu^2 + 14 \nu ^3\right]\right) \nonumber\\
   & \qquad\qquad\qquad+\varepsilon^2
   \left(\frac{3}{8} \nu  - \frac{17}{8} \nu^2 +\frac{9}{2} \nu^3 -\frac{15}{4} \nu^4\right. \nonumber\\
   &\qquad\qquad\qquad\quad \left.+\frac{1}{j}\left[-\frac{13}{2}+ \nu \left(\frac{79369}{1600}+\frac{1551}{2048} \pi^2\right)+ \nu^2\left(-\frac{1200311}{7200}-\frac{2735}{4096} \pi^2\right) +\frac{7741}{48} \nu^3 -42 \nu^4\right]\right)\Biggr\} \\
\label{seq:I4_inTermsOf_varepsilon_j}
   I_4 &= \frac{G^5 j^{3/2} m^5}{c^9 \varepsilon ^{3/2}} \Biggl\{-\frac{45}{8}\nu  + \frac{45}{4} \nu ^2 -7 \nu
   ^3 \nonumber\\*
   &\qquad\qquad +\varepsilon  \left(-\frac{63}{8} \nu +\frac{2761}{64} \nu ^2 - \frac{3035}{48} \nu ^3 +\frac{49}{2}\nu^4\right.\nonumber\\*
   &\qquad\qquad\qquad  \left. +\frac{1}{j}\left[\frac{259}{16}+ \nu \left(-\frac{41689}{3200}-\frac{30729 \pi
   ^2}{4096}\right) + \nu ^2\left(\frac{1843321}{28800}-\frac{76973}{24576}\pi^2 \right) -\frac{3341}{24} \nu ^3+42 \nu^4\right]\right)\Biggr\} \\
\label{seq:I5_inTermsOf_varepsilon_j}
   I_5 &= \frac{G^6 j^{5/2} m^6 \nu }{c^{11} \varepsilon
   ^{5/2}} \Biggl\{\frac{3}{4} \nu ^2+\varepsilon  \left(\frac{27}{16}\nu ^2 -3 \nu ^3 \right. \nonumber\\*
   & \qquad\qquad\qquad\qquad\qquad\quad \left.+\frac{1}{j}\left[-\frac{43897}{800}+\frac{1083}{512} \pi^2+ \nu\left(-\frac{4361}{96}
+\frac{3891}{1024} \pi ^2\right) 
   +\frac{10727}{96} \nu ^2 - 36 \nu ^3\right]\right)\Biggr\} \\
\label{seq:I6_inTermsOf_varepsilon_j}
   I_6 &= \frac{G^7 j^{5/2} m^7 \nu}{c^{13} \varepsilon ^{5/2}}  \left\{-\frac{283}{32}+\frac{3585}{128} \nu -\frac{9281}{384} \nu ^2 +9 \nu ^3\right\} \\
\label{seq:I7_inTermsOf_varepsilon_j}
   I_7 &= \frac{15 G^8 j^{7/2} m^8\nu ^3}{32 c^{15  }
   \varepsilon ^{7/2}} \left\{1-\frac{4}{3}\nu \right\}
\end{align}
\end{subequations}

\subsection{$P^\text{loc}$ and $K^\text{loc}$ in terms of $A$, $B$, $C$, $D_n$, $F$, and $I_n$}
\label{subapp:P_K_of_ABC}

The local pieces of the radial period $P^\text{loc}$ and the periastron advance $K^\text{loc}$ are  expressed in terms of 
 $A$, $B$, $C$, $D_n$, $F$, and $I_n$ hereafter.
The expression for these coefficients are given explicitly in~Sec.~\ref{subapp:ABCDFI}. Through the choice of factorization, this result is organized by PN orders for legibility; note that the 1PN coefficient of $P^\text{loc}$ in terms of these coefficients is vanishing. The expression for $I_r^\text{loc}$ in terms of $(\mathcal{A},\mathcal{B},\mathcal{C},\mathcal{D}_n)$ was given in the main text; see Eq.~\eqref{eq:Irloc_inTermsOf_ABCD}.

\begin{subequations}
\label{eq:P_K_loc_inTermsOf_ABCDFI}
\begin{align}
\label{seq:P_loc_inTermsOf_ABCDFI}
\frac{P^{\text{loc}}}{2\pi} &=   \frac{B}{(-A)^{3/2}}+\frac{3 D_1^2-4 C D_2+12 B D_3}{8 (-C)^{5/2}} \nonumber\\
& \quad +\frac{1}{16 (-C)^{9/2}}\biggl[35 B D_1^3-60 B C D_1 D_2+210 B^2 D_1 D_3-30 A C D_1 D_3  \nonumber\\
& \qquad\qquad\quad\qquad -60 B^2 C D_4+12 A C^2 D_4+140 B^3 D_5-60 A B C D_5\biggr]  \nonumber\\
& \quad +\frac{3}{256 (-C)^{13/2}}\biggl[1155 B^2 D_1^4-105 A C D_1^4-2520 B^2 C D_1^2 D_2+280 A C^2 D_1^2 D_2+560 B^2 C^2 D_2^2-80 A C^3 D_2^2 \nonumber\\
& \qquad\qquad\quad\qquad +9240 B^3 D_1^2 D_3-2520 A B C D_1^2 D_3-3360 B^3 C D_2 D_3+1120 A B C^2 D_2 D_3+4620 B^4
   D_3^2 \nonumber\\
& \qquad\qquad\quad\qquad -2520 A B^2 C D_3^2+140 A^2 C^2 D_3^2-3360 B^3 C D_1 D_4+1120 A B C^2 D_1 D_4+9240 B^4 D_1 D_5 \nonumber\\
& \qquad\qquad\quad\qquad -5040 A B^2 C D_1 D_5+280 A^2 C^2 D_1 D_5-1680 B^4 C D_6+1120 A B^2 C^2 D_6-80 A^2 C^3 D_6\nonumber\\
& \qquad\qquad\quad\qquad +3696 B^5
   D_7-3360 A B^3 C D_7+560 A^2 B C^2 D_7\biggr]  \\
\label{seq:K_loc_inTermsOf_ABCDFI}
K^{\text{loc}} &= \frac{F}{\sqrt{-C}}+\frac{B (3 D_1  F-2 C I_1)}{2 (-C)^{5/2}}\nonumber\\
& \quad +\frac{1}{16 (-C)^{9/2}}\biggl[105 B^2 D_1 ^2 F-15 A C D_1 ^2 F-60 B^2
   C D_2 F+12 A C^2 D_2 F+140 B^3 D_3 F-60 A B C
   D_3 F \nonumber\\*
   &\qquad\qquad\qquad\quad -60 B^2 C D_1  I_1+12 A C^2 D_1 
   I_1+24 B^2 C^2 I_2-8 A C^3 I_2-40 B^3 C
   I_3+24 A B C^2 I_3 \biggr] \nonumber\\
& \quad +\frac{1}{32 (-C)^{13/2}}\biggl[1155 B^3
   D_1 ^3 F-315 A B C D_1 ^3 F-1260 B^3 C D_1  D_2
   F+420 A B C^2 D_1  D_2 F+3465 B^4 D_1  D_3
   F \nonumber\\*
   &\qquad\qquad\qquad\quad -1890 A B^2 C D_1  D_3 F +105 A^2 C^2 D_1  D_3
   F-630 B^4 C D_4 F+420 A B^2 C^2 D_4 F-30 A^2 C^3
   D_4 F \nonumber\\*
   &\qquad\qquad\qquad\quad +1386 B^5 D_5 F-1260 A B^3 C D_5 F+210 A^2 B
   C^2 D_5 F-630 B^3 C D_1 ^2 I_1+210 A B C^2
   D_1 ^2 I_1 \nonumber\\*
   &\qquad\qquad\qquad\quad +280 B^3 C^2 D_2 I_1-120 A B C^3
   D_2 I_1-630 B^4 C D_3 I_1+420 A B^2 C^2
   D_3 I_1-30 A^2 C^3 D_3 I_1 \nonumber\\*
   &\qquad\qquad\qquad\quad +280 B^3 C^2
   D_1  I_2-120 A B C^3 D_1  I_2-630 B^4 C
   D_1  I_3+420 A B^2 C^2 D_1  I_3-30 A^2 C^3
   D_1  I_3 \nonumber\\*
   &\qquad\qquad\qquad\quad +140 B^4 C^2 I_4-120 A B^2 C^3 I_4+12
   A^2 C^4 I_4-252 B^5 C I_5+280 A B^3 C^2 I_5-60 A^2
   B C^3 I_5 \biggr] \nonumber\\
& \quad +\frac{1}{1024 (-C)^{17/2}}\biggl[225225 B^4 D_1 ^4
   F-90090 A B^2 C D_1 ^4 F+3465 A^2 C^2 D_1 ^4 F-360360 B^4
   C D_1 ^2 D_2 F \nonumber\\*
   &\qquad\qquad\qquad\quad +166320 A B^2 C^2 D_1 ^2 D_2
   F-7560 A^2 C^3 D_1 ^2 D_2 F+55440 B^4 C^2 D_2^2
   F-30240 A B^2 C^3 D_2^2 F\nonumber\\*
   &\qquad\qquad\qquad\quad  +1680 A^2 C^4 D_2^2 F+1081080
   B^5 D_1 ^2 D_3 F-720720 A B^3 C D_1 ^2 D_3
   F+83160 A^2 B C^2 D_1 ^2 D_3 F \nonumber\\*
   &\qquad\qquad\qquad\quad -288288 B^5 C D_2
   D_3 F+221760 A B^3 C^2 D_2 D_3 F-30240 A^2 B C^3
   D_2 D_3 F+360360 B^6 D_3^2 F \nonumber\\*
   &\qquad\qquad\qquad\quad -360360 A B^4 C
   D_3^2 F+83160 A^2 B^2 C^2 D_3^2 F-2520 A^3 C^3
   D_3^2 F-288288 B^5 C D_1  D_4 F \nonumber\\*
   &\qquad\qquad\qquad\quad+221760 A B^3 C^2
   D_1  D_4 F-30240 A^2 B C^3 D_1  D_4 F+720720
   B^6 D_1  D_5 F-720720 A B^4 C D_1  D_5
   F \nonumber\\*
   &\qquad\qquad\qquad\quad +166320 A^2 B^2 C^2 D_1  D_5 F-5040 A^3 C^3 D_1 
   D_5 F-96096 B^6 C D_6 F+110880 A B^4 C^2 D_6
   F \nonumber\\*
   &\qquad\qquad\qquad\quad -30240 A^2 B^2 C^3 D_6 F+1120 A^3 C^4 D_6 F+205920 B^7
   D_7 F-288288 A B^5 C D_7 F \nonumber\\*
   &\qquad\qquad\qquad\quad +110880 A^2 B^3 C^2 D_7
   F-10080 A^3 B C^3 D_7 F-120120 B^4 C D_1 ^3
   I_1+55440 A B^2 C^2 D_1 ^3 I_1\nonumber\\*
   &\qquad\qquad\qquad\quad-2520 A^2 C^3
   D_1 ^3 I_1+110880 B^4 C^2 D_1  D_2
   I_1-60480 A B^2 C^3 D_1  D_2 I_1+3360 A^2 C^4
   D_1  D_2 I_1\nonumber\\*
   &\qquad\qquad\qquad\quad-288288 B^5 C D_1  D_3
   I_1+221760 A B^3 C^2 D_1  D_3 I_1-30240 A^2 B
   C^3 D_1  D_3 I_1+44352 B^5 C^2 D_4
   I_1  \nonumber\\*
   &\qquad\qquad\qquad\quad -40320 A B^3 C^3 D_4 I_1+6720 A^2 B C^4
   D_4 I_1-96096 B^6 C D_5 I_1+110880 A B^4 C^2
   D_5 I_1 \nonumber\\*
   &\qquad\qquad\qquad\quad -30240 A^2 B^2 C^3 D_5 I_1+1120 A^3
   C^4 D_5 I_1+55440 B^4 C^2 D_1 ^2 I_2-30240 A
   B^2 C^3 D_1 ^2 I_2 \nonumber\\*
   &\qquad\qquad\qquad\quad +1680 A^2 C^4 D_1 ^2
   I_2-20160 B^4 C^3 D_2 I_2+13440 A B^2 C^4
   D_2 I_2-960 A^2 C^5 D_2 I_2+44352 B^5 C^2
   D_3 I_2 \nonumber\\*
   &\qquad\qquad\qquad\quad-40320 A B^3 C^3 D_3 I_2+6720 A^2 B
   C^4 D_3 I_2-144144 B^5 C D_1 ^2 I_3+110880 A
   B^3 C^2 D_1 ^2 I_3\nonumber\\*
   &\qquad\qquad\qquad\quad -15120 A^2 B C^3 D_1 ^2
   I_3+44352 B^5 C^2 D_2 I_3-40320 A B^3 C^3
   D_2 I_3+6720 A^2 B C^4 D_2 I_3\nonumber\\*
   &\qquad\qquad\qquad\quad -96096 B^6 C
   D_3 I_3+110880 A B^4 C^2 D_3 I_3-30240 A^2
   B^2 C^3 D_3 I_3+1120 A^3 C^4 D_3 I_3 \nonumber\\*
   &\qquad\qquad\qquad\quad+44352
   B^5 C^2 D_1  I_4-40320 A B^3 C^3 D_1  I_4+6720
   A^2 B C^4 D_1  I_4-96096 B^6 C D_1  I_5 \nonumber\\*
   &\qquad\qquad\qquad\quad+110880
   A B^4 C^2 D_1  I_5-30240 A^2 B^2 C^3 D_1 
   I_5+1120 A^3 C^4 D_1  I_5+14784 B^6 C^2
   I_6 \nonumber\\*
   &\qquad\qquad\qquad\quad -20160 A B^4 C^3 I_6 +6720 A^2 B^2 C^4 I_6-320
   A^3 C^5 I_6-27456 B^7 C I_7 \nonumber\\*
   &\qquad\qquad\qquad\quad +44352 A B^5 C^2
   I_7-20160 A^2 B^3 C^3 I_7+2240 A^3 B C^4 I_7\biggr]
\end{align}
\end{subequations}


\bibliography{references.bib}

\end{document}